\documentclass[12pt]{iopart}

\usepackage{iopams}
\usepackage{cite}
\usepackage{color}
\expandafter\let\csname equation*\endcsname\relax
\expandafter\let\csname endequation*\endcsname\relax
\usepackage{amsmath}
\usepackage[colorlinks=true,pdfstartview=FitV,linkcolor=black,citecolor=black,urlcolor=blue]{hyperref}
\usepackage{soul}
\usepackage{bm}
\usepackage{bbold}
\usepackage{graphicx}
\usepackage{tcolorbox}
\usepackage{subcaption}
\usepackage{float}
\usepackage{graphics} 

\newcommand{\be}{\begin{equation}}
\newcommand{\ee}{\end{equation}}
\newcommand{\ba}{\begin{align}}
\newcommand{\ea}{\end{align}}
\def\im{\mathrm{i}}
\def\ud{\mathrm{d}}
\def\ex{\mathrm{e}}
\def\w{\omega}

\def\hc{\text{h.c.}}

\newcommand{\alphaBold}{{\bm \alpha}}
\newcommand{\OmegabBold}{{\bm \bar{\Omega}}}
\newcommand{\DBold}{{\bm D}}
\newcommand{\dBold}{{\bm d}}
\newcommand{\PiBold}{{\bm \Pi}}
\newcommand{\rhoBold}{{\bm \rho}}

\newcommand{\SigmaBold}{{\bm \Sigma}}

\newcommand{\pb}{{\bm{p}}}
\newcommand{\qb}{{\bm{q}}}

\newcommand{\lb}{{\bf{l}}}

\newcommand{\sbar}{{\bar{s}}}

\newcommand{\rb}{{\bar{r}}}

\newcommand{\Gcal}{\mathcal{G}}
\newcommand{\Gcalh}{{G}}

\newcommand{\Sigmacal}{\mathit{\Sigma}}

\renewcommand{\S}{\Sigmacal}

\newcommand{\HF}{{\rm{HF}}}

\newcommand{\Dcal}{\mathcal{D}}

\newcommand{\taub}{\bar{\tau}}

\newcommand{\phih}{\hat{\phi}}

\newcommand{\uh}{\hat{u}}
\newcommand{\ph}{\hat{p}}

\newcommand{\dif}{{\rm d}}
\newcommand{\kb}{\bm{k}}

\newcommand{\mub}{{\bar{\mu}}}
\newcommand{\nub}{{\bar{\nu}}}

\newcommand{\oneb}{{\bf 1}}
\newcommand{\Eq}[1]{Eq.~\eqref{#1}}

\newcommand{\zb}{\bar{z}}

\newcommand{\tb}{\bar{t}}

\newcommand{\alphab}{{\bar{\alpha}}}
\renewcommand{\alphab}{{\bm \alpha}}

\renewcommand\Re{\operatorname{Re}}
\renewcommand\Im{\operatorname{Im}}

\renewcommand{\kb}{{\bm{k}}}
\renewcommand{\lb}{{\bm{l}}}
\newcommand*\tcircle[1]{
  \raisebox{-0.5pt}{
    \textcircled{\fontsize{7pt}{0}\fontfamily{phv}\selectfont #1}
  }
}

\begin{document}

\title[A many-body approach to transport in quantum systems]{A many-body approach to transport in quantum systems: From the transient regime to the stationary state}

\author{M Ridley$^{1,2}$, N W Talarico$^{3,4}$, D Karlsson$^5$, N Lo Gullo$^{6,7}$, and R Tuovinen$^{8,9}$}
\address{$^1$ School of Physics and Astronomy, Tel-Aviv University, Tel-Aviv 69978, Israel}
\address{$^2$ Quantum Research Centre, Technology Innovation Institute, Abu Dhabi, UAE}
\address{$^3$ QTF Centre of Excellence, Center for Quantum Engineering, Department of Applied Physics, Aalto University School of Science, FIN-00076 Aalto, Finland}
\address{$^4$ Algorithmiq Ltd, Kanavakatu 3c, 00160 Helsinki, Finland}
\address{$^5$ Department of Physics, Nanoscience Center, University of Jyv{\"a}skyl{\"a}, 40014 Jyv{\"a}skyl{\"a}, Finland}
\address{$^6$ CSC-IT Center for Science, P.O. Box 405, 02101 Espoo, Finland}
\address{$^7$ Quantum algorithms and software, VTT Technical Research Centre of Finland Ltd}
\address{$^8$ QTF Centre of Excellence, Turku Centre for Quantum Physics, Department of Physics and Astronomy, University of Turku, 20014 Turku, Finland}
\address{$^9$ QTF Centre of Excellence, Department of Physics, P.O. Box 64, 00014 University of Helsinki, Finland}

\ead{riku.tuovinen@helsinki.fi}

\begin{abstract}
We review one of the most versatile theoretical approaches to the study of time-dependent correlated quantum transport in nano-systems: the non-equilibrium Green's function (NEGF) formalism.  Within this formalism, one can treat, on the same footing, inter-particle interactions, external drives and/or perturbations, and coupling to baths with a (piece-wise) continuum set of degrees of freedom. After a historical overview on the theory of transport in quantum systems, we present a modern introduction of the NEGF approach to quantum transport. We discuss the inclusion of inter-particle interactions using diagrammatic techniques, and the use of the so-called embedding and inbedding techniques which take the bath couplings into account non-perturbatively. In various limits, such as the non-interacting limit and the steady-state limit, we then show how the NEGF formalism elegantly reduces to well-known formulae in quantum transport as special cases. We then discuss non-equilibrium transport in general, for both particle and energy currents. Under the presence of a time-dependent drive -- encompassing pump--probe scenarios as well as driven quantum systems -- we discuss the transient as well as asymptotic behavior, and also how to use NEGF to infer information on the out-of-equilibrium system. As illustrative examples, we consider model systems general enough to pave the way to realistic systems. These examples encompass one- and two-dimensional electronic systems, systems with electron--phonon couplings, topological superconductors, and optically responsive molecular junctions where electron--photon couplings are relevant.
\end{abstract}

%
\vspace{2pc}
\noindent{\it Keywords}: quantum transport, many-body correlation, non-equilibrium Green's function

\submitto{\jpa}
%
%
%

\setcounter{secnumdepth}{2}
\setcounter{tocdepth}{2}
\tableofcontents
\markboth{A many-body approach to transport in quantum systems}{}

\section{Introduction}\label{sec:intro}

The study of quantum transport describes the dynamics of energy and charge in constricted systems. Inherently, the studied phenomena are time dependent; the systems are far from equilibrium, with no guarantee to relax to stationary-like states on negligible time scales. Starting from the proposition of a single organic molecule functioning as a one-way conductor of electric current~\cite{Aviram1974}, the technological advances in electronic components and devices, following roughly the prediction by Moore~\cite{Moore1965}, ultimately leads to the present day hot topic of nanotechnology~\cite{Vance2015}. In particular, quantum transport from atomic to nanometer scale systems has been under intensive research, both experimental and theoretical, for the past couple of decades~\cite{Reed1997, Smit2002, Cuniberti2006, di2008electrical, Dubi2011, Thoss2018, Cohen2020}. In addition to the massive speed-up in processing power that arises from reduced spatial dimensions, molecular junctions can produce a massive speedup in clock speed by harnessing THz intramolecular transport processes~\cite{wang_-chip_2018, blair_clock_2018}. The advanced temporal resolution of molecular devices also has applications in frequency doublers and detectors~\cite{iniguez-de-la-torre_three-terminal_2010, cheng_graphene_2017} and switches for fast memory storage~\cite{zhang_photoinduced_2012, guo_actively_2018, behnia_light-driven_2020, reimann_two-dimensional_2021}. Experimental advances in miniaturization of integrated devices in electrical circuits, potentially utilizing the next-generation nanoscale components operational in the GHz--THz regime, have prompted remarkable interest in the theoretical description of these systems. From a technological perspective, the semiconductor and microelectronics industry is currently at a unique crossroads, where the standard Moore's law is indeed fast approaching the atomic limit in dimensions~\cite{Sun2014}, where quantum effects, e.g., direct source--drain electron tunneling makes reliable fabrication of sub-$20$~nm devices progressively more difficult or even impossible~\cite{Loertscher2013,khan_science_2018}. These effects are becoming particularly more relevant as state-of-the-art transport measurements are already able to access these processes at the atomic scale~\cite{Ruby2015, Groenendijk2017, Llinas2017, Drost2017, Farinacci2018, Malavolti2018, Rizzo2018, Choi2019, Khajetoorians2019, Liebhaber2020, Mishra2020, Peters2020} and in real-time with the temporal resolution being brought down to the sub-picosecond regime~\cite{Prechtel2012, Cocker2013, Hunter2015, Rashidi2016, Cocker2016, Arielly2017, Jelic2017, McIver2020, delaTorre2021}.

Due to very nature of these nanoscale devices, simulating their correlated quantum dynamics is a challenging task. It requires a proper and simultaneous treatment of strong correlations and coupling, non-equilibrium phenomena particularly in the transient regime, and strong external perturbations. In principle, we \emph{only} need to solve the time-dependent Schr{\"o}dinger equation describing the dynamical evolution of the entire system of interest. In practice, however, this is not possible since the number of equations to be solved, even numerically by using highly efficient supercomputers, is simply out of reach. Nonetheless, many interesting properties of many-particle systems being observed in experiments involve observables related to only a few particles such as densities and currents. Evaluating these quantities from the underlying theory to be compared with experiments could be possible in terms of some \emph{reduced quantity}. Then, we are able to theoretically describe an observable and predict its properties by looking at less complicated objects than, e.g., the full and complicated state vector or the wave function for every particle of the system.

What is then a suitable reduced quantity? To approach this question, the research field of quantum transport has matured significantly during the past couple of decades. Today there are several theoretical methodologies constructed on different reduced quantities, each with its own communities and nomenclatures, which may result in somewhat unclear connections between them. Our goal here is to review one of the most versatile theoretical approaches to the study of time-dependent correlated quantum transport in nanosystems: the non-equilibrium Green's function (NEGF) formalism. Within this formalism, one can treat, on an equal footing, inter-particle interactions, external drives and/or perturbations, and coupling to baths with infinitely many degrees of freedom. Importantly,
we are able to see in a transparent way how the NEGF approach connects to different theoretical methods commonly used for simulating quantum transport. In various limits, such as the non-interacting limit and the steady-state limit, the NEGF formalism elegantly reduces to well-known formulae in quantum transport as special cases.

In this introductory section we will first outline the key milestones in the development of quantum transport theory, particularly focusing on the time-resolved approaches. This will bring our focus onto the NEGF approach and on the most important achievements in its development. We will close the section by describing our plan and logical flow for the rest of this review.

\subsection{Historical overview on quantum transport theories in the transient regime}\label{sec:history}

The formalism developed by Landauer~\cite{Landauer1957} and B{\"u}ttiker~\cite{Buttiker1986} provides an intuitive physical framework of the particle current flowing through a junction composed of leads and a central conducting device. 
The key ingredient is the current $I_{\alpha\beta}$ flowing from the lead $\alpha$ to the lead $\beta$ which is computed by summing over all possible {\it available} scattering states of the central region which have a finite overlap with both leads. This information is embodied by the transmission function which carries information about the spectral function of the central region as well as how it is coupled to the leads. To compute the particle current into the lead $\alpha$ we then need to sum over all possible $\beta\neq\alpha$ leads the contribution $I_{\beta\alpha}-I_{\alpha\beta}$ which accounts for all incoming and outgoing particles into/from the lead $\alpha$.
The final result is the celebrated, and broadly applicable, Landauer--B{\"u}ttiker formula for the current $I_\alpha$ in the lead $\alpha$ for a system in a \emph{stationary} state. Going beyond the first cumulant of the current, the Landauer--B{\"u}ttiker framework was extended to an $S$-matrix theory for the fluctuations of the quantum current, described in terms of the correlator $C_{\alpha\beta}=\left\langle \Delta I_{\alpha}\Delta I_{\beta}\right\rangle$, where $\Delta I_{\alpha}$ is the deviation of the current in lead $\alpha$ from its mean value \cite{buttiker_scattering_1990,landauer_noise_1998,blanter_shot_2000}. The resulting theory of quantum shot and thermal noise reveals additional physical information not contained in the current, allowing for the determination of effective quasiparticle charges \cite{beenakker_quantum_2003,kumar_detection_2012}, transmission probabilities \cite{djukic_shot_2006}, absorption/emission spectra \cite{onac_using_2006} and high sensitivity to electronic correlations \cite{senkpiel_dynamical_2020}. Again, this formalism is valid for static systems in the steady state regime.

However, typical nanoscale electronic devices are operating at high frequencies (THz) so the systems do not necessarily relax to a stationary configuration instantly. In contrast, there are transient effects depending on, e.g., the system's geometry or topological character~\cite{Khosravi2009, Vieira2009, Perfetto2010, Rocha2015, Wang2015, Tuovinen2019NJP} its predisposition to external perturbations or thermal gradients~\cite{Kurth2010, Foieri2010, Ness2011, Arrachea2012, Eich2016, Tuovinen2016PRB, Covito2018} and the physical properties of the transported quanta and their mutual interactions~\cite{Wijewardane2005, Verdozzi2006, Myohanen2008, Uimonen2011, Myohanen2012, Latini2014, Talarico2019, Talarico2020, Tuovinen2020}. The determination of the AC current response to an external periodic electromagnetic field is thus another important subfield in quantum transport. In the process known as photon-assisted tunneling (PAT), first described theoretically by Tien and Gordon \cite{tien1963multiphoton} irradiated tunnel junctions acquire additional peaks in their conductance \cite{shibata2012photon} and noise \cite{schoelkopf1998observation} spectra. PAT has been experimentally demonstrated as an additional transport channel in a variety of structures, beginning with tunneling between oxide films in superconductors in 1962 \cite{dayem1962quantum}. PAT through quantum dots formed by small conducting subsystems in semiconductor heterostructures has also been observed \cite{platero2004photon}, in frequency regimes again typically lying somewhere in the microwave \cite{blick1995photon,oosterkamp1998microwave} to infrared \cite{drexler1995photon} region. 

Based on the time-dependent Schr{\"o}dinger equation, Caroli et al. presented in 1971 a microscopic derivation of a tunneling current in a transport setup~\cite{Caroli1971a,Caroli1971b}. In these works, the leads are considered to be initially disconnected from the central conducting device and at equilibrium at different chemical potentials. Then the contact is switched on suddenly, and the Landauer--B{\"u}ttiker formula is recovered as the long-time limit, $t\to\infty$, of the expectation value of the current operator. As this approach of suddenly switching on a contact in a junction was too much of an idealization of a real setup, in 1980 Cini proposed an alternative approach~\cite{Cini1980}, where the whole system is considered to be initially contacted and in equilibrium at a given chemical potential. Then the system is driven out of equilibrium by an applied bias voltage (difference of chemical potential) to the leads. Interestingly, even if the initial setups in these two approaches are different, the same Landauer--B{\"u}ttiker formula is recovered in the long-time limit since the way the system is prepared does not affect the stationary properties -- a key finding which was addressed formally only later in 2004 by Stefanucci and Almbladh~\cite{Stefanucci2004}.

The Landauer--B{\"u}ttiker formalism may also be derived from the perspective of non-equilibrium Green's functions (NEGF), as was shown by Meir and Wingreen in 1992~\cite{Meir1992}. This mathematical tool when applied to quantum transport in multi-terminal junctions provides a natural framework to calculate the current \emph{at all times}, and it is not limited to the stationary state. This results in several studies of generalizing the Landauer--B{\"u}ttiker formula to transient regime: Around the same time as the work of Meir and Wingreen, in 1991, Pastawski derived a formula using NEGFs for the time-dependent tunneling current using the partitioned approach of Caroli's in the linear response and adiabatic regime~\cite{Pastawski1991}. Also, in 1994 a calculation of the time-dependent tunneling current was done by Jauho et al. where the partitioned approach 
by Caroli et al.~\cite{Caroli1971a,Caroli1971b}
was also used to write the tunneling current as a double integral over time and energy; the corresponding integrand consists of a combination of Green's functions in the central region~\cite{Jauho1994}. Later, in 2004, Stefanucci and Almbladh further demonstrated that when the leads are described within wide-band approximation and when the central conducting device consists of only a single level, it is possible to perform the time integral analytically and obtain an explicit time-dependent extension to the Landauer--B{\"u}ttiker formula~\cite{Stefanucci2004}. In addition, they used the partition-free approach introduced by Cini, thus confirming the loss of memory of the initial preparation in the steady state. A step forward in deriving a time-dependent Landauer--B{\"u}ttiker formula for arbitrary junctions was done by Perfetto et al. in 2008 where also the spin was added to the single-level junction~\cite{Perfetto2008}. It turns out this logic could be extended even further, and some of the present authors have further contributed to the development of time-dependent Landauer--B{\"u}ttiker (TD-LB) formalism for arbitrary junction shapes and sizes~\cite{Stefanucci2013Book, Tuovinen2013, Tuovinen2014}, time-dependent driving mechanisms~\cite{Ridley2015, Ridley2016, Ridley2016JPC, Ridley2017GNR}, current-correlations~\cite{Ridley2017,Ridley2019Entropy}, thermoelectric transport~\cite{Eich2016, Covito2018}, phononic heat transport~\cite{Tuovinen2016}, and superconducting junctions~\cite{Tuovinen2016PNGF,Tuovinen2019NJP}.

While, arguably the most used technique in time-dependent quantum transport is the NEGF formalism with its natural connection to the traditional Landauer approach, let us briefly comment on other theoretical approaches and computational methods for this task. The time-dependent density in a system of interest may be evaluated from the time-dependent Kohn--Sham equation according to the Runge--Gross theorem~\cite{KohnSham1965, RungeGross1984}. From the Kohn--Sham orbitals it is possible to define a (Kohn--Sham) current density which is equivalent to the true current density when considered via a surface integral relating to the system's geometry~\cite{Stefanucci2004EPL, Kurth2005, Kwok2014}. This is the essential starting point for a partition-free scheme 
(reminiscent of Cini's work~\cite{Cini1980})
based on the time-dependent density-functional theory (TD-DFT) to treat the time-dependent current response, even in fully interacting systems.
The emphasis for the partition-free setup is even more important in the case of TD-DFT since the voltage switch-on is a local potential linearly coupled to the particle density~\cite{Stefanucci2004EPL, Kurth2005, Stefanucci2007}. In contrast, the coupling process in the partitioned approach is not local in space, and the standard TD-DFT formulation is not straightforward in this situation.
Transient dynamics in quantum correlated lattice networks can also be resolved by using the dynamical mean-field theory~(DMFT)~\cite{Georges1996,Aoki2014}, which uses a local impurity model as its starting point, and constructs an expansion in terms of the coupling between the impurity and its dynamical environment. This expansion can also be applied in the context of quantum transport~\cite{Arrigoni2013, Balzer2015, Titvinidze2018}. Another widely used method is the density matrix renormalization group (DMRG) and its time-dependent extension~\cite{White2004, Vidal2004, Daley2004}. DMRG is, also in the context of time-dependent problems, a powerful technique especially in case of one-dimensional interacting quantum systems. It has been successfully used in the study of real-time dynamics, as in time-dependent quantum transport, where a certain hierarchy of equations for the time-evolution operator and the density matrix is solved using specific approximations~\cite{White2004, Schmitteckert2004, Daley2004, Schoellwoeck2006, Schmitteckert2013}. In one-dimensional systems there are fewer connections between the relevant basis states, and together with DMRG a time-evolving block decimation algorithm is often employed~\cite{Vidal2007, Orus2008}. These ideas have recently been further extended to matrix-product and tensor-network states for more efficient computation~\cite{Paeckel2019, Fishman2020}.

Even though in some cases the non-interacting picture might be a sufficient description, e.g., monolayer graphene devices have revealed ballistic transfer lengths ranging from hundreds of nanometers to even micrometers at low temperatures~\cite{Miao2007,Lin2009}, the electron--electron and electron--phonon interaction will, in principle, influence the transport mechanisms~\cite{Galperin2007a, Swenson2012, Hartle2013, Ridley2019JCP}. In the regime in which perturbation theory can be applied, i.e., when the interaction is weak, the transport setup can be described in terms of the one-particle Green's function~\cite{Galperin2006b, Souza2008, Myohanen2009, Lynn2016, Tang2017a, Miwa2017, Tang2018, Cabra2018}. At the moment of writing this review the case of strong interaction cannot yet be included in these approaches, at least in realistic device structures, since considerably more complicated and numerically expensive methods are required in this case~\cite{Gull2011, Cohen2013, Cohen2014, Cohen2015, Ridley2018PRB, Ridley2019PRB}.
Nevertheless, it is often the case that introducing a quasi-particle picture might help modeling the system with an effective interaction which turns out to be weak. This is the case of the polaron transformation~\cite{Ovchinnikov2020} which allows to treat strong electron--phonon interactions.
We note here that, recently, two excellent reviews in Refs.~\cite{Thoss2018,Cohen2020} have thoroughly covered applications and method development in the study of quantum transport in molecular junctions, respectively, although their emphasis was more in the stationary regime.

\subsection{Why non-equilibrium Green's functions?}

While the previous subsection shows an impressive list of theoretical approaches to time-resolved quantum transport, each of them comes with its own constraints and practical limitations. The applicability of TD-DFT depends on how accurately the exchange--correlation potential can be approximated. (This is similar to the construction of the correlation self-energy in the NEGF approach as we will see in the next section.) A systematic construction of the exchange--correlation potential, particularly with the inclusion of transient and memory effects, is problematic. Instead, in the limit of strong correlation DMFT becomes exact. However, DMFT is a local method where the perturbative expansion is constructed by embedding the local impurity in an environment. Therefore, by construction, the description of transport in the strong-coupling regime becomes problematic within DMFT. While DMRG is able to ameliorate these issues to some degree, it remains computationally feasible only in low-dimensional systems as the reduction of the effective basis becomes more problematic for higher dimensions.
Moreover the external leads are typically modeled as a part of the system which is then traced out at the end of the simulation~\cite{Bohr2006, Bohr2007, HeidrichMeisner2009}. This has two main disadvantages: On the one hand, computational power needs to be employed to solve the equations for a system whose properties are not of interest. On the other hand, this introduces a finite recurrence time-scale for the leads, which may result in spurious physical effects.
We will comment on an example of this in Sec.~\ref{sec:applications}; see Fig.~\ref{fig:mbpt}.

The non-equilibrium Green's function approach is an \emph{ab initio} method suitable for both bosonic and fermionic, correlated quantum systems in and out of equilibrium~\cite{Danielewicz1984, Stefanucci2013Book, Balzer2013book}. There are no special requirements for the consideration of the system's dimensionality, the strength of external perturbations, zero or elevated temperatures, or between the transient regime and the stationary state. The NEGF approach has profound mathematical foundations in quantum-field theory~\cite{matsubara_new_1955, konstantinov_graphical_1960, Baym1961, Baym1962, Kadanoff1962, Keldysh1964, chou_equilibrium_1985, Cornean2018, Cornean2019}. The $N$-particle Green's function is an object of $2N$ time variables defined on the complex Keldysh time contour, and this object obeys the hierarchical equations of motion by Martin and Schwinger~\cite{Martin1959,vanLeeuwen2013}. In practice, the hierarchy is truncated by introducing a self-energy kernel which may be thought of as an effective medium (in macroscale) or as a scattering potential (in microscale). Importantly, the construction of the self-energy kernel can be systematically carried out using diagrammatic rules imposing macroscopic conservation laws~\cite{Baym1961}. In the lowest order of the hierarchy, for the single-particle Green's function the equations of motion are known as the Kadanoff--Baym equations. While there exists a few fairly recent and thorough reviews on the NEGF approach for fermionic lattice systems~\cite{Schluenzen2016}, modeling various physical systems~\cite{Hirsbrunner2019}, and a systematic construction and comparison of many-body self-energies~\cite{Schluenzen2019}, our treatment here focuses more on the time-dependent aspects of quantum transport using the NEGF approach. Therefore, for now, we settle for citing only a few highlights in a wide selection of many-body physics phenomena ranging from heavy-ion collisions to quantum thermodynamics~\cite{Yamamoto2006, Jackson2012, Esposito2015, Kurkela2019, Schluenzen2020, Dendzik2020, Dutta2020}, through which the reader can appreciate the versatility of the NEGF approach.

\subsection{Plan for this review}

We start with the description of the time-dependent formulation and the quantum transport Hamiltonian in Sec.~\ref{sec:hamiltonian}. We then list the basic equations of the non-equilibrium Green function formalism in Sec.~\ref{sec:theory} with particular emphasis on the quantum transport setting. After this, we also outline recent developments within the generalized Kadanoff--Baym ansatz in Sec.~\ref{sec:gkbasec}. In Sec.~\ref{sec:chap3}, we show how different limiting cases can be obtained from the underlying equations of motion for the NEGF, and we review the possibilities for various physical mechanisms captured by the NEGF approach. In Sec.~\ref{sec:calc}, we provide a brief look into the problem of performing actual calculations with NEGF formalism in the time-domain. Then, in Sec.~\ref{sec:applications}, we discuss timely and quantum-technologically relevant applications of, e.g., one- and two-dimensional electronic systems, systems with electron--phonon coupling, superconducting systems, and optical response in nanojunctions where electron--photon couplings are important. Finally, in Sec.~\ref{sec:summary}, we summarize the reviewed topics and discuss future prospects what could be achieved with the NEGF approach.

\renewcommand{\pb}{ p}
\renewcommand{\qb}{ q}
\renewcommand{\sb}{ s}
\renewcommand{\rb}{ r}
\renewcommand{\lb}{ l}
\renewcommand{\kb}{ k}

\newcommand{\Gcalb}{{\bm \Gcal}}
\newcommand{\Dcalb}{\bm \Dcal}
\newcommand{\vBold}{{\bm v}}
\newcommand{\wBold}{{\bm w}}
\newcommand{\abar}{{\bar{a}}}
\newcommand{\bbar}{{\bar{b}}}
\newcommand{\rhoe}{\rho}
\newcommand{\rhob}{\rhoBold_{\rm{b}}}
\newcommand{\Ib}{{\bm I}_{\rm{b}}}
\newcommand{\pbar}{\bar{p}}
\newcommand{\qbar}{\bar{q}}
\newcommand{\rbar}{\bar{r}}

\section{Theoretical description of quantum transport}\label{sec:hamiltonian}

\subsection{Time-dependence of observables}
The mathematical description of a many-body quantum system requires to build a wavefunction in a Hilbert space $\mathcal{H}_{\text{MB}}$ by means of the wavefunctions of the constituting subsystems  $S_{1},S_{2},...,S_{N}$. The total Hilbert space is the product-state of the individual ones:
\begin{equation}
\mathcal{H}_{\text{MB}}=\underset{i=1}{\overset{N}{\otimes}}\mathcal{H}_{S_{i}}.
\end{equation}
In second quantization, each Hilbert space $\mathcal{H}_{S_{i}}$ is the direct sum of Fock spaces with fixed number of particles. This representation allows for describing systems with varying number of particles, which turns particularly useful in the description of transport phenomena. The information on the nature of the particles, which may be fermionic or bosonic, is carried by the (anti-)commutation relation of the field operators acting on the Fock space.

The study of many-body processes out-of-equilibrium may be defined in terms of the dynamical response of a quantum subsystem $S_{i}$ to a quench at time $t=t_0$ which brings the system out from its equilibrium Hamiltonian $\hat{H}_0 = \hat{H}\left(t<t_{0}\right)$, i.e. 
\begin{equation}
\hat{H}\left(t\geq t_{0}\right)\neq\hat{H}_0.
\end{equation}
The time evolution of any physical observable $O(t)$ can then be computed in terms of processes which transfer energy and/or particles between $S$ and other subsystems making up the entire closed quantum system. The time evolution of the full system is a unitary mapping between $t_0$ and other times $t$ at which strong measurements are carried out such that the quantum state possesses definite values for the measured quantity. Setting the Planck constant $\hbar=1$, the mapping between quantum states defined at any pair of times $t_{1}, t_{2}$ can be written in terms of a time-ordered integral
\begin{equation}\label{U_propagator}
\hat{U}\left(t_{2},t_{1}\right)=\begin{cases}
{\text T}\exp\left[-\im\int_{t_{1}}^{t_{2}}\ud\bar{t}\hat{H}\left(\bar{t}\right)\right], & t_{2}>t_{1}\\
\tilde{\text T}\exp\left[-\im\int_{t_{1}}^{t_{2}}\ud\bar{t}\hat{H}\left(\bar{t}\right)\right], & t_{2}<t_{1}
\end{cases}
\end{equation}
where ${\text T}$ and $\tilde{\text T}$ denote chronological and anti-chronological time-ordering, respectively.
Given an initial density operator $\hat{\rho}_{0}$ at $t_0$,
which may represent either a mixed or pure state, $O(t)$ is given in terms of its corresponding Hermitian operator $\hat{O}$ as
\begin{equation}\label{O_t}
O\left(t\right)=\textrm{Tr}\left[\hat{\rho}_{0}\hat{U}\left(t_{0},t\right)\hat{O}\hat{U}\left(t,t_{0}\right)\right]\equiv\left\langle \hat{O}_{\text{H}}\left(t\right)\right\rangle ,
\end{equation}
where the Heisenberg representation $\hat{O}_{\text{H}}\left(t\right)\equiv\hat{U}\left(t_{0},t\right)\hat{O}\hat{U}\left(t,t_{0}\right)$ is used. Note that whereas the trace in Eq. \eqref{O_t} runs over all degrees of freedom in the system, $\hat{O}$ may be defined on any proper subspace of $\mathcal{H}_{\text{MB}}$. 

In many applications, the system is initially assumed to have equilibrated prior to the quench time, such that $\hat{\rho}_{0}$ has the grand-canonical form
\begin{equation}\label{eq_rho}
\hat{\rho}_{0}=\frac{\ex^{-\beta\left(\hat{H}_{0}-\mu\hat{N}\right)}}{Z},
\end{equation}
where $\beta=1/k_{B}T$ is the inverse temperature, $\mu$ the chemical potential, and 
$Z=\textrm{Tr}\left[\ex^{-\beta\left(\hat{H}_{0}-\mu\hat{N}\right)}\right]$ the partition function. 

\subsection{The ordered time-contour}
\begin{figure}[t]
\begin{center}
    	\includegraphics[width=.8\textwidth]{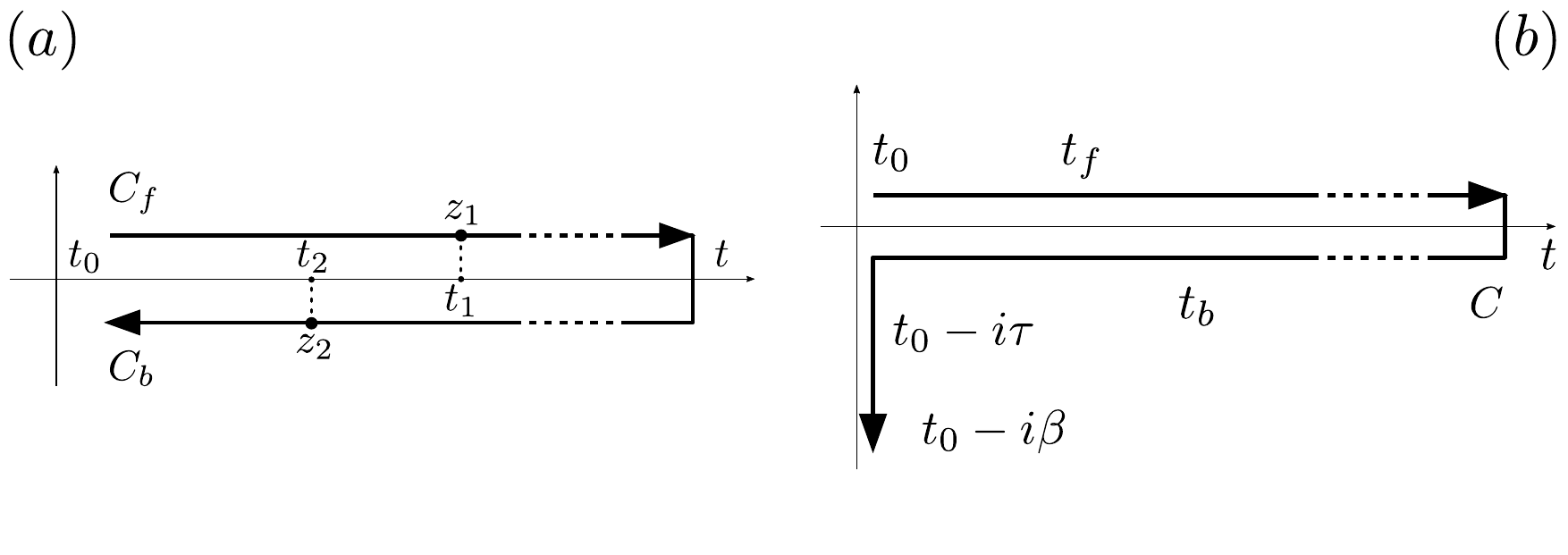}
\end{center}
		\caption{(a) Oriented Keldysh real-time contour $C=C_{f} \oplus C_{b}$ made of forward and backward branches. The arrow denotes the direction of the contour, and we introduce the notation $z_1 < z_2$ between contour times, meaning $z_2$ is later than $z_1$ on the contour. The projections of $z_1$ and $z_2$ are the real times $t_1$ and $t_2$.
		(b) The Konstantinov--Perel' contour $C=C_{f} \oplus C_{b} \oplus C_{M}$ made of forward, backward and Matsubara branches, in this case, the latest time is on the vertical track of the complex contour. Observables along the forward and backward branches are assumed to be equal $O(z=t_{f/b}) = O(t)$.}
\label{fig:time_contours}
\end{figure}

In the 1960s, it was independently noticed by Schwinger and Keldysh that Eq. \eqref{O_t} is a product of pairs of amplitudes describing processes with opposite time orientations connecting fixed points in time at $t_0$ and $t$. This lead them to formulate many-body perturbation theory (MBPT) on a directed time-contour composed of two copies of the time domain $[t_{0},t]$, as shown in Fig.~\ref{fig:time_contours}~(a). 
In this figure, forwards-time propagation is given in terms of times $t^{f}$ on the `upper' branch $C_{f}$ running from $t_{0}$ to $t$, and backwards-time propagation is described on the `lower' branch $C_{b}$ running from $t$ to $t_{0}$, with corresponding time labels $t^{b}$ and $t_{0}^{b}$ for the backwards--oriented time branch. The full Keldysh time-contour is the direct sum of the two branches $C=C_{f} \oplus C_{b}$. 

It was noticed by Matsubara~\cite{matsubara_new_1955} that the formal structure of the numerator in Eq.~\eqref{eq_rho} is equivalent to a time-propagation $\ex^{-\im\hat{H}^{M}\left(t_{2}-t_{1}\right)}$ where $\hat{H}^{M}\equiv\hat{H}_{0}-\mu\hat{N}$ is the Matsubara Hamiltonian and the evolution is extended into the complex-time direction from $t_{1}=t_{0}$ to $t_{2}=t_{0}-\im\beta$. This idea was extended to time-dependent systems by Konstantinov and Perel'~\cite{konstantinov_graphical_1960}.
We therefore rewrite the ensemble average in Eq.~\eqref{O_t} as
\begin{equation}\label{O_t_KP}
O\left(t\right)=\frac{\textrm{Tr}\left[\hat{U}\left(t_{0}-\im\beta,t_{0}\right)\hat{U}\left(t_{0}^{b},t^{b}\right)\hat{O}\hat{U}\left(t^{f},t_{0}^{f}\right)\right]}{\textrm{Tr}\left[\hat{U}\left(t_{0}-\im\beta,t_{0}\right)\hat{U}\left(t_{0}^{b},t^{b}\right)\hat{U}\left(t^{f},t_{0}^{f}\right)\right]}
\end{equation}
which is a propagation over the extended Konstantinov--Perel' contour $\gamma\equiv C_{f}\oplus C_{b}\oplus C_{M}$ consisting of the Keldysh contour with an additional Matsubara branch $C_{M}\equiv\left[t_{0},t_{0}-\im\beta\right]$ shown in Fig.~\ref{fig:time_contours}~(b). This motivates the introduction of a generic contour time $z\in\gamma$. For instance, to obtain the equilibrium value of the
observable $O(z)$, we can evaluate $O(z_M)$, for any contour time $z \in C_M$.
Historically, MBPT was first developed on the vertical Matsubara branch alone to compute quantum expectation values for systems in equilibrium at finite temperature~\cite{matsubara_new_1955}.

We now introduce the concept of a contour-time-ordering operator ${\text T}_{\gamma}$, defined by the relation
\begin{equation}\label{time_ordering}
{\text T}_{\gamma}\left[\hat{O}\left(z_{1}\right)\ldots\hat{O}\left(z_{N}\right)\right]\equiv\underset{P}{\sum}\left(\pm\right)^{\zeta_{P}}\theta_{\gamma}\left(z_{P_{1}},z_{P_{2}}\right)\ldots\theta_{\gamma}\left(z_{P_{N-1}},z_{P_{N}}\right)\hat{O}\left(z_{P_{1}}\right)\ldots\hat{O}\left(z_{P_{N}}\right)
\end{equation}
where the summation runs over permutations of the $N$ contour times with parity $\zeta_{P}$, $\pm$ refers to bosonic/fermionic operators, and the contour step function is defined as
$\theta_\gamma(z_1,z_2) = 1$ if $z_1 > z_2$ contour-wise (see Fig.~\ref{fig:time_contours}), and 0 otherwise. This enables us to define a single 
time-evolution operator
along the full contour as a generalization of Eq.~\eqref{U_propagator}
\begin{equation}\label{U_KP}
\hat{U}\left(z_{2},z_{1}\right)\equiv \begin{cases}
{\text T}_{\gamma}\exp\left[-\im\int_{z_1}^{z_2}\ud z\hat{H}\left(z\right)\right], & z_2 > z_1 \\
\tilde{\text T}_{\gamma}\exp\left[+\im\int_{z_2}^{z_1}\ud z\hat{H}\left(z\right)\right], & z_2 < z_1
\end{cases}
\end{equation}
with which we can reformulate Eq. \eqref{O_t_KP} into the pleasingly simple expression:
\begin{equation}\label{O_z_propagator}
O\left(z\right)=\frac{\textrm{Tr}\left[\hat{U}\left(t_{0}-\im\beta,z\right)\hat{O}\left(z\right)\hat{U}\left(z,t_{0}^{f}\right)\right]}{\textrm{Tr}\left[\hat{U}\left(t_{0}-\im\beta,t_{0}^{f}\right)\right]}.
\end{equation}
For the construction of diagrammatic perturbation theory, it is convenient to bring the operator under the contour-ordering:
\begin{equation}\label{O_z_MBPT}
O\left(z\right)=\frac{\textrm{Tr}\left[{\text T}_{\gamma}\left\{  
\exp\left[-\im\int_\gamma\ud
z\hat{H}\left(z\right)\right]\hat{O}\left(z\right)\right\} \right]}{\textrm{Tr}\left[{\text T}_{\gamma}\left\{ 
\exp\left[-\im\int_\gamma\ud
z\hat{H}\left(z\right)\right]\right\} \right]},
\end{equation}
where the integrals are understood contour-wise from $t_0^f$ to $t_0-\im\beta$.

\subsection{The quantum transport Hamiltonian}
\begin{figure}[t]
\centering
    	\includegraphics[width=.6\textwidth]{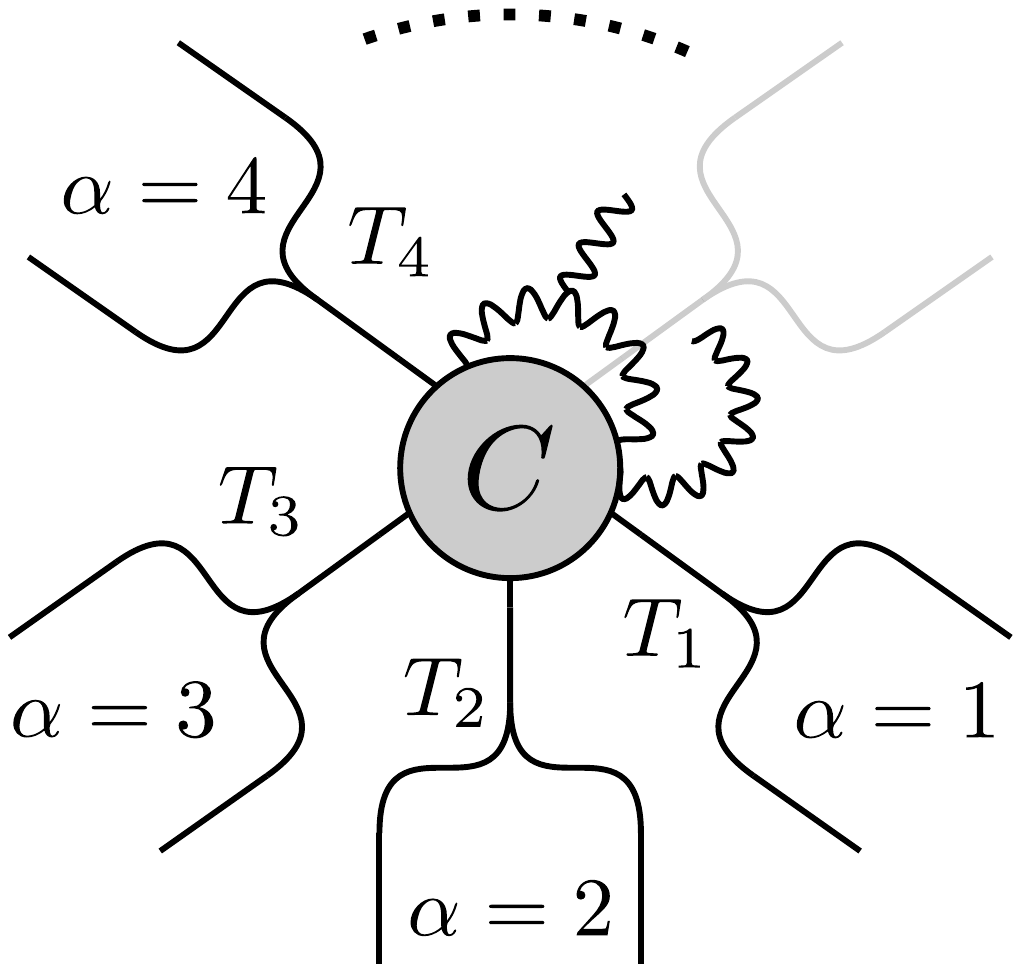}
		\caption{  Schematic representation of the multi-terminal transport model described by Eq.~\eqref{eq:hamtot}. A central correlated quantum system ($C$) tunnel-coupled ($T_\alpha$) to an arbitrary number of macroscopic leads.} 
\label{fig:sys}
\end{figure}

In the most general case, to describe quantum transport in nanoscale devices, it is required to construct a quantum-mechanical model that takes into account the relative positions, the energy configurations and the interactions between many identical particles confined in well-defined structures. 
In the exact quantum theory formulation based on the solution of the time-dependent Schr\"{o}dinger equation~\cite{Schrodinger_1926, Griffiths_2005}, this is absolutely non-trivial, as one needs to deal directly with many-particle wavefunctions.
The second quantization, or occupation number, representation~\cite{Fetter_1971,Mattuck_1976,Mahan_1990,Dirac_1927} can be used to simplify the description in terms of creation and annihilation operators that add or remove particles into/from the system of interest.
The statistical mechanics of the different particle species, Fermi (Bose) statistics, are built-in and furthermore, it is granted a powerful tool to treat systems with variable number of particles, a very frequent situation in non-equilibrium phenomena. In addition, this formalism enables us to describe complex multi-time processes and measurement sequences on a many-body wavefunction.
A typical quantum transport setup consists of two or more non-interacting macroscopic electron reservoirs (leads) and a microscopic scattering region (e.g. a quantum dot, quantum wire, or a molecular system) which is attached to these reservoirs and can exchange particles and energy with them.  
A schematic representation of this model system is shown in Fig.~\ref{fig:sys}.

The general Hamiltonian describing this transport setup consists of
\begin{align}
\label{eq:hamtot}
\hat{H}(z) = \hat{H}_C(z) + \sum_{\alpha=1}^{N} \hat{H}_{\alpha}(z) + \sum_{\alpha=1}^{N}\hat{V}_{\alpha C}(z), \end{align}
with the subscript $C$, $\alpha$, $\alpha C $ indicating the central region, the $\alpha$-th reservoir and the hybridization among this systems respectively, and where $z$ is a contour time $C_\gamma$~\cite{Stefanucci2013Book, Balzer2013book, Haug_2008, Keldysh1964}, see Fig.~\ref{fig:time_contours}. Notice that even though we deal with real-time operators, like the Hamiltonian, we can always express them along the time-contour provided that they are the same on the forward and backward branches of the contour $\hat{H} (z=t_{f/b}) = \hat{H} (t)$, with $t_{f/b} \in C_\gamma $. This is, however, not a critical restriction and it can be generalized to branch-dependent Hamiltonians~\cite{kantorovich_generalized_2020}. In turn, the Hamiltonian for the central correlated region or active region consists of the following terms
\begin{align}
 \hat{H}_C(z) =  \underbrace{ \hat{H}_{\text{el}}(z) + \hat{H}_{\text{bos}}(z)}_{\hat{H}_0 (z)}  +  \underbrace{\hat{H}_{\text{el-el}}(z) + \hat{H}_{\text{el-bos}}(z)}_{\hat{V} (z)}
\end{align}
where the non-interacting part $\hat{H}_0$ consists of one-body operators and the interaction part $\hat{V}$ consists of higher-order operators describing scattering among electrons and electrons with bosonic (quasi-)particles (e.g., phonons or photons) in the active region. \\
The second-quantized form of the electron Hamiltonian is given by 
\begin{align}
\hat{H}_{\text{el}}(z_1) = \int \ud \mathbf{x}_1\; \hat{\psi}^\dagger (1) h(1) \hat{\psi} (1),
\end{align}
where $1= (\mathbf{x}_1, z_1)$ is a collective index for the position--spin coordinate  $\mathbf{x}_1= (\mathbf{r}_1,\sigma_1) $ and contour-time $z_1$. We have also introduced the short-hand notation $\int \ud \mathbf{x} = \sum_{\sigma} \int \ud \mathbf{r}$. 
The single-particle Hamiltonian in the central region, $h$, contains the kinetic energy and a general time-dependent external potential.
The electron--electron interaction is given by the following second quantized Hamiltonian:
\begin{align}
\hat{H}_{\text{el-el}}(z_1) = \frac{1}{2}\int \ud \mathbf{x}_1 \ud 1'  \;\hat{\psi}^\dagger (1) \hat{\psi}^\dagger (1') v(1, 1') \hat{\psi} (1') \hat{\psi} (1).
\end{align}
Here,
$v(1,2)= \delta_\gamma(z_1,z_2) v(\mathbf{x}_1, \mathbf{x}_2, z_1)$
is a generic two-body interaction between pairs of electrons in the scattering region. The interaction can be taken, e.g., to be the familiar Coulomb interaction, or a local interaction where 
$v(\mathbf{x}_1, \mathbf{x}_2, z_1) = \delta(\mathbf{r}_1 -\mathbf{r}_2) v(\sigma_1,\sigma_2,z_1)$,
ubiquitous in model systems used to describe experimental platforms with ultra-cold atomic gases~\cite{Bloch_2008}.
The fermion-field operators satisfy the usual anti-commutation relations $\{\hat{\psi}^\dagger (\mathbf{x}), \hat{\psi} (\mathbf{x'}) \} = \delta(\mathbf{x} -\mathbf{x'})$ and $\{\hat{\psi}^\dagger (\mathbf{x}), \hat{\psi}^\dagger (\mathbf{x'}) \} = \{\hat{\psi} (\mathbf{x}), \hat{\psi} (\mathbf{x'}) \} = 0$. These relations reflect the antisymmetry property of the many-body wave function and thus its statistics.

Similarly, the non-interacting particles in the $\alpha$-th lead are described by
\begin{align}
\hat{H}_\alpha (z_1)= \int \ud \mathbf{x}_1 \hat{\psi}^\dagger_\alpha (1)  h_\alpha(1)\hat{\psi}_\alpha (1)
\end{align}
with $h_\alpha$ the single particle Hamiltonian of the $\alpha$-th lead and similar anti-commutation relations as before for their fermionic-field operators.
The Hamiltonian accounting for the tunneling between the interacting region and the leads is given by
\begin{align}
\hat{V}_{\alpha C} (z_1) = \int \ud \mathbf{x}_1 \left( \hat{\psi}^\dagger (1) T_{\alpha}(1) \hat{\psi}_\alpha (1)+ \text{h.c.}\right),
\end{align}
with $T_\alpha$
the corresponding tunneling amplitude.

A general way to express the free boson Hamiltonian is
\begin{align}\label{eq:nonInteractingBosons}
    \hat{H}_{\text{bos}}(z_1) = \sum_{\mub \nub} \Omega_{\mub \nub}(z_1) \phih_\mub \phih_\nub,
\end{align}
where the multi-component $\hat{\phi}$-fields contain the displacement $\phih_{\mu,1} = \uh_\mu$ and the momentum operator $\phih_{\mu,2} = \ph_\mu$ for the mode $\mu$, and where we have introduced a composite index $\mub = (\mu,\xi_\mu)$ with $\xi_\mu = 1,2$.
Accordingly, Eq.~\eqref{eq:nonInteractingBosons} consists of `kinetic' and `force constant' contributions of the harmonic-oscillator type: $\hat{H}_{\text{bos}}(z_1)=\sum_\mu\hat{p}_\mu^2/2+\sum_{\mu\nu}K_{\mu\nu}(z_1)\hat{u}_\mu\hat{u}_\nu/2$. For example, for phononic systems the force constant matrix $K$ could be defined according to some underlying potential energy surface $E(\mathbf{R})$ as $K_{\mu\nu}=\partial^2 E /(\partial{R_\mu}\partial{R_\nu})$ with $\mathbf{R}$ being the nuclear coordinates. In Eq.~\eqref{eq:nonInteractingBosons}, the matrix elements are thus obtained as $\Omega_{\mub\nub}=\delta_{\xi_\mu \xi_\nu}(\delta_{\xi_\mu 1}K_{\mu\nu}+\delta_{\xi_\mu 2}\delta_{\mu\nu})/2$.
The canonical commutation relations of the displacement and momentum operators, $[\hat{u}_\mu,\hat{p}_\nu]=\im\delta_{\mu\nu}$, 
are included in the commutator of the composite $\hat{\phi}$ operators as follows:
\begin{equation}\label{eq:bosAlpha}
\bigg[\phih_\mub,\phih_\nub \bigg] = \alphaBold_{\mub \nub}, \quad \text{with} \quad \alphaBold_{\mub \nub} = \delta_{\mu \nu}
 \begin{pmatrix}
  0 &  \im \\
  - \im & 0
 \end{pmatrix}_{\xi_\mu \xi_\nu}.
 \end{equation}
For future reference, we note that $\alphab$ fulfills the idempotency relation $\alphab \alphab = \oneb$.
Consistently with this notation, the electron--boson interaction is given by 
\begin{align}\label{elbos}
    \hat{H}_{\text{el-bos}}(z_1) = \sum_{\mub}  \int \ud \mathbf{x}_1 \hat{\psi}^\dagger (1)  \lambda_\mub (1) \hat{\psi} (1) \phih_\mub,
\end{align}
with $\lambda_\mub$ representing the coupling amplitude between electrons and bosons.
While restricting ourselves to linear electron--boson interactions,
the general interaction of Eq.~\eqref{elbos} includes many commonly encountered interactions, such as interactions between electrons and phonons, plasmons, magnons, and cavity--quantum-electrodynamical photons. 

The representation of the total Hamiltonian in terms of the field operators is a useful and powerful tool to carry out calculations and derivations along the time contour. Nonetheless, for practical numerical calculations it is convenient to expand the field operators into a suitable single-particle basis. The latter is chosen accordingly to the problem at hand in such a way to retain only the most relevant degrees of freedom for the specific physical scenario considered.
A possible and very general choice is an orthonormal spin--orbital basis
$\varphi_{i \tau}(\mathbf{x})=\varphi_i(\mathbf{r}) \delta_{\tau \sigma}$, consisting of an orbital index $i$ and a spin index $\tau$, and we choose the spin-projection axis the same as in the $\mathbf{x}=(\mathbf{r}\sigma)$ basis. We can then express the creation and annihilation operators $\hat{d}_{n}^\dagger,\hat{d}_{n}$ for a generic quantum number $n=(i \tau)$ as a linear combination of the field operators at different position--spin coordinates
\begin{align}
\label{eq:ddag}
&\hat{d}_{n}^\dagger \equiv \int \ud \mathbf{x} \;\varphi_{n} (\mathbf{x}) \hat{\psi}^\dagger (\mathbf{x}), \\ 
\label{eq:d}
&\hat{d}_{n} \equiv \int \ud \mathbf{x} \;\varphi_{n}^* (\mathbf{x}) \hat{\psi} (\mathbf{x}).
\end{align} 
The operators $\hat{d}^\dagger $ and $\hat{d}$ inherit the anti-commutation rules from the field operators $\hat{\psi}^\dagger $ and $\hat{\psi}$. If the set $\{\varphi_n\}$ was not complete in the one-particle Hilbert space, then Eqs.~\eqref{eq:ddag} and~\eqref{eq:d} would be an approximate expansion as the overlap matrix $S_{nn'}=\int\ud\mathbf{x}\varphi_n^*(\mathbf{x})\varphi_{n'}(\mathbf{x})$ may have non-vanishing off-diagonal elements.

Using Eqs.~\eqref{eq:ddag} and~\eqref{eq:d}, together with their anti-commutation relations, it is possible to express all the terms of the total Hamiltonian, Eq.~\eqref{eq:hamtot}, in the chosen single particle basis as:
\begin{align}
\label{eq:intsys}
\hat{H}_C (z) & = \hat{H}_0 (z) + \hat{V} (z) , \\
\hat{H}_0 (z) & = \sum_{ij, \sigma} h_{ij} (z) \hat{d}_{i \sigma}^\dagger \hat{d}_{j \sigma} +\sum_{\mub \nub} \Omega_{\mub \nub}(z) \phih_\mub \phih_\nub , \label{eq:intsysh0}\\
\hat{V} (z) & = \frac{1}{2} \sum_{ijkl} \sum_{\sigma \sigma'} v_{ijkl}(z) \hat{d}_{i \sigma}^\dagger \hat{d}_{j \sigma'}^\dagger \hat{d}_{k \sigma'} \hat{d}_{l \sigma} + \sum_{ij, \sigma} \sum_{\mub} \lambda_{ij}^{\mub} (z) \hat{d}_{i \sigma}^\dagger \hat{d}_{j \sigma} \phih_\mub ,  \label{eq:intsysv}\\
\hat{H}_\alpha (z) & = \sum_{ij \in \alpha , \sigma} h_{\alpha ij} (z) \hat{c}_{i \sigma}^\dagger \hat{c}_{j \sigma} , \label{eq:intsyshlead}\\
\hat{V}_{\alpha C}(z) & = \sum_{i \in C , j \in \alpha, \sigma} \left( T_{\alpha ij} (z) \hat{d}_{i \sigma}^\dagger \hat{c}_{j \sigma} + \text{h.c.}\right) , \label{eq:intsyshcoupl}
\end{align} 
where the matrix elements
of the electronic one-body terms are given in the spin--orbital basis by
\begin{equation}
h_{ij}(z)=  \int \ud \mathbf{r} \;\varphi_{i}^* (\mathbf{r}) h(\mathbf{r},z) \varphi_{j} (\mathbf{r}),
\end{equation}
\begin{equation}
h_{\alpha ij}(z)=  \int \ud \mathbf{r} \;\varphi_{i}^* (\mathbf{r}) h_\alpha(\mathbf{r},z) \varphi_{j} (\mathbf{r}),
\end{equation}
\begin{equation} 
T_{\alpha ij} (z) = \int \ud \mathbf{r} \;\varphi_{i}^* (\mathbf{r}) T_{\alpha} (\mathbf{r},z) \varphi_{j} (\mathbf{r}).
\end{equation}
See the discussion below Eq.~\eqref{eq:nonInteractingBosons} for the non-interacting bosonic part.
As the lead partition is non-interacting, it is common to cast it in a diagonal form $\hat{H}_\alpha(z)=\sum_{k\sigma}\epsilon_{k\alpha}(z)\hat{c}_{k\sigma}^\dagger\hat{c}_{k\sigma}$, where $\epsilon_{k\alpha}(z)$ describes the lead energy dispersion with possible time dependence arising from, e.g., a bias-voltage and/or thermal-gradient profile for times on the horizontal branches:
\begin{align}\label{lead_energies}
\epsilon_{k\alpha}(z)=\begin{cases}[\epsilon_{k\alpha}+V_\alpha(t)][1+\vartheta_\alpha(t)], & z\in C_f \oplus C_b \\ \epsilon_{k\alpha}-\mu , & z\in C_M , \end{cases}
\end{align}
where $V_\alpha$ represents the bias voltage and $\vartheta_\alpha=(T_\alpha-T)/T$ is the thermo-mechanical field, related to a relative variation between the lead temperature $T_\alpha$ and a reference temperature $T$~\cite{Eich2014}.
Similarly, the matrix elements describing the electron--electron and electron--boson interaction are 
\begin{align}
v_{ijkl}(z) & = \int \ud \mathbf{r} \ud \mathbf{r'}\; \varphi_{i}^* (\mathbf{r}) \varphi_{j}^* (\mathbf{r'}) v(\mathbf{r}, \mathbf{r'},z) \varphi_{k} (\mathbf{r'}) \varphi_{l} (\mathbf{r}), \label{coulomb} \\
\lambda_{ij}^{\mub}(z) & = \int \ud \mathbf{r} \; \varphi_{i}^* (\mathbf{r}) \lambda_\mub (\mathbf{r}, z) \varphi_{j} (\mathbf{r}). \label{eq:elbosint}
\end{align}
To simplify the model,  we have assumed that both the single particle Hamiltonian $h$ and the interaction $v$ are spin independent,
even so it is easy to generalize the above expressions in the case where a magnetic field and/or spin orbit coupling is present.

While the orbital-basis representation is useful in practical calculations, the formulation of the non-equilibrium Green's functions and the many-body perturbation theory can be expressed in terms of the generic field operators. We will take this approach next.

\section{The non-equilibrium Green's function formalism}\label{sec:theory}

In this section we review briefly the main elements of the non-equilibrium Green's function theory (NEGF), as well as the key ingredients of the many-body perturbation theory (MBPT) used to describe weakly interacting system and couplings of such a system with external macroscopic reservoirs.

The Keldysh Green's function theory~\cite{Kadanoff1962, Keldysh1964, Mahan_1990} includes as limiting cases the zero-temperature Green's function (time-ordered) and the Matsubara formalism which are recovered by specific choices of the time contour.
Furthermore, it retains the formal structure of the many-body perturbation theory extending it to non-equilibrium phenomena. It allows for a systematic study of time-dependent expectation values and steady-state properties when electron--electron (electron--phonon) interaction is present. The correlation effects of the interaction are included via an integral kernel called self-energy that can be methodically constructed via a diagrammatic expansion of selected Feynman diagrams.
We start the section by introducing the definition and properties of the NEGF as well as the equations that govern their dynamics on the time-contour and then we outline the steps that led to the MBPT in the two-time plane.
Our aim is to give a brief characterization and few examples of the most used electronic self-energies approximations which embody both the role of many-body interactions and the coupling with external leads. 

\subsection{Single-particle electron Green's functions}\label{sec:spgf}
The starting point in the theory of the non-equilibrium Green's function together with the many-body perturbation expansion is the definition of the (fermionic) single-particle Green's function (SPGF) as the expectation value of the contour-ordered product of the creation and annihilation operators 
\begin{equation}
\label{eq:green}
G(1,1') = -\im \left\langle {\text T}_\gamma \left[ \hat{\psi}_{\text{H}}(1) \hat{\psi}_{\text{H}}^\dagger (1') \right] \right\rangle 
\end{equation}
here the subscript $\text{H}$ denotes the Heisenberg picture and the indices $1=(\mathbf{x}_{1},z_1)$ and $1'=(\mathbf{x}_{1}',z_{1}')$ are collective indices for position, spin and contour-time. 
Furthermore, $z$ and $z'$ are contour-time variables and ${\text T}_\gamma$ orders the operators along the Keldysh contour $C_\gamma$ by arranging the operators with later contour-times to left Fig.~\ref{fig:time_contours}. 
Similarly, we denote by $h$ the matrix elements of the first quantized Hamiltonian $\hat{h}$
in the position, spin and contour-time indices. By applying $\left[ \im \partial_{z_1} - h(z) \right] $ and using the Heisenberg equations of motion for the field operators $\hat{\psi}$ and $\hat{\psi}^\dagger$ under the evolution given by $\hat{H}_C(z)$ on the time contour, one obtains the first equation of the Martin--Schwinger hierarchy (MSH):
\begin{align}
\label{eq:oneMS}
& \Big( i \partial_{z_1} - h(1) \Big)  G(1 , 1') = \delta(1, 1') + \im \int \dif  \bar{1}  v(1,\bar{1})  G_2 (1,\bar{1}; 1',\bar{1}^+) 
\end{align}
where  
\begin{equation}
G_2 (1,2; 1',2')=(-\im)^2 \left\langle {\text T}_\gamma \left[ \hat{\psi}_{\text{H}} (1) \hat{\psi}_{\text{H}} (2) \hat{\psi}_{\text{H}}^\dagger (2')\hat{\psi}_{\text{H}}^\dagger (1') \right] \right\rangle     
\end{equation}
is the two-particle Green's function.
Here, the integral is $\int \dif  \bar{1} = \int \dif  \bar{\mathbf{x}}_1 \int_{\gamma} \dif  \bar{z}_1$ and $1^+= (\mathbf{x}_1, z_1+\delta)$ denotes a time with an infinitesimally small shift $\delta$ on the Keldysh contour $C_\gamma$. A similar equation is obtained by acting to the left with the operator $\left( -\im \partial_{z_1'} - h(1') \right)   $.
With the contour Heisenberg equations and contour calculus as described above, one can derive the equations of motion for the two-particle and higher-order Green's functions, and find equations which couple the $N$-particle one to the $(N\pm 1)$-particle Green's function.  
In many practical situations, 
the knowledge of the single-particle Green's function is sufficient to describe the physical problem at hand. In this case, it is suitable to introduce the many-body self-energy $\Sigma_{\text{MB}}$,
which allows one to (formally) decouple the time-evolution of the Green's function from those of the $(N>1)$-particle Green's functions and obtain a closed equation in terms of time-local quantities. It is implicitly defined by:
\begin{align}
\label{eq:SEMS}
& \int \dif  \bar{1}\Sigma_{\text{MB}}(1,\bar{1})  G(\bar{1} , 1') = \im\int \dif  \bar{1}  v(1,\bar{1})  G_2 (1,\bar{1}; 1',\bar{1}^+) .
\end{align}
The physical meaning of $\Sigma_{\text{MB}}$ is to introduce an effective function
which accounts for the two-particle scattering encoded into the two-particle Green's function $G_2$.
Thanks to the definition of $\Sigma_{\text{MB}}$ with Eq.~\eqref{eq:SEMS} one obtains the following pair of equations on the time contour for the single-particle Green's functions:
\begin{align}
\Big( \im \partial_{z_1} - h(1) \Big) G(1,1') & = \delta(1,1')+ \int \dif  \bar{1} \Sigma_{\text{MB}}(1, \bar{1})  G (\bar{1},1'), \label{eq:KBeq}\\
G(1,1')\Big( \im \overset{\leftarrow}{\partial}_{z_1'} - h(1') \Big) & = \delta(1,1')+ \int \dif  \bar{1} G (1,\bar{1}) \Sigma_{\text{MB}}(\bar{1},1'), \label{eq:KBeq_2}
\end{align}
which has to be solved with the Kubo--Martin--Schwinger (KMS) boundary conditions $G(t_0, z_1') = - G(t_0 - \im \beta , z_1')$, 
following directly from Eq.~\eqref{eq:green} and the cyclic property of the trace.    
By means of the Langreth and Wilkins rules~\cite{Langreth1976,Stefanucci2013Book}, see~\ref{app:langreth} for details, it is possible to project these equations of motion 
for the single-particle Green's function onto the real and imaginary time axis.
The resulting set of equations for those components are called the Kadanoff--Baym equations~\cite{Baym1961, Kadanoff1962} and represent, 
together with the initial conditions, the standard way to completely determine the single-particle Green's functions once a choice for the self-energy is made.
The numerical implementation of the solution of such equations has been extensively explored and requires fine and elegant schemes for the two-times propagation~\cite{Stan2009, Balzer_2010, Stefanucci2013Book, Balzer2013book}.

An alternative approach to find the interacting Green's function in Eq.~\eqref{eq:green} is to formally expand the evolution operator in powers of the interaction
\begin{align}
G(1,1') = -\im \frac{ \sum_{k=0}^\infty \frac{(-\im)^k}{k!} \int \dots \int_\gamma  \left\langle {\text T}_\gamma \left[ \tilde{V}(z_1) \dots  \tilde{V}(z_k) \tilde{\psi}(1) \tilde{\psi}^\dagger (1') \right] \right\rangle _0}{\sum_{k=0}^\infty \frac{(-i)^k}{k!} \int \dots \int_\gamma  \left\langle {\text T}_\gamma \left[ \tilde{V}(z_1) \dots  \tilde{V}(z_k)  \right] \right\rangle _0}
\end{align}
where the tilde denotes operators in the interaction picture
and the short-hand notation $\langle {\text T}_\gamma \ldots \rangle_0 = \text{Tr}\{{\text T}_\gamma \exp{[-\im\int_\gamma \ud z \hat{H}_0(z)]} \ldots \}$ includes averaging with respect to the non-interacting Hamiltonian.
That is, if we are able to compute contour time-ordered products then we have a general and powerful way to obtain the SPGF.
This can be done for some particular cases, for example by means of the Wick's theorem~\cite{Stefanucci2013Book} we can write down a series expansion for the interacting single-particle Green's function in terms of the non-interacting one $G_0 (1,1')$, which satisfies the following equations:
\begin{align}
\label{eq:nnintgreen}
& \Big(\im \partial_{z_1} - h(1)\Big)  G_0(1, 1') = \delta(1, 1'),\\
&G_0(1, 1') \Big(-\im\overset{\leftarrow}{\partial}_{z_1'} - h(1')\Big) = \delta(1, 1').
\end{align}
By collecting the wanted terms of this series expansion and systematically representing them as Feynman diagrams one can built a proper self-energy and obtain the Dyson equation for the SPGF:
\begin{align}
\label{eq:dysongreen}
G(1,1') &= G_0 (1,1') + \int \dif  \bar{1} \, \dif  \bar{2} \, G_0 (1,\bar{1}) \Sigma_{\text{MB}}(\bar{1}, \bar{2}) G(\bar{2}, 1').
\end{align}
The Dyson equation is the formal solution of the Martin--Schwinger hierarchy for the one-particle Green's function, as one can easily verify by applying $ \left(\im  \delta(1,1') \partial_{z_1} - h(1',1)\right) $ to Eq.~\eqref{eq:dysongreen} to obtain Eq.~\eqref{eq:KBeq} with the help of Eq.~\eqref{eq:nnintgreen}. 
Because of that, the Dyson equation is formally equivalent to Eq.~\eqref{eq:KBeq} and it contains the same physical information about the dynamics of the single-particle Green's function through the double time integral on the r.h.s. of Eq.~\eqref{eq:dysongreen}. The last integration is performed along the Keldysh time-contour $C_\gamma$ and thus encompasses information of the statistical physics (vertical track) as well as the dynamics (horizontal branches) (see Fig.~\ref{fig:time_contours}) that the system is subjected to. Moreover, the correlation effects of the interaction are included via the integral kernel represented by the self-energy functional. 
Introducing now the two-particle exchange--correlation function~\cite{Stefanucci2013Book}
\begin{align}\label{eq:2pex}
    L\left(1,2;1',2'\right)\equiv G_{2}\left(1,2;1',2'\right)-G\left(1,1'\right)G\left(2,2'\right),
\end{align}
Starke and Kresse derived an analogous Dyson-like equation for the one-particle Green's function \cite{starke_self-consistent_2012}:
\begin{align}\label{eq:2pex2}
    G\left(1,1'\right)=G_{0}\left(1,1'\right)+\im\int\dif \bar{1}\dif \bar{2}G_{0}\left(1,\bar{1}\right)v\left(\bar{1},\bar{2}\right)L\left(\bar{2},\bar{2};\bar{1},1'\right).
\end{align}

\subsection{Perturbation theory}\label{sec:prtrbt}
As we mentioned in the previous section, in order to account for the effects of many-body interactions in the dynamics of the single-particle Green's function one needs to define a self-energy which describes these effects at the single particle level.
Formally, the self-energy arises either as a way to truncate the Martin--Schwinger hierarchy or, in the diagrammatic expansion of the evolution operator, as a way to choose which physical processes are relevant to describe the physical system.
\begin{figure}[b!]
\begin{tcolorbox}
	\begin{center}
		\includegraphics[width=.8\textwidth]{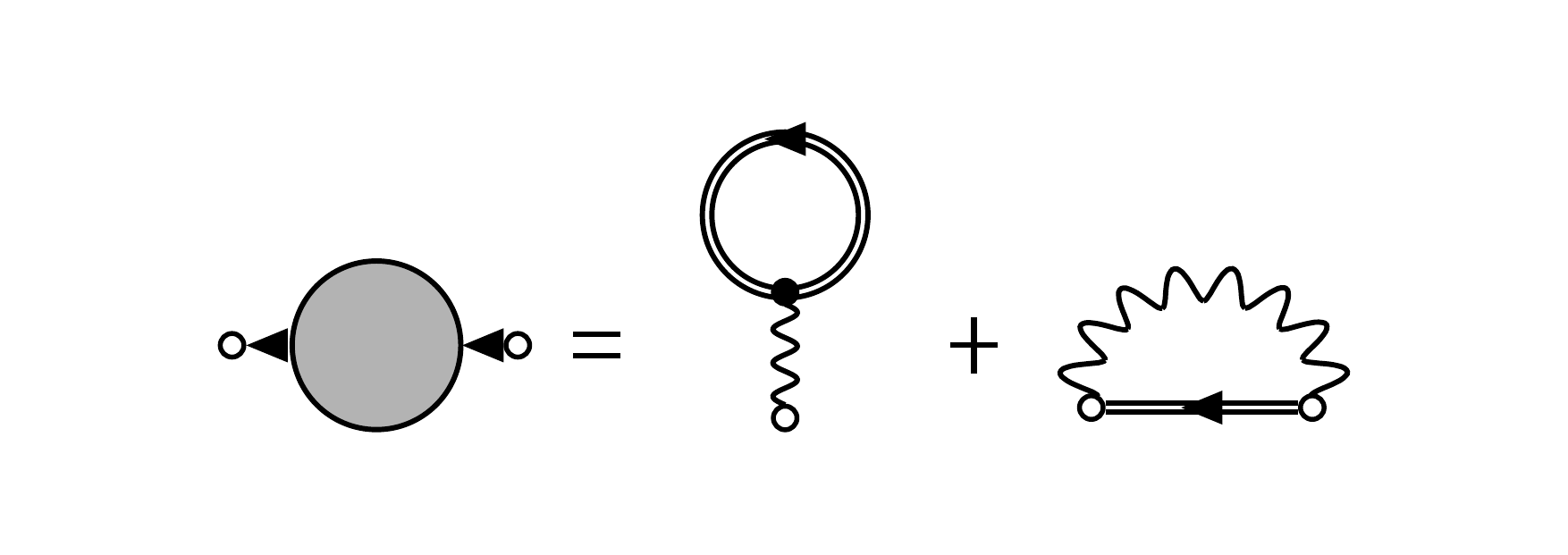}
		\caption{Diagrammatic representation of the Hartree--Fock self-energy Eq.~\eqref{eq:HF}. It is made of the tadpole or Hartree diagram (first term) and the first order exchange or Fock diagram (second term). The wavy line represents the electron--electron interaction.}
		\label{fig:HF}
	\end{center}
\end{tcolorbox}
\end{figure}
In the first case one obtains the Kadanoff--Baym equations, whereas in the second one the Dyson equation or, in a more general framework, the set of Hedin equations~\cite{Hedin1965, Stefanucci2013Book}.
This set, besides the Dyson equation, also contains other four equations for the (correlation) self-energy, the dressed interaction, the polarization diagram and the vertex function. As in the more simpler case of the Dyson equation, these five equations need to be solved self-consistently up to a certain order in the perturbation theory.
Regardless which equations one wants to solve, Kadanoff--Baym, Dyson or Hedin, the choice of the self-energy is, to some extent, left to the needs of the problem addressed, meaning that the choice of the diagrams to be included in the self-energy depends only upon the physical processes which are understood to contribute the most to the specific case at hand. 

However, there are some general restrictions that need to be taken into account, in particular those imposed by macroscopic conservation laws.
A convenient way to guarantee the conservation of particle number, momentum, energy, and angular momentum,
is to ensure that the self-energy is the functional derivative of a Luttinger--Ward functional $\Phi[G]$~\cite{Stefanucci2013Book, Luttinger_1960, Baym1962}. This object
can be obtained
using standard diagrammatic techniques when the nature of the scattering process (the interaction) is known~\cite{Stefanucci2013Book}. The self-energies which are functional derivatives of some $\Phi$ functional are called $\Phi$-derivable:
\begin{equation}\label{eq:phisigma}
\Sigma_{\text{MB}} (1,1') = \frac{\delta \Phi[G]}{\delta G(1',1)}.
\end{equation} 
In this case the resulting single-particle Green's function is guaranteed to fulfill macroscopic conservation laws.  Indeed, these conservation laws can be shown to follow directly from the invariant property of $\Phi$ under the relevant transformations of the Green's function~\cite{Stefanucci2013Book, Luttinger_1960, Baym1962}. 
These conservation laws are of particular relevance for quantum transport:  for instance, the conservation of particles is intimately related to the continuity equation.

Despite all of that, this scenario introduces a non-trivial problem if one attempts to find the interacting  Green's function.
Because of its  $\Phi$-derivable property, the self-energy is a functional of the {\it interacting} single-particle Green's function itself which in turn can be found only through the knowledge of the self-energy. Therefore, conservation laws are satisfied if and only if a self-consistent procedure is employed.
Because of the self-consistent nature of the problem, it is reasonable to resort to numerical techniques to tackle it~\cite{Stan2009, Balzer_2010, Talarico2019}.
Approximations to the full $\Phi$-functional are obtained by including a subset of skeleton diagrams. Examples of such approximations to treat the electron--electron interaction are provided by the Hartree--Fock (HF), Second Born (2B) and $GW$ approximations. Self-consistent solutions to the Dyson equations which use any of these approximations to the self-energy are automatically conserving. Below, we present the diagrammatic representation of these approximations and we briefly recall their properties. 

The first-order approximation for the self-energy, 
i.e. the HF approximation, has the following functional form
\begin{align}
\label{eq:HF}
\Sigma_{\text{HF}} [G](1,1') &= -\im \delta(1,1') \int \dif  \bar{1} v(1, \bar{1}) G(\bar{1},\bar{1}^+) +\im G (1,1') v(1^+,1').
\end{align}
Its diagrammatic representation is depicted in Fig.~\ref{fig:HF}.
This approximation describes how a particle moves freely under the influence of an effective potential which depends on all the other particles, that is the HF self-energy includes the effects of the interaction through a mean-field approximation.
\begin{figure}[t]
\begin{tcolorbox}
	\begin{center}
		\includegraphics[width=.9\textwidth]{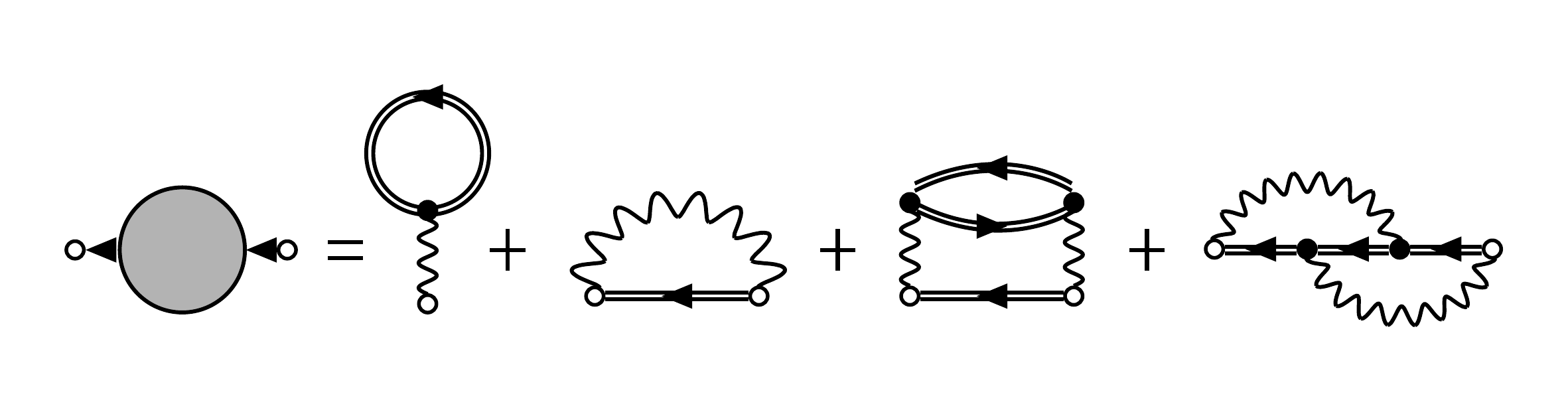}
		\caption{Diagrammatic representation of the second Born self-energy Eq.~\eqref{eq:secB}. Beside the HF terms, it contains the first order bubble diagram (third term) and the second order exchange diagram (fourth term).}
		\label{fig:2B}
	\end{center}
\end{tcolorbox}
\end{figure}
The HF self-energy requires only one integral over the spatial degrees of freedom and not over time due to the delta like structure of $v(z,z')$. This approximation is mostly used to take into account the mean-field effect like the global shift of the single-particle energies due to the many-body interaction.

Up to the second-order approximation for the self-energy, the first example which one encounters is the 2B: 
\begin{align}
\label{eq:secB}
\Sigma_{\text{2B}} [G](1,1') &= \Sigma_{\text{HF}} [G](1,1') -\im^2  G(1,1') \int \dif  \bar{1} \dif  \bar{2}v(1, \bar{1} )G(\bar{1}, \bar{2})  G( \bar{2},\bar{1}) v(1',\bar{2}) \nonumber \\
&+\im^2 \int \dif  \bar{1} \dif \bar{2} v(1, \bar{1} )G(1,\bar{2})G(\bar{2}, \bar{1})  G( \bar{2}, 1') v(1',\bar{2}) .
\end{align}
Here, in addition to the time-local part of the self-energy ($\Sigma_{\text{HF}}$),  we have terms up to the second order in the Coulomb interaction $v(1,2)= v(\mathbf{x}_1, \mathbf{x}_2) \delta(z_1,z_2)$. 
The first term after the HF-part of the self-energy is generally denoted as the
bubble diagram, it describes propagation of a particle (or hole) 
while interacting with particle--hole pair, i.e., it includes
effects of the polarization of the media due to uneven density distribution of particles and holes. The last term is nothing but the second order correction to the Fock term, second term in the r.h.s of Eq.~\eqref{eq:HF}, see Fig.~\ref{fig:2B}. 
\begin{figure}[b!]
\begin{tcolorbox}
	\begin{center}
		\includegraphics[width=.9\textwidth]{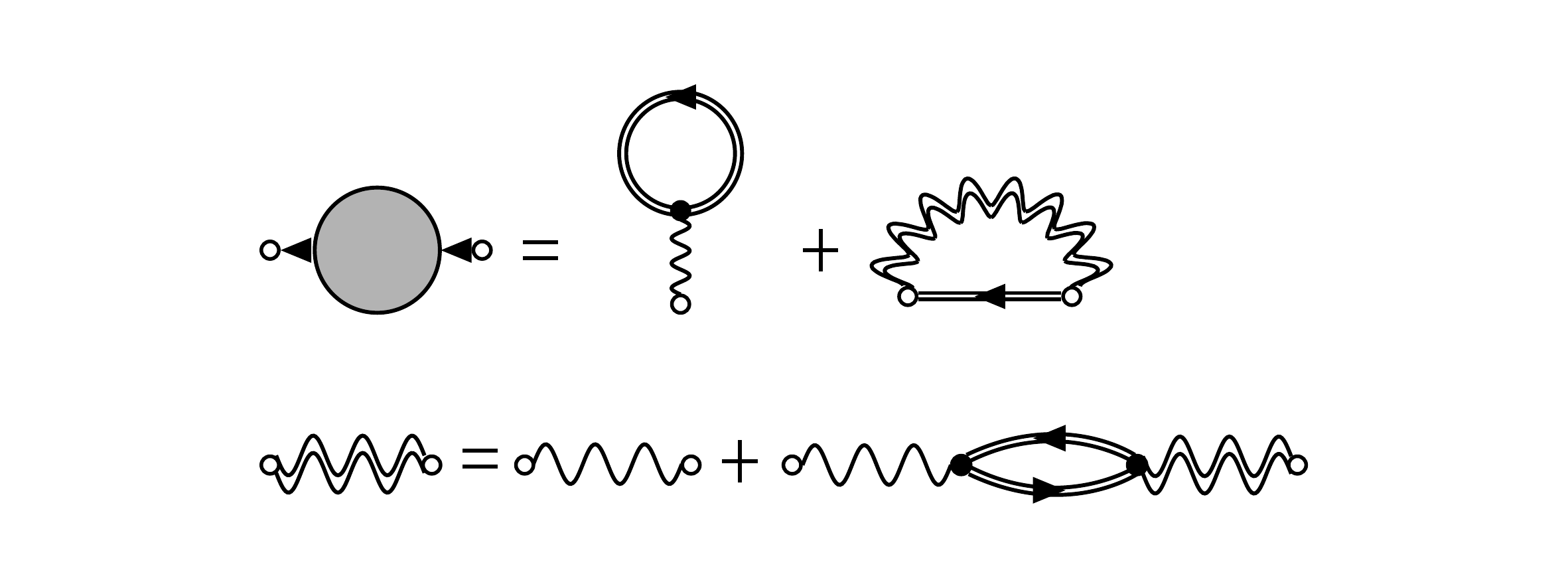}
		\caption{Diagrammatic representation of the $GW$ self-energy Eq.~\eqref{eq:GW} (top). In this approximation the Fock term is calculated with the dynamically screened interaction $W$ (double wavy solid line). In the random phase approximation the latter quantity satisfies its own Dyson equation Eq.~\eqref{eq:W} (bottom).  }
		\label{fig:GW}
	\end{center}
\end{tcolorbox}
\end{figure}
Notice that also the 2B does not require integration over time, once again due to the delta-like structure of the two-body interaction potential.

In the $GW$ approximation the electronic self-energy takes the form 
\begin{equation}
\label{eq:GW}
\Sigma_{GW}[G] (1,1') = \Sigma_{\text{H}}(1,1') + \im G(1,1')  W(1,1')
\end{equation}
with $\Sigma_{\text{H}}$ being the Hartree part of the self-energy, the first term of the r.h.s of Eq. \eqref{eq:HF}, and where the dynamically screened interaction $W$ satisfies the Dyson equation 
\begin{equation}
\label{eq:W}
W(1,1') = v(1,1') + \int \dif  \bar{1} \dif  \bar{2}  v(1,\bar{1}) P(\bar{1},\bar{2}) W(\bar{2}, 1'),
\end{equation}
where the polarization $P$ is usually approximated in the random-phase approximation (RPA) as $P(1,1') = -\im G(1,1')G(1',1)$, see Fig.~\ref{fig:GW}.
The $GW$ approximation can be seen as a dynamically screened exchange approximation able to describe the effects of long-range interaction.
From the computational point of view it is more complex than the HF or the 2B as it requires to solve the equation for the dressed interaction $W$, which is a Dyson-like equation and it involves integration in both time and spatial degrees of freedom.
Other choices for the polarization diagram are possible~\cite{Stefanucci2013Book, Schluenzen2019} but, even if they result in more accurate and precise approximations, 
they typically make the computation more demanding.
In fact, this corresponds to include in the set of equations to be solved a vertex functional which as a consequence leads to a more involved set of equations 
than the ones including only the Green's function and the self-energy. 
This set, named Hedin equations~\cite{Hedin1965, Stefanucci2013Book}, is composed of five equations which have to be solved at the same time and self-consistently.
The difficulties arise mostly due to the nature of the vertex functional which has in general a tensor structure in the localized base and it is not trivially manageable in the Keldysh space.

\begin{figure}[t!]
\begin{tcolorbox}
	\begin{center}
			\includegraphics[width=.9\textwidth]{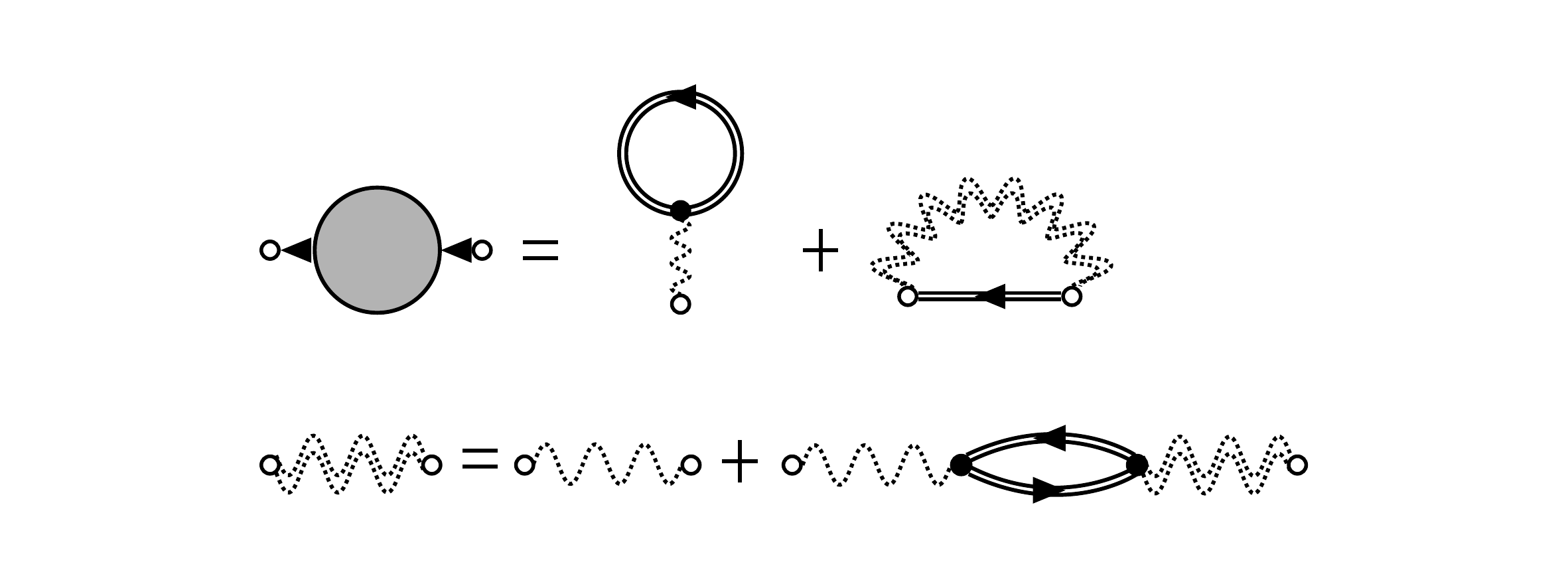}
			\caption{Diagrammatic representation of the $GD$ self-energy Eq.~\eqref{eq:GD} (top). In this approximation the second term is calculated with the fully-dressed boson propagator $D$ (double wavy dotted line). In the random phase approximation the latter quantity satisfies its own Dyson equation Eq.~\eqref{eq:D} (bottom).}
					\label{fig:GD}
	\end{center}
\end{tcolorbox}
\end{figure}

For the description of correlation effects due to the electron--boson interaction, a popular self-energy approximation can be easily obtained by replacing the screening interaction in the $GW$ self-energy (Eq.~\eqref{eq:GW}) with a fully-dressed boson propagator $W(1,1') \to D(1,1') = -\im \left\langle {\text T}_\gamma \left[ \Delta \hat{\phi}(1) \Delta \hat{\phi} (1') \right] \right\rangle $, where $\Delta \hat{\phi}_\mub = \hat{\phi}_\mub - \phi_\mub$ is a fluctuation operator and $\phi_\mub$ the boson field expectation value.
The explicit expression for the electron self-energy is given by
 \begin{equation}
\label{eq:GD}
\Sigma_{GD}[G, D] (1,1') = \Sigma_{\text{H}}(1,1') + \im G(1,1') \lambda(1)  D(1,1') \lambda(1')
\end{equation}
here the first term $\Sigma_{\text{H}}$ is know as the Hartree or Ehrenfest contribution and describes how electrons are influenced at the mean-field level by the potential due to boson fields. The second term 
is a time-nonlocal memory contribution describing single-boson absorption/emission processes. The fluctuation Green's function for the boson field is dressed within the random phase approximation $\Pi(1,1') = - \im \lambda(1) \lambda(1')  G(1,1') G(1',1) $, that describes (at the first order approximation) boson induced electron--hole excitation processes. It fulfills the following Dyson equation
\begin{equation}
\label{eq:D} 
D(1,1') = d(1,1') + \int \dif \bar{1} \dif \bar{2} \, d(1,\bar{1}) \Pi(\bar{1},\bar{2}) D(\bar{2}, 1'),
\end{equation}
where $d$ is the non-interacting boson Green's function, cf.~Eq.~\eqref{eq:dysongreen}.
The diagrammatic representation for the electron--boson self-energy is depicted in Fig.~\ref{fig:GD}.

\subsection{Embedding and inbedding techniques}\label{sec:Embedding}
In this section, we specialize the discussion to open quantum systems by introducing a partitioning into spatially separated subsystems. In quantum transport, one is typically interested in a junction setup like the one shown in Fig. \ref{fig:sys}, described by the Hamiltonian in Eq.~\eqref{eq:intsyshcoupl}. To describe this, we move to the representation of Green's functions in terms of sites localized to the $C$ and $\alpha$ regions:
\begin{align}
G_{ij}\left(z_{1},z_{2}\right) & = -\im\left\langle {\text T}_{\gamma}\left[\hat{a}_{i,\text{H}}\left(z_{1}\right)\hat{a}_{j,\text{H}}^{\dagger}\left(z_{2}\right)\right]\right\rangle , \label{eq:G_1_basis} \\
G_{2}\left(i\left(z_{1}\right),j\left(z_{2}\right);i'\left(z_{1}'\right),j'\left(z_{2}'\right)\right) & = \left(-\im\right)^{2}\left\langle {\text T}_{\gamma}\left[\hat{a}_{i,\text{H}}\left(z_{1}\right)\hat{a}_{j,\text{H}}\left(z_{2}\right)\hat{a}_{j',\text{H}}^{\dagger}\left(z_{2}'\right)\hat{a}_{i',\text{H}}^{\dagger}\left(z_{1}'\right)\right]\right\rangle , \label{eq:G_2_basis}
\end{align}
where the field operators $\hat{a}_{i}$ may refer to states in the $C$ or $\alpha$ regions. Focusing on the one-particle Green's function, one can introduce subsystem-local, $\mathbf{G}_{CC}$ and $\mathbf{G}_{\alpha\alpha}$, as well as subsystem-coupling, $\mathbf{G}_{C\alpha}$ and $\mathbf{G}_{\alpha C}$, block matrices of the Green's function, as follows
\begin{equation}\label{eq:blockmatrixstructure}
    \mathbf{G}\left(z_{1},z_{2}\right)=\left(\begin{array}{cccc}
\mathbf{G}_{11}\left(z_{1},z_{2}\right) & \mathbf{G}_{12}\left(z_{1},z_{2}\right) & \cdots & \mathbf{G}_{1C}\left(z_{1},z_{2}\right)\\
\mathbf{G}_{21}\left(z_{1},z_{2}\right) & \mathbf{G}_{22}\left(z_{1},z_{2}\right) & \cdots & \mathbf{G}_{2C}\left(z_{1},z_{2}\right)\\
\vdots & \vdots & \ddots & \vdots\\
\mathbf{G}_{C1}\left(z_{1},z_{2}\right) & \mathbf{G}_{C2}\left(z_{1},z_{2}\right) & \cdots & \mathbf{G}_{CC}\left(z_{1},z_{2}\right)
\end{array}\right).
\end{equation}
Here, the boldface symbols represent matrix objects with subscripts signifying the partitions in the transport setup. These objects are expanded in terms of the single-particle basis associated with the setup. A special case is the interacting central region, for which we sometimes write simply $G=\mathbf{G}_{CC}$ for brevity. Notice that in the matrix~\eqref{eq:blockmatrixstructure} there are lead--lead coupling terms $\mathbf{G}_{\alpha\beta}$, where $\alpha\neq\beta$, unlike in the Hamiltonian. This is because for all contour times $z$ there will be $\alpha C$ and $C \beta$ terms in the Hamiltonian, so that there must exist a Green's Function describing a lead-to-lead hopping process that is mediated by the central region. We project the general equation of motion, Eq.~\eqref{eq:KBeq} onto the $CC$ block to obtain the corresponding equation of motion for the Green's function $\mathbf{G}_{CC}$ of the central region:
\begin{align}\label{eq:G_CC_diff}
    \left[\im\frac{\ud}{\ud z_{1}}-\mathbf{h}_{CC}\left(z_{1}\right)\right]\mathbf{G}_{CC}\left(z_{1},z_{2}\right) & =\mathbf{1}_{CC}\delta\left(z_{1},z_{2}\right) + \sum_\alpha\mathbf{h}_{\alpha C}\left(z_{1}\right)\mathbf{G}_{\alpha C}\left(z_{1},z_{2}\right) \nonumber \\
    & + \int_{\gamma} \ud \bar{z} \mathbf{\Sigma}_{\text{MB},CC}(z_{1}, \bar{z})  \mathbf{G}_{CC}(\bar{z},z_{2}).
\end{align}

We next seek to express the lead--molecule Green's function $\mathbf{G}_{\alpha C}$ in terms of calculable objects. To this end, we introduce $\mathbf{g}_{\alpha\alpha}\left(z_{1},z_{2}\right)$ as the Green's function of an isolated lead $\alpha$ corresponding to the Hamiltonian with the matrix block $\mathbf{h}_{\alpha\alpha}(z)$:
\begin{equation}\label{eq:g_alpha}
\left[\im\frac{\ud}{\ud z_{1}}-\mathbf{h}_{\alpha\alpha}\left(z_{1}\right)\right]\mathbf{g}_{\alpha\alpha}\left(z_{1},z_{2}\right)=\mathbf{1}_{\alpha\alpha}\delta\left(z_{1},z_{2}\right)
\end{equation}
The delta function on the Konstantinov--Perel' contour satisfies the identities
\begin{equation}
\int_{\gamma}\ud\zb\mathbf{g}_{\alpha\alpha}^{-1}\left(z_{1},\bar{z}\right)\mathbf{g}_{\alpha\alpha}\left(\bar{z},z_{2}\right) = \mathbf{1}_{\alpha\alpha}\delta\left(z_{1},z_{2}\right) =  \int_{\gamma}\ud\zb\mathbf{g}_{\alpha\alpha}\left(z_{1},\bar{z}\right)\mathbf{g}_{\alpha\alpha}^{-1}\left(\bar{z},z_{2}\right) ,
\end{equation}
\begin{equation}
\int_{\gamma}\ud\zb\left[\im\frac{\ud}{\ud z_{1}}-\mathbf{h}_{\alpha\alpha}\left(z_{1}\right)\right]\delta\left(z_{1},\bar{z}\right)\mathbf{g}_{\alpha\alpha}\left(\bar{z},z_{2}\right) = \mathbf{1}_{\alpha\alpha}\delta\left(z_{1},z_{2}\right).
\end{equation}
It therefore follows that
\begin{align}\label{eq:gminus}
\left[\im\frac{\ud}{\ud z_{1}}-\mathbf{h}_{\alpha\alpha}\left(z_{1}\right)\right]\delta\left(z_{1},z_{2}\right)=\mathbf{g}_{\alpha\alpha}^{-1}\left(z_{1},z_{2}\right).
\end{align}
Now, the equation of motion of the $\alpha C$ and $C \alpha$ Green's functions can be obtained by projecting Eqs.~\eqref{eq:KBeq} and~\eqref{eq:KBeq_2} onto the corresponding matrix block:
\begin{align}\label{eq:galphaC}
\left[\im\frac{\ud}{\ud z_{1}}-\mathbf{h}_{\alpha\alpha}\left(z_{1}\right)\right]\mathbf{G}_{\alpha C}\left(z_{1},z_{2}\right)=\mathbf{h}_{\alpha C}\left(z_{1}\right)\mathbf{G}_{CC}\left(z_{1},z_{2}\right).
\end{align}
Substituting Eq.~\eqref{eq:gminus} into Eq.~\eqref{eq:galphaC} leads to
\begin{equation}
\int_{\gamma}\ud\bar{z}\mathbf{g}_{\alpha\alpha}^{-1}\left(z_{1},\bar{z}\right)\mathbf{G}_{\alpha C}\left(\bar{z},z_{2}\right)=\mathbf{h}_{\alpha C}\left(z_{1}\right)\mathbf{G}_{CC}\left(z_{1},z_{2}\right).
\end{equation}
We then multiply Eq.~\eqref{eq:galphaC} from the left-hand side by $\mathbf{g}_{\alpha\alpha}\left(z_{1},\bar{z}'\right)$ and integrate over the full Konstantinov--Perel' contour to give
\begin{equation}\label{eq:GF_aC}
\mathbf{G}_{\alpha C}\left(z_{1},z_{2}\right)=\int_{\gamma}\ud\bar{z}\,\mathbf{g}_{\alpha\alpha}\left(z_{1},\bar{z}\right)\mathbf{h}_{\alpha C}\left(\bar{z}\right)\mathbf{G}_{CC}\left(\bar{z},z_{2}\right).
\end{equation}

Finally, we are able to substitute Eq.~\eqref{eq:GF_aC} into Eq.~\eqref{eq:G_CC_diff} to obtain the equations of motion for $\mathbf{G}_{CC}$ purely in terms of objects defined on the molecular region:
\begin{equation}\label{EOM_GCC1}
\im\frac{\ud\mathbf{G}_{CC}\left(z_{1},z_{2}\right)}{\ud z_{1}}=\mathbf{h}_{CC}\left(z_{1}\right)\mathbf{G}_{CC}\left(z_{1},z_{2}\right)+\mathbf{1}_{CC}\delta\left(z_{1},z_{2}\right)+\int_{\gamma}\ud\bar{z}\mathbf{\Sigma}_{CC}\left(z_{1},\bar{z}\right)\mathbf{G}_{CC}\left(\bar{z},z_{2}\right),
\end{equation}
where $\mathbf{\Sigma}_{CC}=\mathbf{\Sigma}_{\text{em}}+\mathbf{\Sigma}_{\text{MB},CC}$ is the effective self-energy composed of the usual many-body part, and
\begin{align}\label{eq:embedding_self_energy}
\mathbf{\Sigma}_{\text{em}}\left(z_{1},z_{2}\right)=\sum_\alpha\mathbf{\Sigma}_{\text{em},\alpha}\left(z_{1},z_{2}\right)=\sum_\alpha\mathbf{h}_{C\alpha}\left(z_{1}\right)\mathbf{g}_{\alpha\alpha}\left(z_{1},z_{2}\right)\mathbf{h}_{\alpha C}\left(z_{2}\right)
\end{align}
defines the so-called the embedding self-energy. It is worth pointing out that in this representation the embedding self-energy is also a matrix projected onto the $CC$ block. A similar expression exists for the derivative with respect to the second contour time variable:
\begin{equation}\label{EOM_GCC2}
    -\im\frac{\ud\mathbf{G}_{CC}\left(z_{1},z_{2}\right)}{\ud z_{2}}=\mathbf{G}_{CC}\left(z_{1},z_{2}\right)\mathbf{h}_{CC}\left(z_{2}\right)+\mathbf{1}_{CC}\delta\left(z_{1},z_{2}\right)+\int_{\gamma}\ud\bar{z}\mathbf{G}_{CC}\left(z_{1},\bar{z}\right)\mathbf{\Sigma}_{CC}\left(\bar{z},z_{2}\right).
\end{equation}

\begin{figure}[t]
\centering
\includegraphics[width=0.75\textwidth]{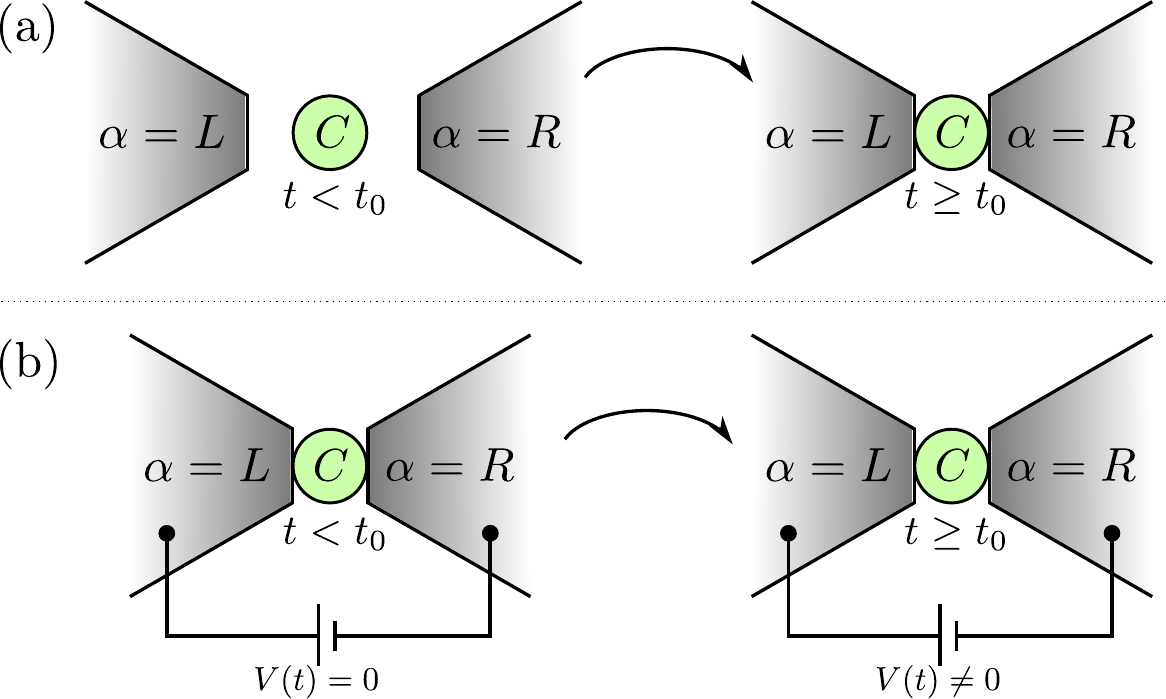}
\caption{Schematic representations of the transport setup with switch-on time $t_0$. (a) Partitioned approach by Caroli et al.~\cite{Caroli1971a,Caroli1971b}: For $t<t_0$ the leads are disconnected from the central region, and they are in separate thermodynamical equilibria (different temperatures and chemical potentials). Once the contact is established for $t\geq t_0$ charge carriers start to flow. (b) Partition-free approach by Cini~\cite{Cini1980}: The whole system is for $t<t_0$ in a global thermodynamical equilibrium (unique temperature and chemical potential). For $t\geq t_0$ a bias is applied in the leads and charge carriers start to flow. From~\cite{Ridley2018JLTP}.}
\label{fig:partition}
\end{figure}

The presence of the $\alpha C$ coupling in the Hamiltonian~\eqref{eq:intsyshcoupl} for $z\in C_{M}$ distinguishes the partition-free approach by Cini~\cite{Cini1980}, which is illustrated schematically in Fig.~\ref{fig:partition}(b), from the partitioned approach by Caroli et al.~\cite{Caroli1971a,Caroli1971b}, shown in  Fig.~\ref{fig:partition}(a). Eq.~\eqref{eq:embedding_self_energy} implies that, in the partitioned approach, the embedding self-energy is always zero when one of $z_{1}, z_{2}$ lies on the vertical branch. This means that the ${M/\urcorner/\ulcorner}$ components of the embedding self energy must vanish in the partitioned approach. 
We finally note that, according to Stefanucci and Almbladh's Theorem of Equivalence, the two types of system preparation always result in the same steady-state properties~\cite{Stefanucci2004}.

So far we have been focused on the SPGF of the microscopic scattering region and on how we could include the influence of correlation effects and couplings to macroscopic external reservoirs on this system. Nonetheless, one could also be interested in the effects that the correlated central system has on the non-interacting leads, thus one could ask the question: Is it possible to infer physical quantities of the reservoirs from the knowledge of the molecular Green's function? In other words, is it possible to explore the back-action that the microscopic scattering system has on the macroscopic electronic leads? The answer is affirmative and we now show how this can be achieved by using the inbedding technique~\cite{Stefanucci2013Book,Myohanen2009, Myohanen2012}. 
We can use the same argument as before to obtain equation of motion for the Green's function $\mathbf{G}_{\alpha \alpha}$ projected into the subspace of the reservoir $\alpha$ 
\be\label{eq:G_alpha_diff}
\left[\im\frac{\ud}{\ud z_{1}}-\mathbf{h}_{\alpha\alpha}\left(z_{1}\right)\right]\mathbf{G}_{\alpha \alpha}\left(z_{1},z_{2}\right)=\mathbf{1}_{\alpha \alpha}\delta\left(z_{1},z_{2}\right) +\mathbf{h}_{ \alpha C }\left(z_{1}\right)\mathbf{G}_{ C \alpha}\left(z_{1},z_{2}\right).
\ee
Notice that in this expression the MB self-energy is missing due to the prior assumption that there is no interaction in the reservoirs.
Using the adjoint of Eq.~\eqref{eq:GF_aC} in this expression we obtain the following integro--differential equation
\begin{align}\label{eq:G_alpha_int}
    \left[\im\frac{\ud}{\ud z_{1}}-\mathbf{h}_{\alpha\alpha}\left(z_{1}\right)\right]\mathbf{G}_{\alpha \alpha}\left(z_{1},z_{2}\right)=\mathbf{1}_{\alpha \alpha}\delta\left(z_{1},z_{2}\right) +
    \int_{\gamma} \ud\bar{z} \mathbf{\Sigma}_{\text{in},\alpha} \left(z_{1},\bar{z}\right)  \mathbf{g}_{\alpha\alpha}\left(\bar{z},z_{2}\right) 
\end{align}
where we define the inbedding self-energy as 
\begin{align}\label{eq:inbedding_self_energy}
\mathbf{\Sigma}_{\text{in},\alpha}\left(z_{1},z_{2}\right)=\mathbf{h}_{\alpha C}\left(z_{1}\right)\mathbf{G}_{C C}\left(z_{1},z_{2}\right)\mathbf{h}_{C \alpha }\left(z_{2}\right),
\end{align}
the latter quantity is fully known once a solution of the KBE has been found for the open interacting system.
The equation for $\mathbf{G}_{\alpha \alpha}$ can be integrated using Eq.~\eqref{eq:g_alpha} and can be cast into the following Dyson equation:
\begin{align}\label{eq:G_alpha_dyson}
   \mathbf{G}_{\alpha \alpha}\left(z_{1},z_{2}\right)=\mathbf{g}_{\alpha \alpha} \left(z_{1},z_{2}\right) +
    \int_{\gamma} \ud\bar{z} \ud\bar{z}' \mathbf{g}_{\alpha\alpha}\left(z_1, \bar{z}\right)  \mathbf{\Sigma}_{\text{in},\alpha} \left(\bar{z}, \bar{z}'\right)  \mathbf{g}_{\alpha\alpha}\left(\bar{z}',z_{2}\right). 
\end{align}

The inbedding technique has been used to obtain physical quantities like density, current, energy, etc. in the reservoirs and is extremely relevant in experimental realizations of these molecular junctions where one has access only to the properties of the reservoirs~\cite{Myohanen2009, Dutta2020}.

\subsection{Two-times scheme: Kadanoff--Baym and Dyson equations}\label{sec:kbet2}


Since the interactions are constrained in the central region ($CC$ block of the previous matrix structures), we consider the equations of motion~\eqref{EOM_GCC1} and~\eqref{EOM_GCC2} in this sub-block and we drop the subscripts for brevity: $G=\mathbf{G}_{CC}$. These objects are thus understood as matrices and they are expanded in the single-particle basis of the problem at hand. For practical calculations, we will also outline some exemplary self-energy expansions explicitly with the orbital indices.

The equations of motion for the lesser $\Gcalh^<$ and greater $\Gcalh^>$ single-particle Green's function are known as the Kadanoff--Baym Equations (KBE)~\cite{Kadanoff1962} and are listed (in matrix form) in Tab.~\ref{tab:kbe-list}. We also list the equations of motion for the other Keldysh components (${M/\urcorner/\ulcorner}$). In Tab.~\ref{tab:dyson-list}, we list the equivalent representations in the Dyson-equation form, cf.~Eq.~\eqref{eq:dysongreen}. The notations $\cdot$ and $\star$ denote the real-time and imaginary-time convolutions:
\begin{align}
\left [A \cdot B\right ] (z_{1},z_{2}) & \equiv \int_{t_0}^\infty \dif \tb\,
A(z_{1},\tb) B(\tb,z_{2}), \label{Convolution_dot} \\
\left [ A \star B \right ](z_{1},z_{2}) & \equiv -\im
\int_{0}^{\beta} \dif \taub A(z_{1},\taub) B(\taub,z_{2}), \label{Convolution_star}
\end{align}
with $\beta$ the inverse temperature.

\begin{table}[t]
\centering
\begin{tcolorbox}
\begin{tabular}{c|c}
Component & Kadanoff--Baym equation \\
\hline
\hline \\
Ret/Adv & $\left[\im{\partial}_{t_{1}}-h_\HF\left(t_{1}\right)\right]\Gcalh^{r/a}\left(t_{1},t_{2}\right)=\mathbb{1}\delta\left(t_{1}-t_{2}\right)+\left[\Sigma^{r/a}\cdot\Gcalh^{r/a}\right]\left(t_{1},t_{2}\right)$ \\ 
& $\Gcalh^{r/a}\left(t_{1},t_{2}\right)\left[-\im\overset{\leftarrow}{\partial}_{t_{2}}-h_\HF\left(t_{2}\right)\right]=\mathbb{1}\delta\left(t_{1}-t_{2}\right)+\left[\Gcalh^{r/a}\cdot\Sigma^{r/a}\right]\left(t_{1},t_{2}\right)$ \\ \\
Lss/Gtr & $\left[\im{\partial}_{t_{1}}-h_\HF\left(t_{1}\right)\right]\Gcalh^{\lessgtr}\left(t_{1},t_{2}\right)=\left [\Sigma^\lessgtr \cdot \Gcalh^{a} + \Sigma^{r} \cdot
\Gcalh^{\lessgtr} + \Sigma^\urcorner \star \Gcalh^\ulcorner
\right](t_{1},t_{2})$ \\ 
& $\Gcalh^{\lessgtr}\left(t_{1},t_{2}\right)\left[-\im\overset{\leftarrow}{\partial}_{t_{2}}-h_\HF\left(t_{2}\right)\right]=\left [\Gcalh^\lessgtr \cdot \Sigma^{a} + \Gcalh^{r} \cdot
\Sigma^\lessgtr  + \Gcalh^\urcorner \star \Sigma^\ulcorner \right
]\left(t_{1},t_{2}\right)$ \\ \\
Right & $\left[\im{\partial}_{t_{1}}-h_\HF\left(t_{1}\right)\right]\Gcalh^{\urcorner}\left(t_{1},\tau_{2}\right)=\left[\Sigma^{r}\cdot\Gcalh^{\urcorner}+\Sigma^{\urcorner}\star\Gcalh^{M}\right]\left(t_{1},\tau_{2}\right)$ \\ \\
Left & $\Gcalh^{\ulcorner}\left(\tau_{1},t_{2}\right)\left[-\im\overset{\leftarrow}{\partial}_{t_{2}}-h_\HF\left(t_{2}\right)\right]=\left[\Gcalh^{\ulcorner}\cdot\Sigma^{a}+\Gcalh^{M}\star\Sigma^{\ulcorner}\right]\left(\tau_{1},t_{2}\right)$ \\ \\
Matsubara & $\left[-{\partial}_{\tau_{1}}-h_\HF^{M}\left(\tau_{1}\right)\right]\Gcalh^{M}\left(\tau_{1},\tau_{2}\right)=\im \mathbb{1} \delta\left(\tau_{1}-\tau_{2}\right)+\left[\Sigma^{M}\star\Gcalh^M\right]\left(\tau_{1},\tau_{2}\right)$ \\ & $\Gcalh^M\left(\tau_{1},\tau_{2}\right)\left[\overset{\leftarrow}{\partial}_{\tau_{2}}-h_\HF^{M}\left(\tau_{2}\right)\right]=\im \mathbb{1} \delta\left(\tau_{1}-\tau_{2}\right)+\left[\Gcalh^{M}\star\Sigma^{M}\right]\left(\tau_{1},\tau_{2}\right)$ \\ 
\end{tabular}
\caption{Representations of the contour-time components of the Kadanoff--Baym equations.
}
\label{tab:kbe-list}
\end{tcolorbox}
\end{table}

\begin{table}[t]
\centering
\begin{tcolorbox}
\begin{tabular}{c|c} 
Component & Dyson equation \\
\hline
\hline \\
Ret/Adv & $\Gcalh^{r/a} (t_1,t_2) = \Big[\Gcalh_0^{r/a} + \Gcalh_0^{r/a} \cdot \Sigma^{r/a} \cdot \Gcalh^{r/a}\Big](t_1,t_2)$ \\ \\
Lss/Gtr & $\Gcalh^{\lessgtr} (t_1,t_2) = \Big[\Gcalh_0^{\lessgtr} + \Gcalh_0^{r} \cdot \Sigma^{\lessgtr} \cdot \Gcalh^{a} + \Gcalh_0^{\lessgtr} \cdot \Sigma^{a} \cdot \Gcalh^{a} +\Gcalh_0^{r} \cdot \Sigma^{r} \cdot \Gcalh^{\lessgtr}$ \\
& $+ \Gcalh_0^{\urcorner} \star \Sigma^{M} \star G^{\ulcorner} + \Gcalh_0^{r} \cdot \Sigma^{\urcorner} \star \Gcalh^{\ulcorner}  + \Gcalh_0^{\urcorner} \star \Sigma^{\ulcorner} \cdot \Gcalh^{a} \Big](t_1,t_2)$ \\ \\
Right & $\Gcalh^{\urcorner} (t_1,\tau_2)$ \\
& $= \Big[\Gcalh_0^{\urcorner} + \Gcalh_0^{r} \cdot \Sigma^{r} \cdot \Gcalh^{\urcorner}+\Gcalh_0^{r} \cdot \Sigma^{\urcorner} \star \Gcalh^{M} +\Gcalh_0^{\urcorner} \star \Sigma^{M} \star \Gcalh^{M} \Big](t_1,\tau_2)$ \\ \\
Left & $\Gcalh^{\ulcorner} (\tau_1,t_2) $ \\
& $= \Big[\Gcalh_0^{\ulcorner} + \Gcalh_0^{\ulcorner} \cdot \Sigma^{a} \cdot \Gcalh^{a} +G_0^{M} \star \Sigma^{\ulcorner} \cdot \Gcalh^{a} +\Gcalh_0^{M} \cdot \Sigma^{M} \star \Gcalh^{\ulcorner} \Big](\tau_1,t_2)$ \\ \\
Matsubara & $ \Gcalh^{M} (\tau_1,\tau_2) = \Big[\Gcalh_0^{M} + \Gcalh_0^{M} \star \Sigma^{M} \star \Gcalh^{M}\Big](\tau_1,\tau_2)$
\end{tabular}
\caption{Representations of the contour-time components of the Dyson equations.
}
\label{tab:dyson-list}
\end{tcolorbox}
\end{table}

The imaginary-time convolutions involve the so-called \emph{mixed} functions, the left $X^\ulcorner(\tau,t)$ and right  $X^\urcorner(t,\tau)$ functions with one real time and one imaginary time. The naming convention `left' and `right' can be motivated by the fact that the imaginary-time argument is to the left or right in the arguments of $X$. The imaginary-time convolution contains information about the initially correlated state, as well as information of the (if present) initial contacts to external reservoirs~\cite{Stefanucci2013Book}. The retarded and advanced functions are defined as
\begin{equation}
 X^{r/a}(t,t') = \pm \theta(\pm(t-t')) \left [ X^>(t,t') - X^<(t,t')
\right ] . \label{RetAdv}
\end{equation}

The quantity $\Sigma$ in the KBE and Dyson equations is the correlation $+$ embedding part of the self-energy. The time-local mean-field or Hartree--Fock (HF) part of the self-energy is incorporated in the HF Hamiltonian $h_\HF$, defined as
\begin{equation}
 h_{\HF,ij}(t) = h_{ij}(t) + \sum_{mn} w_{imnj}(t) \rho_{nm}(t),
 \label{HFhamiltonian}
\end{equation}
where $\rho(t)= -\im \Gcalh^<(t,t)$ is the single-particle density
matrix and we have defined
\be\label{eq:w4}
w_{imnj}(t) \equiv \xi v_{imnj}(t) - v_{imjn}(t),
\ee
where we included a spin-degeneracy factor $\xi$ in front of the direct term (see the bubble diagram in Fig.~\ref{fig:HF})~\cite{Balzer2013book}.

The various equations for the components of the Green's function in Tabs.~\ref{tab:kbe-list} and~\ref{tab:dyson-list} are evaluated once the generic form of the self-energy components are inserted as input. 
Let us consider as an example the second-order Born self-energy in Eq.~\eqref{eq:secB}. For generic contour-time arguments, this quantity reads as, see Eq.~\eqref{eq:secB},
\be
\Sigma_{\text{2B}}(z_1,z_2) = \delta(z_1,z_2)\Sigma_{\text{HF}}(z_1) + \Sigma_{\text{2B}}^{\text{c}}(z_1,z_2) .
\ee
While the HF part is time-local and its contour-time components of two times are zero, the correlation part is expanded as
\begin{align}\label{eq:2b-expanded}
(\Sigma_{\text{2B}}^{\text{c}})_{ij}(z_1,z_2) = \sum_{mnpqrs} v_{irpn}(z_1) v_{mqsj}(z_2) & \left[ \xi G_{nm}(z_1,z_2)G_{pq}(z_1,z_2)G_{sr}(z_2,z_1) \right.\nonumber\\
& \left. - G_{nq}(z_1,z_2)G_{pm}(z_1,z_2)G_{sr}(z_2,z_1) \right] 
\end{align}
with a generic $4$-index Coulomb integral [cf.~Eq.~\eqref{coulomb}]. Also here, the spin-degeneracy factor $\xi$ appears for the direct term, see Fig.~\ref{fig:2B}~\cite{Balzer2013book}. It is also worth noting that the summation in Eq.~\eqref{eq:2b-expanded} over the basis indices can be reorganized for more efficient computation~\cite{Perfetto2019,Schluenzen2019,Tuovinen2019JCP}. While the interaction itself is instantaneous, it is customary to include explicit time-dependence for $v$ to describe, e.g., interaction quenches or adiabatic switching. Using the Langreth rules (see~\ref{app:langreth}), the components of the 2B self-energy are obtained in Tab.~\ref{tab:2b-list}. It is worth noting that the Matsubara component represents an equilibrium setting, where the Green's function depends only on the difference between the two time coordinates: $G^M(\tau_1,\tau_2)= G^M(\tau_1-\tau_2)\equiv G^M(\tau)$. 

\begin{table}[t]
\centering
\begin{tcolorbox}
\begin{tabular}{c|c} 
Component & $\Sigma_{\text{2B}}^{\text{c}}$ \\
\hline
\hline \\
Ret/Adv & $(\Sigma_{\text{2B}}^{\text{c}})_{ij}^{r/a}(t_1,t_2) = \underset{mnpqrs}{\sum}v_{irpn}v_{mqsj}\{\xi[G_{nm}^{r/a}(t_1,t_2)G_{pq}^>(t_1,t_2)G_{sr}^<(t_2,t_1)$\\
& $+G_{nm}^<(t_1,t_2)G_{pq}^<(t_1,t_2)G_{sr}^{a/r}(t_2,t_1)+G_{nm}^<(t_1,t_2)G_{pq}^{r/a}(t_1,t_2)G_{sr}^<(t_2,t_1)]$\\
&$-[G_{nq}^{r/a}(t_1,t_2)G_{pm}^>(t_1,t_2)G_{sr}^<(t_2,t_1)+G_{nq}^<(t_1,t_2)G_{pm}^<(t_1,t_2)G_{sr}^{a/r}(t_2,t_1)$\\
&$+G_{nq}^<(t_1,t_2)G_{pm}^{r/a}(t_1,t_2)G_{sr}^<(t_2,t_1)]\}$\\ \\
Lss/Gtr & $(\Sigma_{\text{2B}}^{\text{c}})_{ij}^\lessgtr(t_1,t_2)=\underset{mnpqrs}{\sum} v_{irpn} v_{mqsj}[ \xi G_{nm}^\lessgtr(t_1,t_2)G_{pq}^\lessgtr(t_1,t_2)G_{sr}^\gtrless(t_2,t_1)$\\
&$-G_{nq}^\lessgtr(t_1,t_2)G_{pm}^\lessgtr(t_1,t_2)G_{sr}^\gtrless(t_2,t_1) ]$ \\ \\
Right & $(\Sigma_{\text{2B}}^{\text{c}})_{ij}^\urcorner(t_1,\tau_2)=\underset{mnpqrs}{\sum} v_{irpn} v_{mqsj}[ \xi G_{nm}^\urcorner(t_1,\tau_2)G_{pq}^\urcorner(t_1,\tau_2)G_{sr}^\ulcorner(\tau_2,t_1)$\\
&$-G_{nq}^\urcorner(t_1,\tau_2)G_{pm}^\urcorner(t_1,\tau_2)G_{sr}^\ulcorner(\tau_2,t_1) ]$ \\ \\
Left & $(\Sigma_{\text{2B}}^{\text{c}})_{ij}^\ulcorner(\tau_1,t_2)=\underset{mnpqrs}{\sum} v_{irpn} v_{mqsj}[ \xi G_{nm}^\ulcorner(\tau_1,t_2)G_{pq}^\ulcorner(\tau_1,t_2)G_{sr}^\urcorner(t_2,\tau_1)$\\
&$-G_{nq}^\ulcorner(\tau_1,t_2)G_{pm}^\ulcorner(\tau_1,t_2)G_{sr}^\urcorner(t_2,\tau_1) ]$\\ \\
Matsubara & $(\Sigma_{\text{2B}}^{\text{c}})_{ij}^M(\tau)=\underset{mnpqrs}{\sum} v_{irpn} v_{mqsj}[ \xi G_{nm}^M(\tau)G_{pq}^M(\tau)G_{sr}^M(-\tau)$\\
&$-G_{nq}^M(\tau)G_{pm}^M(\tau)G_{sr}^M(-\tau) ]$
\end{tabular}
\caption{Various time-contour components of the second-order Born self-energy.}
\label{tab:2b-list}
\end{tcolorbox}
\end{table}

In addition, we can compute the components of the embedding self-energy for a molecular junction, assuming non-interacting leads, depending on where the two `time' arguments in the self-energy $\mathbf{\Sigma}_{\text{em}}\left(z_{1},z_{2}\right)$ in Eq.~\eqref{eq:embedding_self_energy} lie. These require calculation of the corresponding components of the Green's function $\mathbf{g}_{\alpha\alpha}\left(z_{1},z_{2}\right)$ of the isolated lead $\alpha$, which can be done by solving the equations of motion for the relevant field operators. For example, the lesser/greater components are given by 
\begin{align}
    \left[\mathbf{g}_{\alpha\alpha}^{\lessgtr}\left(t_{1},t_{2}\right)\right]_{kk'}=\pm \im\delta_{kk'}f\left(\pm\left(\epsilon_{k\alpha}-\mu\right)\right)\ex^{-\im\epsilon_{k\alpha}\left(t_{1}-t_{2}\right)}\ex^{-\im\psi_{\alpha}\left(t_{1},t_{2}\right)},
\end{align}
where $f(\omega)=\left(\ex^{\beta\omega}+1\right)^{-1}$ is the Fermi function and we introduce the integral phase factor, taking into account the time-dependence of the bias-voltage profile, see Eq.~\eqref{lead_energies}:
\begin{align}\label{eq:phase}
    \psi_{\alpha}\left(t_{1},t_{2}\right)\equiv\int_{t_{2}}^{t_{1}}V_{\alpha}\left(\bar{t}\right)\ud\bar{t} .
\end{align}
All the components of the self-energy can now be obtained by substituting these and similar expressions for the decoupled GFs into the embedding self-energy definition, Eq.~\eqref{eq:embedding_self_energy}. To obtain the retarded component, we Fourier transform that part of the expression which depends only on the time-difference $t_{1}-t_{2}$:
\begin{align}
    \left[\mathbf{\Sigma}_{\text{em}}^{r}\left(t_{1},t_{2}\right)\right]_{mn} & =\sum_{\alpha}\ex^{-\im\psi_{\alpha}\left(t_{1},t_{2}\right)}\int\frac{\ud\omega}{2\pi}\ex^{\im\omega\left(t_{1}-t_{2}\right)}\underset{k}{\sum}\frac{T_{m,k\alpha}T_{k\alpha,n}}{\omega-\epsilon_{k\alpha}+\im \eta} \nonumber \\
    & =\sum_{\alpha}\ex^{-\im\psi_{\alpha}\left(t_{1},t_{2}\right)}\int\frac{\ud\omega}{2\pi}\ex^{\im\omega\left(t_{1}-t_{2}\right)}\left[\Lambda_{\alpha,mn}\left(\omega\right)-\frac{\im}{2}\Gamma_{\alpha,mn}\left(\omega\right)\right],
\end{align}
where $\eta$ is a positive infinitesimal, and we have defined the \emph{level shift}:
\begin{align}\label{eq:shift}
    \Lambda_{\alpha,mn}\left(\omega\right)=\mathcal{P}\int\frac{\ud\omega^{\prime}}{2\pi}\frac{\Gamma_{\alpha,nm}\left(\omega^{\prime}\right)}{\omega-\omega^{\prime}},
\end{align}
where the symbol $\mathcal{P}$ corresponds to the Cauchy principal part, and the \emph{level width}:
\begin{align}\label{eq:width}
    \Gamma_{\alpha,mn}\left(\omega\right)=2\pi\sum_{k}T_{m,k\alpha}T_{k\alpha,n}\delta\left(\omega-\epsilon_{k\alpha}\right).
\end{align}
In a non-interacting system, $\Lambda\left(\omega\right)$ moves the location of molecular eigenmodes in frequency space whereas $\Gamma\left(\omega\right)$ determines the spread of those eigenmodes. The lifetime of eigenmodes typically corresponds to the inverse of the level width, $1/\Gamma$.
In Tab.~\ref{tab:emb-list} we list all components of the embedding self-energy in terms of these well-defined frequency-dependent objects.

A common approach in this context is the wide-band approximation (WBA), where the level width is assumed to be independent of frequency, $\Gamma_\alpha(\w)=\Gamma_\alpha$. This corresponds to the lead density of states being featureless in the energy scale of the molecular system. With WBA, the level-shift matrix vanishes due to Kramers--Kronig relations, and the retarded/advanced embedding self-energy becomes time-local, $\mathbf{\Sigma}_{\text{em}}^{r/a}(t_1,t_2) = \mp \im \sum_\alpha \Gamma_\alpha \delta(t_1-t_2)/2 = \mp \im \Gamma \delta(t_1-t_2)/2$. We will return to this in more detail with analytic solutions to the Kadanoff--Baym equations for non-interacting systems in Sec.~\ref{sec:nonint}.

\begin{table}
\centering
\begin{tcolorbox}
\begin{tabular}{ c|c } 
Component & $\Sigma_{\text{em}}$ \\
\hline
\hline \\
Ret/Adv & $\mathbf{\Sigma}_{\text{em}}^{r/a}(t_{1},t_{2})=\sum_{\alpha}\ex^{-\im\psi_{\alpha}\left(t_{1},t_{2}\right)}\int\frac{\ud\omega}{2\pi}\ex^{-\im\omega\left(t_{1}-t_{2}\right)}\left[\Lambda_{\alpha}\left(\omega\right)\mp\frac{\im}{2}\Gamma_{\alpha}\left(\omega\right)\right]$\\ \\
Lss/Gtr & $\mathbf{\Sigma}_{\text{em}}^{\lessgtr}\left(t_{1},t_{2}\right)=\pm \im\sum_{\alpha}\ex^{-\im\psi_{\alpha}\left(t_{1},t_{2}\right)}\int\frac{\ud\omega}{2\pi}\Gamma_{\alpha}\left(\omega\right)f\left(\pm\left(\omega-\mu\right)\right)\ex^{-\im\omega\left(t_{1}-t_{2}\right)}$ \\ \\
Right & $\mathbf{\Sigma}_{\text{em}}^{\urcorner}\left(t_{1},\tau_{2}\right)=\underset{\alpha}{\sum}\ex^{-\im\psi_{\alpha}\left(t_{1},t_{0}\right)}\int\frac{\ud\omega}{2\pi}\ex^{-\im\omega\left(t_{1}-t_{0}\right)}\Gamma_{\alpha}\left(\omega\right)\frac{\im}{\beta}\underset{q}{\sum}\frac{\ex^{\omega_{q}\tau_{2}}}{\omega_{q}-\omega+\mu}$ \\ \\
Left & $\mathbf{\Sigma}_{\text{em}}^{\ulcorner}\left(\tau_{1},t_{2}\right)=\underset{\alpha}{\sum}\ex^{\im\psi_{\alpha}\left(t_{2},t_{0}\right)}\int\frac{\ud\omega}{2\pi}\ex^{\im\omega\left(t_{2}-t_{0}\right)}\Gamma_{\alpha}\left(\omega\right)\frac{\im}{\beta}\underset{q}{\sum}\frac{\ex^{-\omega_{q}\tau_{1}}}{\omega_{q}-\omega+\mu}$ \\ \\
Matsubara & $\mathbf{\Sigma}_{\text{em}}^{M}\left(\tau_{1},\tau_{2}\right)=\underset{\alpha}{\sum}\int\frac{\ud\omega}{2\pi}\Gamma_{\alpha}\left(\omega\right)\frac{\im}{\beta}\underset{q}{\sum}\frac{\ex^{-\omega_{q}\left(\tau_{1}-\tau_{2}\right)}}{\omega_{q}-\omega+\mu}$
\end{tabular}
\caption{Various time-contour components of the embedding self-energy.}
\label{tab:emb-list}
\end{tcolorbox}
\end{table}

\subsection{Coupling of electrons and bosons}
\label{ssec:el_bs}
Let us outline briefly also the KBE equations for bosons. That is, the equations of motion of the single-particle Green's function $D(12) = D(\mub z;\nub z')$, see Sec.~\ref{sec:prtrbt}.
We note that the KBE for bosons can be written in many ways, see~\cite{Sakkinen2015a,Sakkinen2015c,Schueler2016,Tuovinen2016a, Sakkinen2016}. Here, we focus on the displacement--momentum representation using $\hat{\phi}$-fields, as the resulting equations of motion for the bosonic Green's function become first-order in time\cite{Sakkinen2016, Karlsson2018Handbook, Karlsson2021}:
\begin{align}
 \left [\im \alphaBold \partial_z - \OmegabBold (z) \right ] \DBold(z;z') =
 \oneb \delta(z,z')
 +
 \int_\gamma \dif \zb \ \PiBold(z;\zb) \DBold(\zb;z') \label{KBED1}
 \\
 \DBold(z;z') \left [-\im \alphaBold \partial_{z'} - \OmegabBold(z') \right ]  =
 \oneb\delta(z,z')
 +
 \int_\gamma \dif \zb \ \DBold(z; \zb) \PiBold(\zb;z'). \label{KBED2}
\end{align}
All quantities in boldface are $2N_c \times 2N_c$ matrices, where $N_c$ is the size of the system. The symbol $\oneb$ denotes the unit matrix, $\oneb_{\mub \nub} = \delta_{\mu \nu} \delta_{\xi_\mu \xi_\nu}$, and we remind that $\alphaBold$ contains the commutation relations, see \Eq{eq:bosAlpha}. Furthermore, $\OmegabBold = \Omega + \Omega^T$, and $\Pi$ denotes the bosonic self-energy.

In this work, we will discuss the commonly encountered $GD$ approximation (see Fig.~\ref{fig:GD}), which is first order in the fully dressed bosonic propagator. The explicit form of the electronic self-energy due to the presence of bosons in the $GD$ approximation is given by
\begin{equation}\label{elBosSelfEnergy}
\SigmaBold_{GD,\pbar \qbar} (z,z') = \SigmaBold_{\text{H},\pbar \qbar} (z,z') +
\im \sum_{\mub \nub \rbar \sbar} \lambda^\mub_{\pbar \rbar}(z) D_{\mub \nub}(z,z') \lambda^{\nub}_{\sbar \qbar}(z')  G_{\rbar \sbar}(z,z'),
\end{equation}
where the `bar' over the electronic indices signifies both orbital and spin indices. Here, the mean-field contribution, known as the classical, Hartree, or Ehrenfest contribution, is given by
\begin{equation}
 \Sigma_{\text{H},\pbar \qbar}(z,z') = \delta(z,z') \sum_{\mub}
  \lambda^\mub_{\pbar \qbar}(z) \phi_\mub(z),
 \label{HartreeSigma}
\end{equation}
The amplitude of the $\phi$-field, $\phi(t) = \langle \phih_{\text{H}}(t) \rangle$ is given by
\begin{equation}
 \phi_\mub(z) = \sum_{\nub \pbar \qbar} \int _\gamma \dif \zb \
 d_{\mub \nub}(z,\zb) \lambda^\nub_{\pbar \qbar}(\zb)
 (-\im )G_{\qbar \pbar}(\zb,\zb^+).
 \label{expValueField}
\end{equation}
Note that it is the non-interacting bosonic Green's function $d$ that enters here, and not the interacting $D$, since this would lead to double counting.
The bosonic self-energy for the $GD$ approximation, in turn, is given by 
\begin{equation}\label{eq:bosSelfEnergy}
 \Pi_{\mub \nub}(z,z')
  =
  -\im \sum_{\pbar \qbar \sbar \rbar} \lambda^\mub_{\pbar \qbar} (z) G_{\qbar \sbar}(z,z')
  G_{\rbar \pbar}(z',z) \lambda^\nub_{\sbar \rbar}(z').
\end{equation}
We stress that the $GD$ approximation is fully conserving, given that the electronic and bosonic KBE are solved self-consistently. 

\section{The Generalized Kadanoff--Baym Ansatz}\label{sec:gkbasec}

The NEGF formalism is a powerful one. High-order diagrams can be systematically included to achieve high accuracy, while the formalism offers access to a large variety of observables. Open systems, that is, systems in contact with reservoirs, can be treated non-perturbatively in the system-reservoir coupling.

Nevertheless, for large-scale calculations, such as \emph{ab-initio} calculations where the goal is to perform parameter-free calculations, the computational solution of the NEGF equations, either in Dyson form or the KBE form, is extremely demanding, scaling \emph{cubically} with simulation time (however, see Ref.~\cite{Schlunzen2016} for large-scale simulations of cold atoms in optical lattices and Ref.~\cite{Kaye2021} of the Falicov--Kimball model).

Despite the computationally demanding schemes, the interest of NEGF simulations has increased tremendously during the recent years. One of the reasons of this interest lies in the reinvention of the Generalized Kadanoff--Baym Ansatz (GKBA)~\cite{Lipavsky1986}, which drastically reduces the computational effort. Using the GKBA, the computational scaling is reduced from cubic to \emph{quadratic} scaling with simulation time. 

The fast GKBA has allowed for various research groups to study time-dependent correlated phenomena in a large variety of systems. 
As examples, we mention the study of atoms~\cite{Perfetto2015b}, biologically relevant molecules~\cite{Perfetto2018}, organic compounds~\cite{Pal2011,Bostrom2018} as well as a large class of extended systems~\cite{Sangalli2015,Sangalli2016} including several two-dimensional layered materials~\cite{Pogna2016,Molina-Sanchez2017}. 
The scheme has also been used to study model Hamiltonians, for example systems with Hubbard or extended Hubbard interactions~\cite{Hermanns2012, Hermanns2013, Hermanns2014, Latini2014, BarLev2016, Schluenzen2017}. 
Other interesting examples involve the study of non-equilibrium excitonic behavior~\cite{Tuovinen2018,Murakami2020,Tuovinen2020}, quench dynamics of topological materials~\cite{Schuler2019a,Schuler2020}, real-time dynamics of Auger processes~\cite{Covito2018,Covito2018a}, and ultrafast charge-migration in Glycine~\cite{Perfetto2019}.

\subsection{Isolated systems}

Let us first concentrate on isolated systems, i.e., systems without coupling to the environmental degrees of freedom such as baths. We will see in Sec.~\ref{sec:gkba-open} what is to be added for the description of open systems as in quantum transport. The reduction in scaling from cubic to quadratic stems from studying the time-diagonal of the lesser and greater Green's functions. Thus, the equations relevant for a GKBA treatment are the equations of motion for the density matrix, given by $\rho(t) = -\im G^<(t,t)$. The equation of motion for the density matrix $\rho(t)$ can be obtained from the KBE for the lesser and greater Green's functions, see Tab.~\ref{tab:kbe-list}, by subtracting the second equation from the first, and then letting $t_2 \to t_1 \equiv t$. The end result is 
\be\label{eq:electronicMasterEquation}
\frac{\ud}{\ud t} \rho(t) + \im [h_{\text{HF}}(t),\rho(t)] = - [\mathcal{I}(t) + \mathcal{I}^{\text{ic}}(t) + \hc] ,
\ee
with $h_{\text{HF}}$ being the HF Hamiltonian, and the \emph{collision integrals} are defined as
\begin{eqnarray}
\mathcal{I}(t) & = & \int_{t_0}^t \ud \bar{t} \left[\Sigma^>(t,\bar{t})G^<(\bar{t},t) - \Sigma^<(t,\bar{t})G^>(\bar{t},t)\right] , \label{eq:electronicCollision} \\
\mathcal{I}^{\text{ic}}(t) & = & -\im \int_0^\beta \ud \tau \Sigma^\urcorner(t,\tau) G^\ulcorner (\tau,t) . \label{eq:electronicCollision-ic}
\end{eqnarray}
The abbreviation `ic' here refers to initial correlations, which shall be addressed in Sec.~\ref{sec:gkba-ic}. The equation for the density matrix, \Eq{eq:electronicMasterEquation}, also known as the quantum master equation, is not closed, as the real-time collision integral in \Eq{eq:electronicCollision} depends on the time-off-diagonal lesser and greater Green's function. This is where the GKBA comes in.

The GKBA~\cite{Lipavsky1986} is the following approximation for the lesser and greater Green's function
\be\label{eq:gkba}
G^\lessgtr(t,t') \approx \mp \left[G^r(t,t') \rhoe^\lessgtr(t') - \rhoe^\lessgtr(t)G^a(t,t') \right] 
\ee
where the greater density matrix is given by $\rhoe^>(t) \equiv 1 - \rhoe(t)$, and we write $\rhoe^<(t) \equiv \rhoe(t)$ for short.
This approximation decouples the spectral information ($G^{r/a}$) from the population dynamics ($\rho^\lessgtr$). The form of the retarded/advanced Green's function therefore is no longer determined by the lesser/greater Green's function, and it has to be specified in advance or determined dynamically. The central idea of Eq.~\eqref{eq:gkba} is that the spectral properties of the system are assumed to vary not as strongly with time as the population dynamics, and this type of `decoherence' assumption thus associates, e.g., the quasiparticle life-time being relatively long compared to the average collision time~\cite{Lipavsky1986,Schaefer2001,Kalvova2002,Axt2004}. At equal times, the GKBA automatically obeys $\rho(t) = -\im G^<(t,t)$.

A common approximation for the retarded and advanced Green's functions is the mean-field one. The explicit shape can be obtained from the solution of the KBE for $G^{r/a}$ (Tab.~\ref{tab:kbe-list}) for $\Sigma=0$~\cite{Stefanucci2013Book}, which for the retarded Green's function has the form
\be\label{eq:Gret}
G^r(t,t') = -\im \theta(t-t') {\text T} \rme^{-\im \int_{t'}^{t}h_{\text{HF}} (\tb) \dif \tb},
\ee
where ${\text T}$ is the real-time ordering operator. The advanced Green's function is directly obtained from the symmetry relation $G^a(t,t') = [G^{r}(t',t)]^\dagger$. 
These considerations make $G^{\lessgtr}$ become (time-non-local) functionals of $\rhoe(t)$, and thus also the collision integral in Eq.~\eqref{eq:electronicCollision} (provided an appropriate expression for the self-energy is used). Thus, the use of the GKBA closes the equation of motion for the density matrix, \Eq{eq:electronicMasterEquation}. However, this applies only in the absence of initial correlations, $\mathcal{I}^{\text{ic}}=0$ in Eq.~\eqref{eq:electronicCollision-ic}. This is because the GKBA does not provide an approximation for the mixed functions in Eq.~\eqref{eq:electronicCollision-ic}. In practice, this issue may be circumvented by starting the time-evolution from an initially noncorrelated state and build correlations up adiabatically~\cite{Rios2011, Hermanns2012}. However, the inclusion of the initial correlations is possible also within GKBA~\cite{Semkat2003,Karlsson2018,Hopjan2019,Tuovinen2021}, as we will discuss in section~\ref{sec:gkba-ic}.

It is computationally advantageous to write the retarded propagator (Eq.~\eqref{eq:Gret}) in terms of a time-evolution operator $Y$~\cite{Balzer2013book}
\be
G^r(t,t') = -\im\theta(t-t')Y(t,t'),
\ee
where $Y$ satisfies $Y(t,t')=[Y(t',t)]^\dagger$ and $Y(t,t)=1$. Then, the collision integral in Eq.~\eqref{eq:electronicCollision} can be rewritten as
\be
\mathcal{I}(t) = \int_{t_0}^t \ud \bar{t} [ \Sigma^>(t,\bar{t})G^<(\bar{t},\bar{t}) + \Sigma^<(t,\bar{t})G^>(\bar{t},\bar{t}) ]Y(\bar{t},t) .
\ee
This is useful because instead of consecutive diagonalizations of the Hamiltonian for each pair of times $(t,t')$, the following recurrence relation can be used for the construction of $Y$ instead
\be
Y(\bar{t},t) = Y(\bar{t},t-\delta)U^\dagger(t-\delta) ,
\ee
where $\delta$ is the time-step length and $U(\delta) \equiv \ex^{-\im h_{\text{HF}} \delta}$ is the standard evolution operator for a step of length $\delta$. This procedure is justified for comparatively small step lengths. In this representation, for example, the lesser/greater second-order Born self-energy (see Tab.~\ref{tab:2b-list}) is written as
\begin{align}\label{eq:2b-gkba}
(\Sigma_{\text{2B}})_{ij}^\lessgtr(t_1,t_2) & = \sum_{mnpqrs} v_{irpn}(t_1) v_{mqsj}(t_2)  \left[G^\gtrless(t_2,t_2) Y(t_2,t_1)\right]_{sr} \nonumber \\
& \times\left\{ \xi \left[Y(t_1,t_2)G^\lessgtr(t_2,t_2)\right]_{nm} \left[Y(t_1,t_2)G^\lessgtr(t_2,t_2)\right]_{pq} \right. \nonumber \\
& \left.-\left[Y(t_1,t_2)G^\lessgtr(t_2,t_2)\right]_{nq}\left[Y(t_1,t_2)G^\lessgtr(t_2,t_2)\right]_{pm} \right\} .
\end{align}

The computational cost of solving the integro--differential equation \Eq{eq:electronicMasterEquation} (given the self-energy) is quadratic, as for every time $t$ we need to compute an integral from $t_0$ to $t$ with an integral kernel that explicitly depends on $t$. 
Common procedures to solve the GKBA integro--differential equation is to make use of time-stepping, augmented with predictor-corrector methods~\cite{Stan2009,Balzer2013book,Tuovinen2018,Perfetto2018b}. 

The GKBA scheme has several advantages. 
One is the drastically improved computational efficiency compared to the solution of the KBE. 
The accuracy of the GKBA performs well as compared to the full solution of the KBE. 
The scheme automatically satisfies various conservation laws, such as particle number and energy conservation, under the same conditions as the KBE~\cite{Hermanns2014,Karlsson2021}.

There are also some disadvantages of the GKBA scheme. 
The GKBA scheme is practically usable on the time-diagonal of the Green's function. 
Thus, off-time-diagonal objects, such as the spectral function, are not fully obtainable in the GKBA formalism. In contrast, they are limited by the approximation of the propagators~\eqref{eq:Gret}, which are used to reconstruct the two-time behaviour of the lesser and greater Green's function in Eq.~\eqref{eq:gkba}.

In addition, a phenomenon which is not yet fully understood is that observables from the KBE solution can damp in several cases. Sometimes the damping is artificial, such as the damping observed in a finite few-level system~\cite{Friesen2010}, while in other scenarios, the damping has a physical meaning~\cite{Tuovinen2020}. For the GKBA, this damping does not seem to be present \cite{Hermanns2014, Tuovinen2020, Karlsson2021}.

\subsection{Open systems}\label{sec:gkba-open}

The protocol described above can be directly extended to open quantum systems. In this case, the self-energies appearing in the collision integrals in Eqs.~\eqref{eq:electronicCollision} and~\eqref{eq:electronicCollision-ic} contain both many-body and embedding contributions. Also, the propagators are described for the \emph{contacted} system at the HF level, cf.~Eq.~\eqref{eq:Gret}
\be\label{eq:propagator-gamma}
G^{r}(t,t') = - \im \theta(t-t'){\text T}\ex^{-\im \int_{t'}^t \ud \bar{t} [h_{\text{HF}}(\bar{t}) - \im \Gamma/2]} ,
\ee
where $\Gamma$ is the tunneling rate matrix from the leads to the molecular region. Here we have used the WBA for the retarded/advanced embedding self-energy, in which the tunneling rate appears independent of frequency. This point highlights the key difference between isolated and open systems within the GKBA. The lesser/greater embedding self-energy~\cite{Croy2009,Ridley2015}, cf.~Tab.~\ref{tab:emb-list},
\be\label{eq:lssgtrembsigma}
\Sigma_{\text{em}}^\lessgtr(t,t')  = \pm\im \sum_\alpha \ex^{-\im\psi_\alpha(t,t')} \int \frac{\ud \w}{2\pi}f[\pm(\w-\mu)]\Gamma_\alpha(\w)\ex^{-\im\w(t-t')} ,
\ee
enters explicitly in the collision integral~\eqref{eq:electronicCollision} for which there is no requirement of the WBA. In addition, the frequency dependency of $\Gamma_\alpha(\w)$ is useful for regulating the frequency integral~\cite{Tuovinen2020,Tuovinen2021}, which would not strictly speaking converge within the WBA.  On the other hand, the retarded/advanced embedding self-energy enters also implicitly in the description of propagators. The mathematical structure of Eq.~\eqref{eq:propagator-gamma} is precisely what makes GKBA an attractive approach, because otherwise the propagators would have to be solved from their equations of motion, thus canceling the computational benefit of using GKBA. The use of the WBA is expected to provide an accurate description when the retarded/advanced embedding self-energy depends weakly on frequency around the Fermi level. Even without the embedding self-energy, the form of the propagator (at the HF level) is an approximation, and this approach is sometimes referred to as the ``HF-GKBA'', see Ref.~\cite{Hermanns2014}. Therefore, in closed systems at the HF level, there is no difference between GKBA and full KBE even for the off-diagonal $G^<(t,t')$. However, for open systems the WBA embedding self-energy enters only in the propagators, i.e., between GKBA and full KBE only the diagonal $G^<(t,t)$ are identical~\cite{Latini2014}. In principle, the WBA is therefore not a critical restriction, but the more we move away from the wide-band limit the more the propagators are an approximation.

\subsection{Initial contact and correlations}\label{sec:gkba-ic}

As mentioned above, a drawback in the GKBA approach is that the mixed Keldysh components, with one of the time arguments imaginary and the other one real, are not approximated. These components relate the Matsubara calculation (equilibration) to the out-of-equilibrium one, and therefore a consistent description of the initial contact and correlations can be difficult. It has been shown to be possible, however, to represent the initial correlations collision integral in Eq.~\eqref{eq:electronicCollision-ic} in an equivalent form~\cite{Karlsson2018}
\be\label{eq:collint}
\mathcal{I}^{\text{ic}}(t) = \int_{-\infty}^0 \ud \bar{t} [\Sigma^>(t,\bar{t}) G^<(\bar{t},t) - \Sigma^<(t,\bar{t})G^>(\bar{t},t)] ,
\ee
where the self-energy kernels can, again, be of many-body or embedding type. In principle, Eq.~\eqref{eq:collint} involves a convergence factor $\ex^{\eta \bar{t}}$. In particular, for a contacted system this factor can be left out due to continua of lead states accounting for proper convergence. The equivalence between Eqs.~\eqref{eq:electronicCollision-ic} and~\eqref{eq:collint} is indeed formal: The information about the contacted and/or correlated initial state at time $t$ requires an integration from $-\infty$ to $0$. For this to be useful in practice, an analytic evaluation of Eq.~\eqref{eq:collint} is desired for a particular many-body or embedding self-energy. Importantly, this approach therefore allows for starting the real-time evolution from an initially contacted and/or correlated state. This removes the requirement of the adiabatic preparation of the correlated initial state. Instead, the system can be driven out of equilibrium already at the beginning of the simulation, thereby reducing the computational cost significantly.

Suppose then we have somehow obtained a correlated initial state characterized by an equilibrium density matrix $\rho^{\text{eq}}$. This could be done via adiabatic preparation or, e.g., by solving the Dyson equation for the Matsubara Green function (see Tab.~\ref{tab:dyson-list}). Let us first make the distinction between many-body and embedding self-energy as
\be\label{eq:separate}
\mathcal{I}^{\text{ic}}(t) = \mathcal{I}_{\text{MB}}^{\text{ic}}(t) + \mathcal{I}_{\text{em}}^{\text{ic}}(t) 
\ee
and concentrate first on the many-body part. The embedding part will be described afterwards. For the many-body self-energy described at the second-Born level (see Tab.~\ref{tab:2b-list}) the initial correlation collision integral can be obtained as~\cite{Karlsson2018}
\be\label{eq:ic2b}
\mathcal{I}_{\text{MB}}^{\text{ic}}(t) = \widetilde{\mathcal{I}}(t) G^{a}(0,t),
\ee
with
\be\label{eq:ic2b2}
\widetilde{\mathcal{I}}_{ik}(t) = \im \sum_{npr} \frac{\widetilde{v}_{irpn}(t)\widetilde{w}_{nprk}}{\epsilon_r+\epsilon_k-\epsilon_n-\epsilon_p+\im\eta},
\ee
where
\begin{eqnarray}
\widetilde{v}_{irpn}(t) & = & \sum_{\widetilde{n}\widetilde{p}\widetilde{r}} v_{i\widetilde{r}\widetilde{p}\widetilde{n}}G_{\widetilde{n}n}^{r}(t,0)G_{\widetilde{p}p}^{r}(t,0)G_{r\widetilde{r}}^{r}(0,t) , \\
\widetilde{w}_{nprk} & = & \sum_{mqsj} w_{mqsj}\left(\bar{\rho}_{nm}^{\text{eq}}\bar{\rho}_{pq}^{\text{eq}}\rho_{sr}^{\text{eq}}\rho_{jk}^{\text{eq}}-\rho_{nm}^{\text{eq}}\rho_{pq}^{\text{eq}}\bar{\rho}_{sr}^{\text{eq}}\bar{\rho}_{jk}^{\text{eq}}\right) ,
\end{eqnarray}
and $w_{mqsj}$ is given by Eq.~\eqref{eq:w4}. In Eq.~\eqref{eq:ic2b2}, $\epsilon$ represents the energy eigenvalues of the HF Hamiltonian in terms of the equilibrium density matrix: $h_{\text{HF}}^{\text{eq}}\equiv h_{\text{HF}}[\rho^{\text{eq}}]$. For consistency, the tensors $\widetilde{v}(t)$ and $\widetilde{w}$ are constructed in this basis as well. Importantly, Eq.~\eqref{eq:ic2b} can be evaluated at any time $t$ with minor computational cost, provided that $G^{a}(0,t)$ is already available during the time evolution. It is also possible to obtain similar representations for other many-body self-energies, such as the $GW$ approximation~\cite{Karlsson2018}.

The protocol described above applies for the embedding self-energy (see Tab.~\ref{tab:emb-list}) as well. In Ref.~\cite{Tuovinen2021} it is shown that the initial contact collision integral can be written as
\be\label{eq:freqintegral}
\mathcal{I}_{\text{em}}^{\text{ic}}(t) = \sum_\alpha \ex^{-\im \psi_\alpha(t,0)}\int\frac{\ud \w}{2\pi} \Gamma_\alpha(\w) [f(\w-\mu) - \rho^{\text{eq}}] \frac{\ex^{-\im \w t}}{\w-(h_{\text{HF}}^{\text{eq}}+\im\Gamma/2)} G^{a}(0,t) .
\ee
Here, it is worth noting that the frequency-dependence of $\Gamma_\alpha(\w)$ results from the lesser/greater embedding self-energy in Eq.~\eqref{eq:lssgtrembsigma}, which itself is of general form and does not require the WBA. In contrast, since the equilibrium system is contacted, the frequency-independent $\Gamma$ appearing in the denominator is due to the WBA, in accordance with Eq.~\eqref{eq:propagator-gamma}.
We will get back to this protocol in the context of calculating the current in Section~\ref{sec:obsqt}, and how the initial contacting is taken into account in this regard.

\subsection{Electron--boson coupling}
As in the electronic case, a bosonic density-matrix formalism, amenable to a GKBA, can be obtained from the bosonic KBE by subtracting the equations, \Eq{KBED1} and \Eq{KBED2}, and setting the time arguments equal. The bosonic density matrix is defined as $\rhob(t) \equiv \rhob^<(t)= \im \DBold^<(t,t)$, and its equation of motion is
\begin{eqnarray}\label{eq:bosonMasterEquation}
    \partial_t \rhob(t)
  +\im \left [\alphaBold \OmegabBold \rhob(t) - \rhob(t)\OmegabBold
\alphaBold \right ]  =
 \Ib(t) + \Ib^T(t), 
\end{eqnarray}
where we defined the \emph{bosonic} collision integral
\begin{eqnarray}\label{eq:bosonicCollision}
 \Ib(t)
  &=
 \alphaBold \int_0^t \dif \tb
 \left [\PiBold ^>(t,\tb) \DBold^<(\tb,t)
  - \PiBold ^<(t,\tb)\DBold^>(\tb,t) \right].
\end{eqnarray}

As the electronic and bosonic self-energies depend on both the electronic and bosonic Green's functions, $\Sigma = \Sigma[G,\DBold]$ and $\PiBold = \PiBold[G,\DBold]$, the equations for the electronic and bosonic density matrices are coupled. In the coupled electron--boson case, an electron--boson GKBA is needed to close the equations of motion. This was recently introduced in Ref.~\cite{Karlsson2021}. The explicit shape of the bosonic GKBA is 
\begin{equation}\label{bosonicGKBA}
 \DBold^\lessgtr(t,t') =  \DBold^r(t,t') \alphaBold
\rhob^\lessgtr(t') - \rhob^\lessgtr(t) \alphaBold \DBold^a(t,t'),
\end{equation}
where $\rhob^>(t) = \alphaBold + \rhob(t)$. The propagators $\DBold^{r/a}(t,t')$ are chosen to be the mean-field ones: 
\begin{equation}\label{eq:bosGKBAPropagators}
 \DBold^{r/a}(t,t') = \mp \im \alphaBold \theta[\pm(t-t')] \ex^{-\im \OmegabBold \alphaBold (t-t')}.
\end{equation}
The electronic and bosonic GKBA close the coupled electron--boson equations of motion, allowing for a speedup of NEGF calculation by one order of magnitude, from cubic to quadratic numerical cost.  

\subsection{Linear-time formulation}

Very recently~\cite{Schluenzen2020,Joost2020} it was shown that the integro--differential GKBA equations can be equivalently recast as a set of time-local first-order ordinary differential equations (ODEs). The computational cost of the ODE scheme is \emph{linear} instead  of quadratic, which means that GKBA time evolutions can be performed with the same scaling as the fastest quantum methods available, such as time-dependent density functional theory. Very recently, it was shown that the same technique can be applied to the electron--boson case~\cite{Karlsson2021, Pavlyukh2021a, Pavlyukh2021b}.

The procedure for showing the equivalence between the integro--differential form and the ODE form of the GKBA rests on two steps. The first step comes from taking the time derivative of (essentially) the collision integral, \Eq{eq:electronicCollision}. By the Leibniz rule of differentiation,
\begin{equation}\label{eq:Leibniz}
 \frac{\dif}{\dif t} \Big ( \int_{0}^t \dif \tb \ f(t,\tb)\Big)
 =
 f(t,t) + \int_{0}^t \dif \tb \ \frac{\partial}{\partial t}
f(t,\tb),
\end{equation}
we obtain two terms: A time-local term which is trivial to evaluate, and a time-non-local term. It is for the second term that the second step comes into play. The GKBA can be seen as a non-interacting Ansatz augmented by an interacting density matrix. As such, the GKBA follows a non-interacting equation of motion. Taking the electronic lesser Green's function as an example, the relation satisfied is $\im \partial_t G^<(t,t') = h(t) G^<(t,t')$, and similarly for the adjoint equation. Using this relation in \Eq{eq:Leibniz} allows for relating the time-non-local term to a term similar to the original collision integral, removing the need to perform an integral to solve the GKBA equations. We have then ended up with a larger set of ODEs, with a time-local form. 

The time-linear form of the GKBA has to be established for each many-body approximation. Approximations for which the time-linear form is available are, for example, the 2nd Born approximation, the $GW$ and $T$-matrix approximation\cite{Schluenzen2020}, and the $Gd$ and $GD$ approximation for the electron--boson case~\cite{Karlsson2021, Pavlyukh2021a, Pavlyukh2021b}.

\section{Transport from the transient to the stationary state}\label{sec:chap3}
\subsection{Observables of quantum transport}\label{sec:obsqt}
\subsubsection{Particle and energy currents.}
The time-dependent Green's functions of an open subsystem $S_{i}$ give the time-dependent currents and correlations of energy and particles within that spatial region. The primary observable describing the population of local sites within a subsystem $S_{i}$ is given by the number operator 
\begin{equation}
\hat{N}_{S_{i}}\left(t\right) = \underset{i\in S_{i}}\sum\hat{d}_{i}^{\dagger}\left(t\right)\hat{d}_{i}\left(t\right)
\end{equation}
whose statistical average is obtained from the Green's function of the embedded region
\begin{equation}
\left\langle \hat{N}_{S_{i}}\left(t\right)\right\rangle =-\im\textrm{Tr}\left[G_{S_{i}S_{i}}^{<}\left(t,t\right)\right].
\end{equation}
The associated particle current in this spatial region is the time derivative
\begin{equation}
\hat{I}_{S_{i}}^{p}\left(t\right)=\frac{\ud\hat{N}_{S_i}\left(t\right)}{\ud t}=\im\left[\hat{H},\hat{N}_{S_i}\left(t\right)\right].
\end{equation}
In addition, we may define the energy current
\begin{equation}\label{eq:energycurrent}
\hat{I}_{S_{i}}^{E}\left(t\right)=\frac{\ud \hat{H}_{S_{i}}\left(t\right)}{\ud t}=\im\left[\hat{H},\hat{H}_{S_{i}}\left(t\right)\right].
\end{equation}
Here, we only discuss the energy current as defined above and turn aside further discussions on the proper thermodynamical definition of the heat current~\cite{Ludovico2014,Esposito2015b}.

Now we consider the case of a central molecular system $C$ sandwiched between a set of leads, labelled by $\alpha$, such that the theory developed in Section \ref{sec:Embedding} can be applied to give explicit expressions for the current and fluctuations in terms of Green's functions and self energies of the many body and embedding varieties. To show this, we assume the system Hamiltonian is as in Section \ref{sec:Embedding}. Using the equations of motion~\eqref{EOM_GCC1}, \eqref{EOM_GCC2} and the invariance of the trace under cyclic permutations, we find~\cite{Myohanen2009}
\begin{align}\label{eq:I_C}
    \frac{\ud\left\langle \hat{N}_{C}\left(t\right)\right\rangle }{\ud t} & = -\im\frac{\ud}{\ud t}\textrm{Tr}_{C}\left[\mathbf{G}_{CC}\left(t^{f},t^{b}\right)\right] = -\im\textrm{Tr}_{C}\left[\overset{\rightarrow}{\partial}_{z_{1}}\mathbf{G}_{CC}\left(z_{1},z_{2}\right)+\mathbf{G}_{CC}\left(z_{1},z_{2}\right)\overset{\leftarrow}{\partial}_{z_{2}}\right]^{z_{1}=t^{f}}_{z_{2}=t^{b}} \nonumber \\
    & = -\textrm{Tr}_{C}\left[\int_{\gamma}\ud\bar{z}\left(\mathbf{\Sigma}_{CC}\left(t^{f},\bar{z}\right)\mathbf{G}_{CC}\left(\bar{z},t^{b}\right)-\mathbf{G}_{CC}\left(t^{f},\bar{z}\right)\mathbf{\Sigma}_{CC}\left(\bar{z},t^{b}\right)\right)\right].
\end{align}
The particle current operator in the leads is given by
\begin{align}\label{eq:I_alpha_operator}
    \hat{I}_{\alpha}^p\left(t\right)=2\im q\underset{k,m}{\sum}\left[T_{mk\alpha}\hat{d}_{m}^{\dagger}\left(t\right)\hat{d}_{k\alpha}\left(t\right)-T_{mk\alpha}^{*}\hat{d}_{k\alpha}^{\dagger}\left(t\right)\hat{d}_{m}\left(t\right)\right],
\end{align}
where the factor of $2$ comes from an implicit spin summation. The expectation value of this is expressed, using the expression in Eq.~\eqref{eq:GF_aC}, as follows
\begin{align}\label{eq:I_alpha}
    I_{\alpha}^p\left(t\right) & =4q\mbox{Re}\textrm{Tr}_{C}\left[\mathbf{h}_{C\alpha}\mathbf{G}_{\alpha C}\left(t^{f},t^{b}\right)\right] \nonumber \\
    &=2q\textrm{Tr}_{C}\left[\int_{\gamma}\ud \bar{z}\left(\mathbf{\Sigma}_{\text{em},\alpha}\left(t^{f},\bar{z}\right)\mathbf{G}_{CC}\left(\bar{z},t^{b}\right)-\mathbf{G}_{CC}\left(t^{f},\bar{z}\right)\mathbf{\Sigma}_{\text{em},\alpha}\left(\bar{z},t^{b}\right)\right)\right],
\end{align}
where, on the first line, another factor of $2$ appeared due to the complex-number identity $z+z^*=2\Re z$, which was explicitly written out on the second line.
This can be replaced with convolution integrals taken on the horizontal and vertical branches of $\gamma$ to give
\begin{align}\label{eq:I_alpha2}
    I_{\alpha}^p\left(t\right)=4q\mbox{Re}\,\mbox{Tr}_{C}\left[\left(\mathbf{\Sigma}_{\text{em},\alpha}^{<}\cdot\mathbf{G}_{CC}^{a}+\mathbf{\Sigma}_{\text{em},\alpha}^{r}\cdot\mathbf{G}_{CC}^{<}+\mathbf{\Sigma}_{\text{em},\alpha}^{\urcorner}\star\mathbf{G}_{CC}^{\ulcorner}\right)\left(t,t\right)\right].
\end{align}

Noting that the reservoir currents only explicitly contain the embedding self-energy in the integral kernel, we combine Eqs. \eqref{eq:I_alpha} and \eqref{eq:I_C} to evaluate the rate of change of all particles in the system
\begin{align}\label{eq:conservation}
    & \frac{\ud\left\langle \hat{N}_{C}\left(t\right)\right\rangle }{\ud t}+\sum_\alpha I_{\alpha}^p\left(t\right) = I_{\text{MB}}^p (t),
\end{align}
where we have defined 
\begin{align}
 I_{\text{MB}}^p (t) = -\textrm{Tr}_{C}\left[\int_{\gamma}\ud \bar{z}\left(\mathbf{\Sigma}_{\text{MB},CC}\left(t^{f},\bar{z}\right)\mathbf{G}_{CC}\left(\bar{z},t^{b}\right)-\mathbf{G}_{CC}\left(t^{f},\bar{z}\right)\mathbf{\Sigma}_{\text{MB},CC}\left(\bar{z},t^{b}\right)\right)\right];
\end{align}
this integral is exactly zero when particle number is conserved. 
However, as a result of the $\Phi$-derivability of the MB self-energy [see Eq.~\eqref{eq:phisigma}], the macroscopic conservation laws hold only if a self-consistent scheme is implemented, and in this case we expect that $I_{\text{MB}}(t) \to 0$ and thus one can make use of this quantity as good figure of merit for the convergence of the self-consistency procedure. This is indeed a consequence of the invariance of the $\Phi$-functional under gauge transformations~\cite{Myohanen2009,Stefanucci2013Book}.

An important special case of the time-dependent Meir--Wingreen formula~\eqref{eq:I_alpha2} is in the context of the GKBA. As the GKBA does not provide an approximation for the mixed functions ($\ulcorner, \urcorner$) -- a limitation circumvented via the adiabatic preparation of the correlated initial state -- the imaginary-time branch is simply left out:
\begin{equation}\label{eq:I_alpha3}
I_\alpha^p(t) = 4q\mathrm{Re}\int_{t_0}^t \mathrm{d}\bar{t}\mathrm{Tr}_C\left[\mathbf{\Sigma}_{\text{em},\alpha}^>(t,\bar{t})\mathbf{G}_{CC}^<(\bar{t},t)-\mathbf{\Sigma}_{\text{em},\alpha}^<(t,\bar{t})\mathbf{G}_{CC}^>(\bar{t},t)\right],
\end{equation}
where we used the symmetry relation~\eqref{RetAdv}.
With this form, there is also an important connection with Section~\ref{sec:gkba-ic}, where we outlined a procedure to evaluate explicitly the initial contacting collision integral with the GKBA. It is important to notice that the case of initial contact affects not only the collision integral but also the Meir--Wingreen current formula~\eqref{eq:I_alpha3}.
An adjustment for including the effect from the initial contacting can be obtained by writing Eq.~\eqref{eq:freqintegral} as $\mathcal{I}_{\text{em}}^{\text{ic}}(t) = \sum_\alpha \mathcal{I}_{\text{em},\alpha}^{\text{ic}}(t)$, and then transforming the Meir--Wingreen formula~\eqref{eq:I_alpha3} as
\be
I_\alpha^p(t) \longrightarrow I_\alpha^p(t) + 4q \text{Re} \text{Tr}_C \mathcal{I}_{\text{em},\alpha}^{\text{ic}}(t) .
\ee

Similarly as was done above for the particle current, starting from Eq.~\eqref{eq:energycurrent}, we may derive a time-dependent Meir--Wingreen formula for the energy current. The energy current at the interface between a lead $\alpha$ and the central scattering region $C$ is defined as the rate of change of the energy of lead $\alpha$:
\begin{equation} 
I_\alpha^E(t) = \left\langle \frac{\mathrm{d} \hat{H}_\alpha(t)}{\mathrm{d} t} \right\rangle = -\mathrm{i}\left\langle\sum_{mk\sigma}\epsilon_{k\alpha}\left(-T_{mk\alpha}\hat{d}_{m\sigma}^\dagger(t)\hat{d}_{k\alpha\sigma}(t)+T_{k\alpha m}\hat{d}_{k\alpha\sigma}^\dagger(t)\hat{d}_{m\sigma}(t)\right)\right\rangle.
\end{equation}
Also here, the definitions of the lesser Green's function and the embedding self-energy may be used to rewrite the energy current as convolutions over the time contour~\cite{Eich2014}
\begin{equation}\label{eq:IEalpha}
I_\alpha^E(t) = 4 \mathrm{Im}\mathrm{Tr}_C \left[ \left(\mathbf{\dot{\Sigma}}_{\text{em},\alpha}^{<}\cdot\mathbf{G}_{CC}^{a}+\mathbf{\dot{\Sigma}}_{\text{em},\alpha}^{r}\cdot\mathbf{G}_{CC}^{<}+\mathbf{\dot{\Sigma}}_{\text{em},\alpha}^{\urcorner}\star\mathbf{G}_{CC}^{\ulcorner}\right)(t,t) \right],
\end{equation}
where $\mathbf{\dot{\Sigma}}_{\text{em},\alpha}^{<,r,\urcorner}(t,z) = \frac{\mathrm{d}}{\mathrm{d}t}\mathbf{\Sigma}_{\text{em},\alpha}^{<,r,\urcorner}(t,z)$ for $z$ either on the horizontal or vertical branches. An alternative representation of this result is similar to that of Eq.~\eqref{eq:I_alpha2} where each term is weighted with the corresponding energy density (Hamiltonian)~\cite{Talarico2020}. Also here, in the context of the GKBA, the time-dependent Meir--Wingreen formula for the energy current can be written by neglecting the imaginary-time branch as~\cite{Perfetto2018b}
\begin{equation}
I_\alpha^E(t) = 4\mathrm{Im}\int_{t_0}^t \mathrm{d}\bar{t}\mathrm{Tr}_C\left[\mathbf{\dot{\Sigma}}_{\text{em},\alpha}^>(t,\bar{t})\mathbf{G}_{CC}^<(\bar{t},t)-\mathbf{\dot{\Sigma}}_{\text{em},\alpha}^<(t,\bar{t})\mathbf{G}_{CC}^>(\bar{t},t)\right],
\end{equation}
where the retarded and advanced functions were again rewritten in terms of the lesser and greater functions using Eq.~\eqref{RetAdv}.

We emphasize that Eqs.~\eqref{eq:I_alpha2} and~\eqref{eq:IEalpha} are general formulas for the time-dependent particle and energy currents, and they implicitly include interaction effects provided that the Green's functions are solved from their respective equations of motion with many-body self-energies.

\subsubsection{Current--current correlations.} Finally, in the theory of quantum noise one is concerned with the computation of the correlations of current fluctuations in possibly distinct subsystems $S_{i}$, $S_{j}$
\begin{align}\label{eq:C_correlation}
    C_{S_{i}S_{j}}^{\mu \nu}\left(t_{1},t_{2}\right)=\left\langle \left(\hat{I}_{S_{i}}^{\mu}\left(t_{1}\right)-\left\langle \hat{I}_{S_{i}}^{\mu}\left(t_{1}\right)\right\rangle \right)\left(\hat{I}_{S_{j}}^{\nu}\left(t_{2}\right)-\left\langle \hat{I}_{S_{j}}^{\nu}\left(t_{2}\right)\right\rangle \right)\right\rangle, 
\end{align}
where $\mu, \nu = p,E$.
We note that higher-order moments of energy and particle currents can also be accessed from a variety of different methods, including the $S$-matrix approach of Refs~\cite{levitov_electron_1996,schonhammer2007full,hassler2008wave}, master equation methods~\cite{braggio2006full,maisi2014full}, the path-integral formulation of Refs.~\cite{Tang2014a,Tang2014b}, and quantum Monte Carlo simulations~\cite{Ridley2018PRB,Ridley2019JCP,Ridley2019PRB}. However in the present work we focus on the first two moments using the Kadanoff--Baym equation of motion approach.

From its definition in Eq.~\eqref{eq:C_correlation}, the two-time correlator 
obviously satisfies the symmetry property:
\begin{align}\label{eq:C_symmetry}
    C_{S_{i}S_{j}}^{\mu \nu}\left(t_{1},t_{2}\right)^{*}=C_{S_{j}S_{i}}^{\nu \mu}\left(t_{2},t_{1}\right).
\end{align}
Since $\Delta\hat{I}^{\mu}_{S_{i}}\left(t_{1}\right)$ and $\Delta\hat{I}^{\nu}_{S_{j}}\left(t_{2}\right)$ do not commute in general, $C_{S_{i}S_{j}}^{\mu \nu}\left(t_{1},t_{2}\right)$ is not guaranteed to be real and so in most studies the symmetrized correlation function
\begin{align}
 P_{S_{i}S_{j}}^{\mu \nu}\left(t_{1},t_{2}\right)\equiv\frac{1}{2}\left\langle \Delta\hat{I}_{S_{i}}^{\mu}\left(t_{1}\right)\Delta\hat{I}_{S_{j}}^{\nu}\left(t_{2}\right)+\Delta\hat{I}_{S_{j}}^{\nu}\left(t_{2}\right)\Delta\hat{I}_{S_{i}}^{\mu}\left(t_{1}\right)\right\rangle =\textrm{Re}\left[C_{S_{i}S_{j}}^{\mu \nu}\left(t_{1},t_{2}\right)\right],   
\end{align}
is preferred \cite{buttiker_scattering_1990,Ridley2017}. However, since $P_{S_{i}S_{j}}^{\mu \nu}\left(t_{1},t_{2}\right)$ is just the real part of $C_{S_{i}S_{j}}^{\mu \nu}\left(t_{1},t_{2}\right)$, knowledge of the latter object is sufficient for a full characterization of the symmetric noise properties of the system. 

The two-time (particle) current correlation function can be expressed in terms of the two-particle exchange correlation function $L$ defined in Eq.~\eqref{eq:2pex}.
Here, we only consider particle current ($I^p$) correlations, and we omit the superscript $pp$ for brevity.
Substituting Eq.~\eqref{eq:I_alpha_operator} into Eq.~\eqref{eq:C_correlation} leads to the rather convenient result:
\begin{align}\label{eq:C_noise_L}
    C_{\alpha\beta}\left(t_{1},t_{2}\right)=4q^{2}\underset{\substack{k\in\alpha,k'\in\beta \\ m,m'\in C}}{\sum} & \left[T_{mk}T_{m'k'}L\left(k\left(t_{1}^{b}\right),k'\left(t_{2}^{f}\right);m\left(t_{1}^{b+}\right),m'\left(t_{2}^{f+}\right)\right)\right.\nonumber\\
    &\left.+T_{mk}^{*}T_{m'k'}^{*}L\left(m\left(t_{1}^{b}\right),m'\left(t_{2}^{f}\right);k\left(t_{1}^{b+}\right),k'\left(t_{2}^{f+}\right)\right)\right.\nonumber\\
    &\left.-T_{mk}T_{m'k'}^{*}L\left(k\left(t_{1}^{b}\right),m'\left(t_{2}^{f}\right);m\left(t_{1}^{b+}\right),k'\left(t_{2}^{f+}\right)\right)\right.\nonumber\\
    &\left.-T_{mk}^{*}T_{m'k'}L\left(m\left(t_{1}^{b}\right),k'\left(t_{2}^{f}\right);k\left(t_{1}^{b+}\right),m'\left(t_{2}^{f+}\right)\right)\right],
\end{align}
where the notation $z^{+}$ refers to a contour time slightly later than $z$ on the time contour. 

To parameterize our system with experimentally relevant variables, we can work in the relative time coordinate system so that $t_{1}=\tau+t$ and $t_{2}=t$, where $\tau\equiv t_{1}-t_{2}$ is the relative time that we wish to take a Fourier transform with respect to and $t$ is the measurement time. Note that, to make the mapping to the Fourier space associated with $\tau$, one needs $\tau$ to take on negative values. However, since both $t_{1}$ and $t_{2}$ must be times occurring chronologically later than $t_{0}$, this means that $\tau$ is restricted to lie in the range $\left[-t+t_{0},t-t_{0}\right]$, as was done in Ref. \cite{Joho2012}. We define the Fourier transform of the correlation with respect to the time difference $\tau\equiv t_{1}-t_{2}$, as a function of a single frequency and the measurement time $t$:
\begin{align}\label{symmetrized}
   P_{\alpha\beta}\left(\Omega,t\right)\equiv\int_{-t+t_{0}}^{t-t_{0}}\ud\tau \ex^{\im\Omega\tau}P_{\alpha\beta}\left(t+\tau,t\right)=\frac{1}{2}\left(C_{\alpha\beta}\left(\Omega,t\right)+C_{\alpha\beta}^{*}\left(-\Omega,t\right)\right),
\end{align}
where $C_{\alpha\beta}\left(\Omega,t\right)$ is the Fourier transform of $C_{\alpha\beta}\left(t+\tau,t\right)$ with respect to $\tau$. Note that the relation
\begin{align}
    P_{\alpha\beta}^{*}\left(\Omega,t\right)	=	P_{\alpha\beta}\left(-\Omega,t\right)
\end{align}
immediately follows. It is sufficient for knowledge of $P_{\alpha\beta}\left(t_{1},t_{2}\right)$ to know the non-symmetrized function $C_{\alpha\beta}\left(t_{1},t_{2}\right)$. In addition to the power spectrum, one can calculate several other useful quantities in terms of the $C_{\alpha\beta}$. For instance, in a junction with two leads $L$, $R$, one may focus on the net current,
\begin{align}
    \hat{I}_{LR}^{\left(-\right)}\left(t\right)=\frac{1}{2}\left(\hat{I}_{L}\left(t\right)-\hat{I}_{R}\left(t\right)\right)  
\end{align}
or on the sum of currents, which by the continuity equation is proportional to the rate of change of charge in the molecule
\begin{align}
    \hat{I}_{LR}^{\left(+\right)}\left(t\right)=\frac{1}{2}\left(\hat{I}_{L}\left(t\right)+\hat{I}_{R}\left(t\right)\right).
\end{align}
The time-dependent noise spectra of these objects can be written as
\begin{align}\label{noise_spectra}
    C^{\left(\mp\right)}\left(\Omega,t\right)=\int \ud\tau \ex^{\im\Omega\tau}\left\langle \Delta\hat{I}_{LR}^{\left(\mp\right)}\left(t+\tau\right)\Delta\hat{I}_{LR}^{\left(\mp\right)}\left(t\right)\right\rangle =\frac{1}{2}\left(C^{\left(\text{auto}\right)}\left(\Omega,t\right)\mp C^{\left(\times\right)}\left(\Omega,t\right)\right).
\end{align}
Here, we have defined the average autocorrelation and cross-correlations:
\begin{align}
    C^{\left(\text{auto}\right)}\left(t+\tau,t\right) & \equiv\frac{1}{2}\left(C_{LL}\left(t+\tau,t\right)+C_{RR}\left(t+\tau,t\right)\right) , \\
    C^{\left(\times\right)}\left(t+\tau,t\right) & \equiv\frac{1}{2}\left(C_{LR}\left(t+\tau,t\right)+C_{RL}\left(t+\tau,t\right)\right) .
\end{align}
In general, $C^{\left(\text{auto}\right)}$ and $C^{\left(\times\right)}$ are complex quantities and so cannot be observed. However, due to the symmetry property~\eqref{eq:C_symmetry}, they are both real at the equal observation time point $\tau=0$. This fact was exploited in Ref. \cite{Feng2008}, where the equal time autocorrelation in the left lead, $C_{LL}\left(t,t\right)$, was studied in the time domain. Using the identity~\eqref{eq:C_symmetry}, one can show that the real parts of these functions are always symmetric in the $\tau=0$ line:
\begin{align}
    \textrm{Re}\left[C^{\left(\text{auto}/\times\right)}\left(t+\tau,t\right)\right]=\textrm{Re}\left[C^{\left(\text{auto}/\times\right)}\left(t,t+\tau\right)\right]
\end{align}
whereas the imaginary parts are always antisymmetric about this line:
\begin{align}
    \textrm{Im}\left[C^{\left(\text{auto}/\times\right)}\left(t+\tau,t\right)\right]=-\textrm{Im}\left[C^{\left(\text{auto}/\times\right)}\left(t,t+\tau\right)\right].
\end{align}

Similarly here, we wish to emphasize that Eq.~\eqref{eq:C_noise_L}, and the resulting expressions for the noise power spectra, etc. are general formulas, implicitly including interaction effects, as long as the two-particle exchange correlation function $L$ is resolved from its equation of motion~\eqref{eq:2pex2}.

\subsubsection{Traversal times.} A fundamental variable which limits the operational frequency of a molecular device is the traversal time $\tau_{\text{traversal}}$ describing the time taken for an electronic signal to be transferred across the terminals of the molecular junction \cite{dragoman_time_2011}. For example, recent work has shown the cutoff frequency $f_{\text{max}}$ to be related to the traversal time as $f_{\text{max}}=1/2\pi\tau_{\text{traversal}}$ in graphene \cite{Lin2009}. In the present context, the dynamical problem of the time taken for a signal to propagate across a nanosized device is just the problem of measuring temporally non-local correlations in signal variations between the subsystems of the device. The problem of calculating the traversal time is complicated by the fact that in standard quantum mechanics, time is usually not given the same status as a dynamical variable such as the energy or particle position (although many proposals have been made to remedy this defect of the theory~\cite{olkhovsky_time_1974, kijowski_time_1974, bauer_time_1983, page_evolution_1983, bauer_tunneling_2017, maccone_quantum_2020}). Correspondingly, much debate has centered around the correct definition of the traversal time by considering tunneling through a potential barrier~\cite{buttiker_traversal_1982,hauge_tunneling_1989}, as well as the relation of this quantity to the dwell time in the molecular region~\cite{collins_quantum_1987}, the Larmor clock time~\cite{rybachenko_time_1998} and the group delay time~\cite{winful_tunneling_2006}. 

In Refs.~\cite{Ridley2017,Ridley2019Entropy} it was demonstrated that the timescales associated with electron traversal times and internal reflection processes could be seen as resonances in the real part of symmetrized cross-lead correlations $C^{\left(\times\right)}\left(t+\tau,t\right)$ as a function of the relative time $\tau$. The traversal time $\tau_{\text{traversal}}$ may therefore be defined by the following relation
\begin{align}\label{eq:traversaltime}
\textrm{max}\left|\textrm{Re}\left[C^{\left(\times\right)}\left(t+\tau,t\right)\right]\right|\equiv\left|\textrm{Re}\left[C^{\left(\times\right)}\left(t\pm\tau_{\text{traversal}},t\right)\right]\right| .
\end{align}
The idea of this approach to is to quantify the traversal time for electronic information to cross the system by looking directly at the temporal correlations between the electronic signals in spatially separated subsystems. The procedure is exemplified in Sec.~\ref{sec:applications}; see Fig.~\ref{fig:Extended_systems}. This is preferable to approaches which use an indirect definition of traversal times from the calculation of transmission probabilities. The definition of traversal time here is closely related to the definition of Pollak and Miller, which makes use of flux--flux correlation functions~\cite{pollak_new_1984}.

\begin{figure}[t]
\centering
\includegraphics[width=0.85\textwidth]{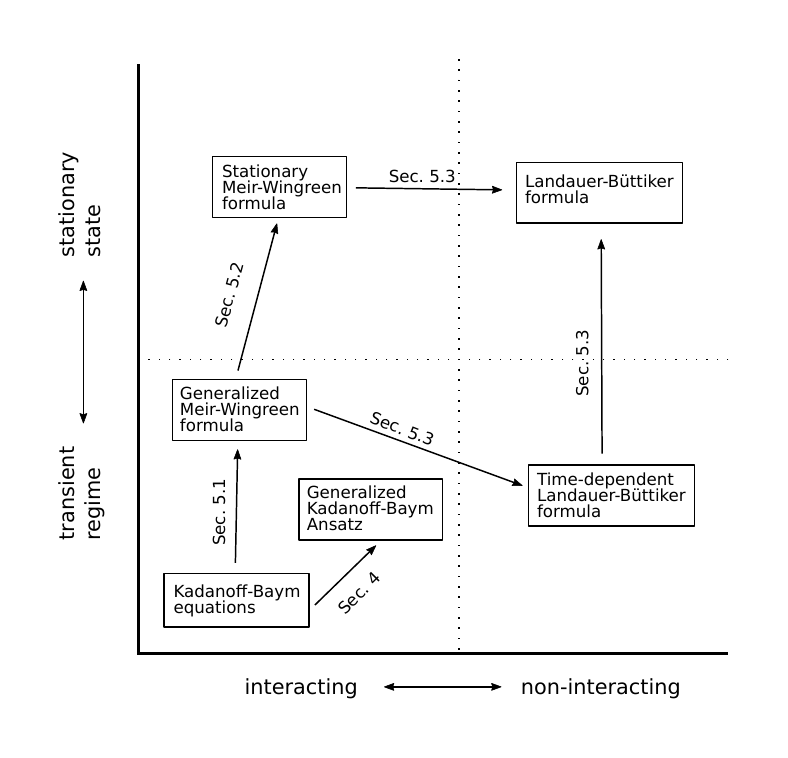}
\caption{Logical flowchart of connections between different approaches for characteristic interaction strengths (horizontal axis) and characteristic time scales (vertical axis). The corresponding section numbers are identified where each connection is discussed.}
\label{fig:flow}
\end{figure}

As we have now established the basis of calculating quantum-transport observables from the Green's functions, we will now show how the underlying NEGF approach with the generalized Meir--Wingreen formula~\eqref{eq:I_alpha2} reduces to known results at the stationary state and at the limit of no interactions; see Fig.~\ref{fig:flow}.

\subsection{Interacting systems}

\subsubsection{Current fluctuations.}
We see instantly from Eq.~\eqref{eq:C_noise_L} that the quantum noise of the current vanishes in the Hartree approximation, as Eq.~\eqref{eq:2pex} implies $L\left(i\left(z_{1}\right),j\left(z_{2}\right);i'\left(z_{1}'\right),j'\left(z_{2}'\right)\right)=0$ in this case. Physically, this means that the non-interacting quantum noise arises directly from the superposition of distinct trajectories for a system of identical particles. 

Expanding $G_{2}$ to $0$-th order in the two-particle interaction using the prescription in Eqs.~\eqref{eq:SEMS} and~\eqref{eq:HF}, the minimal conserving approximation is given by the Hartree--Fock expansion (more specifically, only the Fock or the exchange term contributes here)
\begin{align}
    L_{\textrm{HF}}\left(i\left(z_{1}\right),j\left(z_{2}\right);i'\left(z_{1}'\right),j'\left(z_{2}'\right)\right)= G\left(i\left(z_{1}\right);j'\left(z_{2}'\right)\right)G\left(j\left(z_{2}\right);i'\left(z_{1}'\right)\right),
\end{align}
where the one-particle Green's functions are evaluated self-consistently with the appropriate HF many body self energy, Eq.~\eqref{eq:HF}. Substituting this into Eq.~\eqref{eq:C_noise_L}, we can express the current correlator in terms of Kadanoff--Baym components of the Green's function of the junction:
\begin{align}\label{correlation_HF_GF}
    C_{\textrm{HF},\alpha\beta}\left(t_{1},t_{2}\right)=-4q^{2}\textrm{Tr}_{C} & \left[\mathbf{h}_{C\alpha}\left(t_{1}\right)\mathbf{G}_{\alpha C}^{>}\left(t_{1},t_{2}\right)\mathbf{h}_{C\beta}\left(t_{2}\right)\mathbf{G}_{\beta C}^{<}\left(t_{2},t_{1}\right)\nonumber\right.\\
    &\left.+\mathbf{G}_{C\beta}^{>}\left(t_{1},t_{2}\right)\mathbf{h}_{\beta C}\left(t_{2}\right)\mathbf{G}_{C\alpha}^{<}\left(t_{2},t_{1}\right)\mathbf{h}_{\alpha C}\left(t_{1}\right)\right.\nonumber\\
    &\left.-\mathbf{h}_{C\alpha}\left(t_{1}\right)\mathbf{G}_{\alpha\beta}^{>}\left(t_{1},t_{2}\right)\mathbf{h}_{\beta C}\left(t_{2}\right)\mathbf{G}_{CC}^{<}\left(t_{2},t_{1}\right)\right.\nonumber\\
    &\left.-\mathbf{G}_{CC}^{>}\left(t_{1},t_{2}\right)\mathbf{h}_{C\beta}\left(t_{2}\right)\mathbf{G}_{\beta\alpha}^{<}\left(t_{2},t_{1}\right)\mathbf{h}_{\alpha C}\left(t_{1}\right)\right].
\end{align}

In addition to correlation functions describing particle hopping events between the leads and the molecule, Eq.~\eqref{correlation_HF_GF} also contains contributions from lead--lead electron transfer and `circular'~\cite{ochoa_pumpprobe_2015} currents involving internal electronic transport processes between molecular sites. Using the Langreth rules for the lead--molecule coupling Green's functions, we can expand the current correlator into an expression that only contains explicit reference to the Green's function of the central molecule and the embedding self-energy. Defining the object
\begin{align}
\mathbf{\Lambda}_{\alpha}^{\pm}\left(t_{1},t_{2}\right)\equiv\left(\mathbf{\Sigma}_{\text{em},\alpha}^{\gtrless}\cdot\mathbf{G}_{CC}^{a}+\mathbf{\Sigma}_{\text{em},\alpha}^{\urcorner}\star\mathbf{G}_{CC}^{\ulcorner}\right)\left(t_{1},t_{2}\right)
\end{align}
we obtain the exact relation 
\begin{align}\label{eq:correlation_HF_GF_embedding}
    C_{\textrm{HF},\alpha\beta}\left(t_{1},t_{2}\right) & = -4q^{2}\textrm{Tr}_{C}\left[\delta_{\alpha\beta}\left(\mathbf{\Sigma}_{\text{em},\alpha}^{>}\left(t_{1},t_{2}\right)\mathbf{G}_{CC}^{<}\left(t_{2},t_{1}\right)+\mathbf{G}_{CC}^{>}\left(t_{1},t_{2}\right)\mathbf{\Sigma}_{\text{em},\alpha}^{<}\left(t_{2},t_{1}\right)\right)\nonumber\right.\\
    & \left. + \left[\mathbf{\Lambda}_{\alpha}^{+}\cdot\mathbf{\Sigma}_{\text{em},\beta}^{a}+\mathbf{\Sigma}_{\text{em},\alpha}^{r}\cdot\left(\mathbf{G}_{CC}^{>}\cdot\mathbf{\Sigma}_{\text{em},\beta}^{a}-\left(\mathbf{\Lambda}_{\beta}^{+}\right)^{\dagger}\right)\right]\left(t_{1},t_{2}\right)\mathbf{G}_{CC}^{<}\left(t_2,t_1\right)\right.\nonumber\\
    & \left. + \mathbf{G}_{CC}^{>}\left(t_{1},t_{2}\right)\left[\mathbf{\Lambda}_{\beta}^{-}\cdot\mathbf{\Sigma}_{\text{em},\alpha}^{a}+\mathbf{\Sigma}_{\text{em},\beta}^{r}\cdot\left(\mathbf{G}_{CC}^{<}\cdot\mathbf{\Sigma}_{\text{em},\alpha}^{a}-\left(\mathbf{\Lambda}_{\alpha}^{-}\right)^{\dagger}\right)\right]\left(t_2,t_1\right)\right.\nonumber\\
    & \left. - \left(\mathbf{\Lambda}_{\alpha}^{+}+\mathbf{\Sigma}_{\text{em},\alpha}^{r}\cdot\mathbf{G}_{CC}^{>}\right)\left(t_{1},t_{2}\right)\left(\mathbf{\Lambda}_{\beta}^{-}+\mathbf{\Sigma}_{\text{em},\beta}^{r}\cdot\mathbf{G}_{CC}^{<}\right)\left(t_{2},t_{1}\right)\right.\nonumber\\
    & \left. - \left(\mathbf{G}_{CC}^{>}\cdot\mathbf{\Sigma}_{\text{em},\beta}^{a}-\left(\mathbf{\Lambda}_{\beta}^{+}\right)^{\dagger}\right)\left(t_{1},t_{2}\right)\left(\mathbf{G}_{CC}^{<}\cdot\mathbf{\Sigma}_{\text{em},\alpha}^{a}-\left(\mathbf{\Lambda}_{\alpha}^{-}\right)^{\dagger}\right)\left(t_{2},t_{1}\right)\right].
\end{align}
Since this expression contains no explicit reference to the many-body self energy, it is also valid for non-interacting systems, as seen in Ref.~\cite{Ridley2017}. Information about correlations is implicit in $\mathbf{G}_{CC}$, which is solved for using the Hartree--Fock expansion of $\mathbf{\Sigma}_{\text{MB}}$.

\subsubsection{From the transient to the stationary regime.}

Eqs.~\eqref{eq:I_alpha} and~\eqref{eq:correlation_HF_GF_embedding} are valid for all times following the quench at $t_{0}$. However, in most experimental situations, moments of the electron current are measured in the steady state regime only, so that we may investigate the current and noise in the limit of $t_{0}\rightarrow-\infty$. 

In this limit, the dependency of the initial state is washed out and we may take $\mathbf{\Sigma}_{\text{em},\alpha}^\rceil \star \mathbf{G}_{CC}^\lceil \to 0$. The resulting equation for the time-dependent particle current becomes
\be
I^p_\alpha(t) = 4q\text{Re}\text{Tr}_C \int_{t_0}^\infty\ud\tb\left[\mathbf{\Sigma}_{\text{em},\alpha}^<(t,\tb) \mathbf{G}_{CC}^a(\tb,t) + \mathbf{\Sigma}_{\text{em},\alpha}^{r}(t,\tb) \mathbf{G}_{CC}^<(\tb,t)\right].
\ee
At the long-time limit, the Green's functions and self-energies depend on the time-difference only and we can Fourier transform, e.g., $\Sigma^<(t,\tb)=\Sigma^<(t-\tb)=\int_{-\infty}^\infty\frac{\ud\w}{2\pi}\ex^{-\im\w(t-\tb)}\Sigma^<(\w)$. Changing the integration over time with respect to a relative-time coordinate $\tau=t-\tb$, inserting the Fourier transforms into the time-dependent current formula, and using $\int_{-\infty}^\infty \ud\tau\ex^{-\im(\w-\w')\tau}=2\pi\delta(\w-\w')$ leads to the steady-state current
\be
I^p_\alpha = 4q\text{Re}\int_{-\infty}^\infty\frac{\ud\w}{2\pi}\text{Tr}_C\left[\mathbf{\Sigma}_{\text{em},\alpha}^<(\w)\mathbf{G}_{CC}^a(\w)+\mathbf{\Sigma}_{\text{em},\alpha}^r(\w)\mathbf{G}_{CC}^<(\w)\right].
\ee
Using the symmetry relations of the Green's functions and self-energies, and the cyclic property of the trace we obtain
\be
I^p_\alpha = 2\im q \int_{-\infty}^\infty\frac{\ud\w}{2\pi}\text{Tr}_C\left[\mathbf{\Sigma}_{\text{em},\alpha}^<(\w)\mathbf{A}_{CC}(\w)-\mathbf{\Gamma}_\alpha(\w)\mathbf{G}_{CC}^<(\w)\right],\label{eq:mwderiv}
\ee
where we introduced the spectral and tunneling rate functions
\begin{align}
\mathbf{A}_{CC}(\w) & = \im\left[\mathbf{G}_{CC}^r(\w)-\mathbf{G}_{CC}^a(\w)\right] \label{eq:spectral}\\
\mathbf{\Gamma}_\alpha(\w) & = \im\left[\mathbf{\Sigma}_{\text{em},\alpha}^r(\w)-\mathbf{\Sigma}_{\text{em},\alpha}^a(\w)\right].
\end{align}
A typical quantum-transport calculation consists of two leads ($L,R$) and then evaluating the total current through the central region as $I^p = I_R^p - I_L^p$. Inserting a fluctuation--dissipation relation for the embedding self-energy 
$\mathbf{\Sigma}_{\text{em},\alpha}^<(\w) = \im f_\alpha(\w)\mathbf{\Gamma}_\alpha(\w)$, with $f_\alpha$ the Fermi function of the $\alpha$-th lead we arrive at the original Meir--Wingreen formula~\cite{Meir1992}
\begin{align}\label{eq:IMW}
I^p = 2\im q \int_{-\infty}^\infty\frac{\ud\w}{2\pi} \text{Tr}_C & \left\{ \left[f_L(\w)\mathbf{\Gamma}_L(\w) - f_R(\w)\mathbf{\Gamma}_R(\w)\right]\left[\mathbf{G}_{CC}^r(\w)-\mathbf{G}_{CC}^a(\w)\right] \right.\nonumber \\
& \left. + \ \left[\mathbf{\Gamma}_L(\w) - \mathbf{\Gamma}_R(\w)\right]\mathbf{G}_{CC}^<(\w) \right\} .
\end{align}
As this result followed directly as the long-time limit of the time-dependent current, we may indeed regard Eq.~\eqref{eq:I_alpha2} as the time-dependent generalization of the Meir--Wingreen formula, also taking into account the initial correlations.

Similarly, for the energy current in Eq.~\eqref{eq:IEalpha}, the long-time limit is obtained with $\mathbf{\Sigma}_{\text{em},\alpha}^\rceil \star \mathbf{G}_{CC}^\lceil \to 0$ and similar steps as above:
\be\label{eq:mwderiv2}
I_\alpha^E = 2\im\int_{-\infty}^\infty \frac{\ud \w}{2\pi} \w \text{Tr}_C\left[\mathbf{\Sigma}_{\text{em},\alpha}^<(\w)\mathbf{A}_{CC}(\w)-\mathbf{\Gamma}_\alpha(\w)\mathbf{G}_{CC}^<(\w)\right].
\ee
In contrast to Eq.~\eqref{eq:mwderiv}, it is worth noting the overall factor of $\w$ inside the integral, which results from the Fourier transform of the time-derivatives of the self-energies in Eq.~\eqref{eq:IEalpha}. By inserting the same fluctuation--dissipation relation for the lesser embedding self-energy, we may write the total energy current in a two-lead setup, $I^E = I_R^E - I_L^E$, as the Meir--Wingreen formula~\cite{Bergfield2009,Dubi2011}
\begin{align}\label{eq:IEMW}
I^E = 2\im \int_{-\infty}^\infty\frac{\ud\w}{2\pi} \w \text{Tr}_C & \left\{ \left[f_L(\w)\mathbf{\Gamma}_L(\w) - f_R(\w)\mathbf{\Gamma}_R(\w)\right]\left[\mathbf{G}_{CC}^r(\w)-\mathbf{G}_{CC}^a(\w)\right] \right.\nonumber \\
& \left. + \ \left[\mathbf{\Gamma}_L(\w) - \mathbf{\Gamma}_R(\w)\right]\mathbf{G}_{CC}^<(\w) \right\} .
\end{align}
Similarly to the particle current, we may then justify Eq.~\eqref{eq:IEalpha} as the time-dependent generalization of the Meir--Wingreen formula for energy current, properly including the effect of initial correlations via convolutions along the imaginary track.

\subsection{Non-interacting systems}\label{sec:nonint}

\subsubsection{Stationary state and the Landauer--B{\"u}ttiker formula.}

Let us first concentrate on the steady-state limit, where the Meir--Wingreen formula for the particle and energy currents were obtained in Eqs.~\eqref{eq:mwderiv} and~\eqref{eq:mwderiv2}. In the absence of interactions, we may write the lesser Green's function in the frequency domain simply in terms of the embedding self-energy,
\be
\mathbf{G}_{CC}^<(\w) = \mathbf{G}_{CC}^r(\w)\mathbf{\Sigma}_{\text{em}}^<(\w)\mathbf{G}_{CC}^a(\w),
\ee
since the many-body self-energy is zero. In this context, it is also useful to write the spectral function in Eq.~\eqref{eq:spectral} as
\be
\mathbf{A}_{CC}(\w) = \mathbf{G}_{CC}^r(\w)\mathbf{\Gamma}(\w)\mathbf{G}_{CC}^a(\w),
\ee
which follows from a direct expansion of the retarded and advanced Green's functions in the non-interacting case. By introducing the transmission function between leads $\alpha$ and $\alpha'$ as
\begin{align}\label{transmission}
T_{\alpha\gamma}(\w)=\text{Tr}_C\left[\mathbf{\Gamma}_\alpha(\w)\mathbf{G}_{CC}^r(\w)\mathbf{\Gamma}_{\gamma}(\w)\mathbf{G}_{CC}^a(\w)\right],
\end{align}
we may express the steady-state particle and energy currents as the Landauer--B{\"u}ttiker formula
\begin{align}
I_\alpha^p & = 2q\int_{-\infty}^\infty \frac{\ud\w}{2\pi} \sum_{\alpha'} \left[f_{\alpha'}(\w)-f_\alpha(\w)\right]T_{\alpha\alpha'}(\w), \label{eq:LBFormula}\\
I_\alpha^E & = 2\int_{-\infty}^\infty \frac{\ud\w}{2\pi} \sum_{\alpha'} \w \left[f_{\alpha'}(\w)-f_\alpha(\w)\right]T_{\alpha\alpha'}(\w),
\end{align}
respectively, and the summations run over an arbitrary multiterminal setup. In the common two-lead setup, the total currents (cf.~Eqs.~\eqref{eq:IMW} and~\eqref{eq:IEMW}) are further simplified as
\begin{align}
I^p & = 4q\int_{-\infty}^\infty \frac{\ud\w}{2\pi}  \left[f_L(\w)-f_R(\w)\right]\text{Tr}_C\left[\mathbf{\Gamma}_L(\w)\mathbf{G}_{CC}^r(\w)\mathbf{\Gamma}_R(\w)\mathbf{G}_{CC}^a(\w)\right], \\
I^E & = 4\int_{-\infty}^\infty \frac{\ud\w}{2\pi} \w \left[f_L(\w)-f_R(\w)\right]\text{Tr}_C\left[\mathbf{\Gamma}_L(\w)\mathbf{G}_{CC}^r(\w)\mathbf{\Gamma}_R(\w)\mathbf{G}_{CC}^a(\w)\right].
\end{align}

\subsubsection{Time-dependent Landauer--B{\"u}ttiker formula.}\label{sec:TDLB}
For outlining the TD-LB formalism, we now focus on analytic solutions to the Kadanoff--Baym equations which, although approximate, provide extensive insights into the quantum dynamics of molecular structures. In particular, we focus on the case in which the level width in Eq.~\eqref{eq:width} is replaced with an energy-independent expression
\begin{align}\label{eq:width_WBLA}
    \Gamma_{\alpha,mn}\left(\omega\right)\rightarrow\Gamma_{\alpha,mn}=2\pi\sum_k T_{m,k\alpha}T_{k\alpha,n}\delta\left(\epsilon_{\alpha}^{\text{F}}-\epsilon_{k\alpha}\right),
\end{align}
evaluated at the Fermi energy $\epsilon_{\alpha}^{\text{F}}$, which implies that the level shift $\Lambda_{\alpha,mn}\left(\omega\right)\rightarrow 0$. As already mentioned after Eq.~\eqref{eq:width}, this is the wide-band approximation (WBA), and it leads to the localization of the retarded/advanced self-energy components in the time domain:
\begin{align}
    \left[\mathbf{\Sigma}_{CC}^{r/a}\left(t_{1},t_{2}\right)\right]_{mn}=\mp\frac{\im}{2}\delta\left(t_{1}-t_{2}\right)\Gamma_{mn}.
\end{align}
Here, the time-locality is crucial for solving the equations of motion for the Green's functions. In the non-interacting limit, the self-energy kernels appearing inside the collision integrals contain only the embedding part, and this can now be trivially integrated with the help of the delta-function form. This makes it possible to obtain a closed and analytic solution for the Green's functions and, consequently, for the time-dependent currents and current correlators.

The lesser, greater, right, left and Matsubara self-energy components are obtained in a similar manner within the WBA, by substituting Eq.~\eqref{eq:width_WBLA} into the expressions of Tab.~\ref{tab:emb-list}. Since, in Tab.~\ref{tab:kbe-list},
the target Green's function components appear in convolution integrals with retarded and advanced self-energy components, the delta function serves to simplify the Kadanoff--Baym equations such that they are re-expressed in terms of the effective Hamiltonian $\mathbf{{h}}_{CC}^{\text{eff}}=\mathbf{{h}}_{CC}-\im\mathbf{\Gamma}/2$. For instance, the equations of motion for the greater/lesser Green's functions are given by
\begin{align}
    \left[\im\frac{\ud}{\ud t_{1}}-\mathbf{h}_{CC}^{\text{eff}}\right]\mathbf{G}_{CC}^{\gtrless}\left(t_{1},t_{2}\right) & =\left[\mathbf{\Sigma}_{CC}^{\gtrless}\cdot\mathbf{G}_{CC}^{a}+\mathbf{\Sigma}_{CC}^{\urcorner}\star\mathbf{G}_{CC}^{\ulcorner}\right]\left(t_{1},t_{2}\right) , \\
    \mathbf{G}_{CC}^{\gtrless}\left(t_{1},t_{2}\right)\left[-\im\frac{\overset{\leftarrow}{\ud}}{\ud t_{2}}-\left(\mathbf{h}_{CC}^{\text{eff}}\right)^\dagger\right] & =\left[\mathbf{G}_{CC}^{r}\cdot\mathbf{\Sigma}_{CC}^{\gtrless}+\mathbf{G}_{CC}^{\urcorner}\star\mathbf{\Sigma}_{CC}^{\ulcorner}\right]\left(t_{1},t_{2}\right).
\end{align}
These formulae can be mapped to a more tractable form by introducing the substitution
\begin{align}
    \mathbf{G}_{CC}^{\gtrless}\left(t_{1},t_{2}\right)\equiv \ex^{-\im\mathbf{h}_{CC}^{\text{eff}}\left(t_{1}-t_{0}\right)}\mathbf{\widetilde{G}}_{CC}^{\gtrless}\left(t_{1},t_{2}\right)\ex^{\im\left(\mathbf{h}_{CC}^{\text{eff}}\right)^\dagger\left(t_{2}-t_{0}\right)},
\end{align}
and then carrying out a line integral in the two-time plane for $\mathbf{\widetilde{G}}_{CC}^{\gtrless}$, before retrieving $\mathbf{G}_{CC}^{\gtrless}$ from the inverse mapping. The demonstration of path-independence in the two-time plane and full details of the resulting line integral can be found in Refs.~\cite{Ridley2015,ridley_time-dependent_2017b,Ridley2017}. 
This results in the following rather compact and convenient formula
\begin{equation}\label{G_Lesser_Greater}
\mathbf{G}_{CC}^{\lessgtr}\left(t_{1},t_{2}\right)=\pm \im\int\frac{\ud\omega}{2\pi}f\left(\pm\left(\omega-\mu\right)\right)\underset{\beta}{\sum}\mathbf{S}_{\beta}\left(t_{1},t_{0};\omega\right)\Gamma_{\beta}\mathbf{S}_{\beta}^{\dagger}\left(t_{2},t_{0};\omega\right) ,
\end{equation}
where the upper (lower) signs on the right hand side correspond to the lesser (greater) components and we have introduced the matrix
\begin{align}\label{S_alpha}
& \mathbf{S}_{\alpha}\left(t,t_{0};\omega\right)\equiv \ex^{-\im\mathbf{h}_{CC}^{\text{eff}}\left(t-t_{0}\right)}\left[\mathbf{G}_{CC}^{r}\left(\omega\right) -\im\int_{t_{0}}^{t}\ud\bar{t}\ex^{-\im\left(\omega\mathbf{1}-\mathbf{h}_{CC}^{\text{eff}}\right)\left(\bar{t}-t_{0}\right)}\ex^{-\im\psi_{\alpha}\left(\bar{t},t_{0}\right)}\right]
\end{align}
defined in terms of the retarded Green's function $\mathbf{G}_{CC}^{r}\left(\omega\right)=\left(\omega\mathbf{1}-\mathbf{h}_{CC}^{\text{eff}}\right)^{-1}$, where the time-dependent voltage in the leads is contained in the phase factor $\psi_{\alpha}\left(t_{1},t_{2}\right)$ defined in Eq.~\eqref{eq:phase}. We list explicit expressions for all Green's function components in the TD-LB formalism in Tab.~\ref{tab:GF_TDLB-list}.

\begin{table}
\centering
\begin{tcolorbox}
\begin{tabular}{ c|c } 
Component & Green's function \\
\hline
\hline \\
Ret &  $\mathbf{G}_{CC}^{r}\left(t_{1},t_{2}\right) = -\im\theta\left(t_{1}-t_{2}\right)\ex^{-\im\mathbf{h}{}_{CC}^{\text{eff}}\left(t_{1}-t_{2}\right)}$\\ \\
Adv &  $\mathbf{G}_{CC}^{a}\left(t_{1},t_{2}\right) = \im\theta\left(t_{2}-t_{1}\right)\ex^{-\im\left(\mathbf{h}{}_{CC}^{\text{eff}}\right)^{\dagger}\left(t_{1}-t_{2}\right)}$\\ \\
Lss/Gtr & $\mathbf{G}_{CC}^{\lessgtr}\left(t_{1},t_{2}\right)=\pm \im\int\frac{\ud\omega}{2\pi}f\left(\pm\left(\omega-\mu\right)\right)\underset{\beta}{\sum}\mathbf{S}_{\beta}\left(t_{1},t_{0};\omega\right)\Gamma_{\beta}\mathbf{S}_{\beta}^{\dagger}\left(t_{2},t_{0};\omega\right)$ \\ \\
Right & $\mathbf{G}_{CC}^{\urcorner}\left(t_{1},\tau_{2}\right)=\ex^{-\im\mathbf{h}_{CC}^{\text{eff}}\left(t_{1}-t_{0}\right)}\left[\mathbf{G}_{CC}^{M}\left(0^{+},\tau_{2}\right)\right.$ \\
& $\left.-\im\underset{t_0}{\overset{t_1}{\int}}\ud\bar{t}\ex^{\im\mathbf{h}_{CC}^{\text{eff}}\left(\bar{t}-t_{0}\right)}\left(\mathbf{\Sigma}_{CC}^{\urcorner}\star\mathbf{G}_{CC}^{M}\right)\left(\bar{t},\tau_{2}\right)\right]$ \\ \\
Left & $\mathbf{G}_{CC}^{\ulcorner}\left(\tau_{1},t_{2}\right)=\left[\mathbf{G}_{CC}^{M}\left(\tau_{1},0^{+}\right)\right.$ \\
& $\left.+\im\underset{t_0}{\overset{t_2}{\int}}\ud\bar{t}\left(\mathbf{G}_{CC}^{M}\star\mathbf{\Sigma}_{CC}^{\ulcorner}\right)\left(\tau_{1},\bar{t}\right)\ex^{-\im\left(\mathbf{h}_{CC}^{\text{eff}}\right)^{\dagger}\left(\bar{t}-t_{0}\right)}\right]\ex^{\im\left(\mathbf{h}_{CC}^{\text{eff}}\right)^{\dagger}\left(t_{2}-t_{0}\right)}$ \\ \\
Matsubara & $\mathbf{G}_{CC}^{M}\left(\tau_{1},\tau_{2}\right)=\frac{\im}{\beta}\sum_{q}\ex^{-\omega_{q}\left(\tau_{1}-\tau_{2}\right)}\mathbf{G}_{CC}^{M}\left(\omega_{q}\right),$  \\ 
& $\omega_{q}=\im\left(2q+1\right)\pi/\beta,$ \\
& $\mathbf{G}_{CC}^{M}\left(\omega_{q}\right)=\begin{cases}
\left[\left(\omega_{q}+\mu\right)\mathbf{1}_{CC}-\mathbf{h}_{CC}^{\text{eff}}\right]^{-1}, & \Im\left(\omega_{q}\right)>0\\
\left[\left(\omega_{q}+\mu\right)\mathbf{1}_{CC}-\left(\mathbf{h}_{CC}^{\text{eff}}\right)^\dagger\right]^{-1}, & \Im\left(\omega_{q}\right)<0
\end{cases}$ \\ \\
\end{tabular}
\caption{Green's functions in the TD-LB formalism for the case of an arbitrary time-dependent bias.}
\label{tab:GF_TDLB-list}
\end{tcolorbox}
\end{table}

The quantum statistical expectation value of the current operator, setting the electronic charge $q=-1$, may also be expressed in terms of the auxiliary matrix $\mathbf{S}_\alpha$ in Eq.~\eqref{S_alpha}, as~\cite{Ridley2017}:
\begin{align}\label{Current_alpha}
I_{\alpha}\left(t\right) = \frac{1}{\pi}\int \ud\omega f\left(\omega-\mu\right)\,\mbox{Tr}_{C} & \left[2\Re\left[\im\Gamma_{\alpha}\ex^{\im\omega\left(t-t_{0}\right)}\ex^{\im\psi_{\alpha}\left(t,t_{0}\right)}\mathbf{S}_{\alpha}\left(t,t_{0};\omega\right)\right] \phantom{\sum_\beta} \right.\nonumber \\
&\left.-\Gamma_{\alpha}\sum_\beta\mathbf{S}_{\beta}\left(t,t_{0};\omega\right)\Gamma_{\beta}\mathbf{S}_{\beta}^{\dagger}\left(t,t_{0};\omega\right)\right].
\end{align}
Using the fact that $N_{C}\left(t\right)=-2\im\mathrm{Tr}_{C}\left[\mathbf{G}_{CC}^{<}\left(t,t\right)\right]$ it can be shown straightforwardly that the conservation law in Eq.~\eqref{eq:conservation} is satisfied, where $I_{\text{MB}}(t)=0$ for the non-interacting case.

We emphasize here that an exact solution to the Kadanoff--Baym equations was obtained for the non-interacting case with the lead couplings being independent of energy, a setting similar to the original works of Wingreen, Jauho, and Meir~\cite{Wingreen1993,Jauho1994}. Indeed, Eqs.~\eqref{G_Lesser_Greater} and~\eqref{Current_alpha} are reminiscent of the ones in Refs.~\cite{Wingreen1993,Jauho1994} (partitioned approach) although we stress that our derivation includes explicitly the imaginary branch of the time contour and the initial coupling between the leads and the molecular region (partition-free). The results have been shown to be formally equivalent~\cite{Stefanucci2004, Odashima2017, Ridley2018JLTP}. We also wish to point out that the model and the starting point are similar to many independent works addressing the simulation of quantum nanoelectronics devices, such as the \texttt{KWANT} software and its time-dependent extension \texttt{TKWANT}, constructed with the scattering wavefunction formalism~\cite{Groth2014, Gaury2014, Kloss2021}.

The result in Eq.~\eqref{G_Lesser_Greater} is both very general and useful. The lesser Green's function at the time diagonal, $t_1=t_2=t$, directly gives us access to the time-dependent density matrix $\rho(t)=-\im G^<(t,t)$. However, these expressions for the lesser and greater Green's functions are explicit solutions to the full Kadanoff--Baym equations, and they are not limited to the time-diagonal only but can be used for investigating the full two-time plane, which allows for calculation of the quantum noise (see below). This closed solution was possible through the wide-band approximation and the non-interacting limit. Similarly, Eq.~\eqref{Current_alpha} is the time-dependent generalization of the Landauer--B{\"u}ttiker formula~\eqref{eq:LBFormula} and it can be used for calculating the time-dependent current at the interface of the $\alpha$-th lead and the central region. It correctly reduces to the form of Ref.~\cite{Jauho1994} when the switch-on time is taken to the remote past. It also provides an extension to the result of Refs.~\cite{Stefanucci2004} and~\cite{Perfetto2008}, where single-level systems, with the addition of the spin degree-of-freedom in the latter, were considered. Here, the matrix structures allow arbitrary geometries for the central region as long as the Hamiltonian is represented in the non-interacting form. The result in Eq.~\eqref{Current_alpha} also reduces to the earlier time-dependent Landauer--B{\"u}ttiker formula of Ref.~\cite{Tuovinen2013,Tuovinen2014} when the bias is maintained at a constant value after the switch-on. Here, $\psi_\alpha(t,t_0)$ enables arbitrary time-dependent modulations of the bias voltage.

We can also extend the result for the first moment of the current within the WBA to calculations of the two-time current correlators. In the WBA, Eq.~\eqref{eq:correlation_HF_GF_embedding} can be expressed in the following form 
\begin{align}
& \label{Correlation_alpha_beta}
C_{\alpha\beta}\left(t_{1},t_{2}\right) \nonumber \\
& =	4q^{2}\mathrm{Tr}_{C}\left\{\delta_{\alpha\beta}\left(\mathbf{\Sigma}_{\alpha}^{>}\left(t_{1},t_{2}\right)\mathbf{G}_{CC}^{<}\left(t_{2},t_{1}\right)+\mathbf{G}_{CC}^{>}\left(t_{1},t_{2}\right)\mathbf{\Sigma}_{\alpha}^{<}\left(t_{2},t_{1}\right)\right)\right. \nonumber \\
& \left. + \mathbf{\Gamma}_{\alpha}\mathbf{G}_{CC}^{>}\left(t_{1},t_{2}\right)\mathbf{\Gamma}_{\beta}\mathbf{G}_{CC}^{<}\left(t_{2},t_{1}\right)+\im\mathbf{G}_{CC}^{>}\left(t_{1},t_{2}\right)\left[\mathbf{\Lambda}_{\beta}^{+}\left(t_{2},t_{1}\right)\mathbf{\Gamma}_{\alpha}+\mathbf{\Gamma}_{\beta}\left(\mathbf{\Lambda}_{\alpha}^{+}\right)^{\dagger}\left(t_{1},t_{2}\right)\right]\right.\nonumber \\
& \left.+\im\left[\mathbf{\Lambda}_{\alpha}^{-}\left(t_{1},t_{2}\right)\mathbf{\Gamma}_{\beta}+\mathbf{\Gamma}_{\alpha}\left(\mathbf{\Lambda}_{\beta}^{-}\right)^{\dagger}\left(t_{2},t_{1}\right)\right]\mathbf{G}_{CC}^{<}\left(t_{2},t_{1}\right) \right. \nonumber \\
& \left.-\mathbf{\Lambda}_{\beta}^{+}\left(t_{2},t_{1}\right)\mathbf{\Lambda}_{\alpha}^{-}\left(t_{1},t_{2}\right)-\left(\mathbf{\Lambda}_{\alpha}^{+}\right)^{\dagger}\left(t_{1},t_{2}\right)\left(\mathbf{\Lambda}_{\beta}^{-}\right)^{\dagger}\left(t_{2},t_{1}\right)\right\},
\end{align}
which is expressible solely in terms of the $\mathbf{S}_{\alpha}$ matrices, using Eq.~\eqref{G_Lesser_Greater} and the following pair of functional identities~\cite{Ridley2017}:
\begin{align}
\mathbf{\Lambda}_{\beta}^{+}\left(t_{2},t_{1}\right) & =	\im\ex^{-\im\psi_{\beta}\left(t_{2},t_{0}\right)}\int\frac{\ud\omega}{2\pi}f\left(\omega-\mu\right)\ex^{-\im\omega\left(t_{2}-t_{0}\right)}\mathbf{\Gamma}_{\beta}\mathbf{S}_{\beta}^{\dagger}\left(t_{1},t_{0};\omega\right),\label{eq:lambda_matrix_1}\\
\mathbf{\Lambda}_{\alpha}^{-}\left(t_{1},t_{2}\right) & =-\im\ex^{-\im\psi_{\alpha}\left(t_{1},t_{0}\right)}\int\frac{\ud\omega}{2\pi}\left(1-f\left(\omega-\mu\right)\right)\ex^{-\im\omega\left(t_{1}-t_{0}\right)}\mathbf{\Gamma}_{\alpha}\mathbf{S}_{\alpha}^{\dagger}\left(t_{2},t_{0};\omega\right),\label{eq:lambda_matrix_2}
\end{align}
such that the $\mathbf{\Lambda}_{\alpha}^{\pm}$ can be thought of as particle/hole propagators in the leads. We therefore interpret the two terms appearing on the second line of Eq. \eqref{Correlation_alpha_beta} as describing processes in which electrons in the leads interfere with holes in the molecular region, or holes in the leads interefere with electrons in the molecule. The terms on the third line of Eq.~\eqref{Correlation_alpha_beta} are cross-lead particle--hole interference terms. Although Eq.~\eqref{Correlation_alpha_beta} is given in terms of analytically tractable expressions, it is not yet in a form convenient for numerical implementation. The details of the numerical evaluation of the noise and current are given in~\ref{app:numerics}.

We now work in the relative time coordinate system described in the discussion prior to Eq.~\eqref{symmetrized}. In studies of high-frequency shot noise for a static bias, the interesting physical observable is usually the static non-symmetrized power spectrum \cite{aguado2000double, zamoum2016nonsymmetrized, crepieux2021electronic}, which is the regular Fourier transform of
$C_{\alpha\beta}\left(\tau\right)\equiv\lim_{t_{0}\rightarrow-\infty}C_{\alpha\beta}\left(t+\tau,t\right)$ defined in Eq.~\eqref{symmetrized}:
\begin{align}
    C_{\alpha\beta}\left(\Omega\right)\equiv\lim_{t_{0}\rightarrow-\infty}C_{\alpha\beta}\left(\Omega,t\right)=\int_{-\infty}^{\infty}\ud\tau \ex^{\im\Omega\tau}C_{\alpha\beta}\left(\tau\right)\equiv\mathcal{F}\left[C_{\alpha\beta}\left(\tau\right);\Omega\right].
\end{align}
For those experiments which do distinguish between absorption and emission processes, the quantity of interest is most often $C_{\alpha\alpha}\left(\Omega\right)$, which in general satisfies the inequality $C_{\alpha\alpha}\left(\Omega\right)\neq C_{\alpha\alpha}\left(-\Omega\right)$. $C_{\alpha\alpha}\left(\Omega\right)$ can therefore be used to describe measurements in which a quantum of energy $\hbar\Omega$ is transferred from the measuring device to the system. By contrast, the symmetrized spectrum obeys $P_{\alpha\alpha}\left(\Omega,t\right)=P_{\alpha\alpha}\left(-\Omega,t\right)$, i.e. it does not distinguish between emission and absorption processes.

However, in the original Landauer--B{\"u}ttiker formalism, the symmetrized noise was the object of study and is, for instance, the object used to compute Fano factors in nanoscale conductors. These formulae are usually given in terms of the transmission matrix~\cite{di2008electrical}
\begin{align}\label{transmission_matrix}
   \mathbf{T}_{CC}^{\left(\alpha\gamma\right)}\left(\omega\right)	\equiv	\left[\mathbf{\Gamma}_{\alpha}\right]^{\frac{1}{2}}\mathbf{G}_{CC}^{r}\left(\omega\right)\left[\mathbf{\Gamma}_{\gamma}\right]^{\frac{1}{2}},
\end{align}
which is related to the transmission function defined in Eq.~\eqref{transmission} as
\begin{align}\label{transmission_2}
T_{\alpha\gamma}(\w)=\text{Tr}_C\left[\mathbf{T}_{CC}^{\left(\alpha\gamma\right)}\left(\omega\right)\mathbf{T}_{CC}^{\dagger\left(\alpha\gamma\right)}\left(\omega\right)\right].
\end{align}
If we now specialize this discussion to the case of a two-lead junction, i.e. a junction in which $\alpha$ may be one of two indices $L$, $R$, it is simple to show (see Ref.~\cite{Ridley2017} for more details) that, neglecting the frequency-dependence of the transmission, the $LL$ component of the symmetrized noise is given by the well-known formula:
\begin{align}\label{transmission_matrix2}
& P_{LL}\left(\Omega\right) \nonumber \\
& =\frac{q^{2}}{\pi}\left\{ \textrm{Tr}_{C}\left[\mathbf{T}_{CC}^{\left(LR\right)}\mathbf{T}_{CC}^{\dagger\left(LR\right)}\mathbf{T}_{CC}^{\left(LR\right)}\mathbf{T}_{CC}^{\dagger\left(LR\right)}\right]2\Omega\coth\left(\frac{\Omega}{2k_{B}T}\right)\right. \nonumber \\
& \left.+\textrm{Tr}_{C}\left[\mathbf{T}_{CC}^{\left(LR\right)}\mathbf{T}_{CC}^{\dagger\left(LR\right)}\left(1-\mathbf{T}_{CC}^{\left(LR\right)}\mathbf{T}_{CC}^{\dagger\left(LR\right)}\right)\right]\right.\nonumber \\
& \left. \times \left[\left(V_{L}-V_{R}-\Omega\right)\coth\left(\frac{V_{L}-V_{R}-\Omega}{2k_{B}T}\right) + \left(V_{L}-V_{R}+\Omega\right)\coth\left(\frac{V_{L}-V_{R}+\Omega}{2k_{B}T}\right)\right]\right\} .
\end{align}
This expresses the interplay of the shot noise, Nyquist noise and quantum vacuum fluctuations in a conductor, and moreover has been verified experimentally for a wide range of mesoscale conductors \cite{schoelkopf1997frequency,zakka2007experimental}. 

In many experiments, $P_{\alpha\alpha}\left(\Omega\right)$ is measured for time intervals greatly exceeding the characteristic timescales of the junction, corresponding to the zero-frequency limit \cite{gabelli2008dynamics}. In this case, we retain the non-trivial frequency-dependent transmission matrices and recover the following well-known results for the thermal and shot noise, respectively:
\begin{align}\label{eq:thermal_noise}
& \underset{\Omega\rightarrow0}{\lim}P_{LL}^{\left(\mathrm{thermal}\right)}\left(\Omega\right) \nonumber \\
& = 4q^{2}\int\frac{\ud\omega}{2\pi}\left\{\left[1-f_{L}\left(\omega-\mu\right)\right]f_{L}\left(\omega-\mu\right)+\left[1-f_{R}\left(\omega-\mu\right)\right]f_{R}\left(\omega-\mu\right)\right\}T_{LR}\left(\omega\right),
\end{align}
\begin{align}\label{eq:shot_noise}
& \underset{\Omega\rightarrow0}{\lim}P_{LL}^{\left(\mathrm{shot}\right)}\left(\Omega\right) = 4q^{2}\int\frac{\ud\omega}{2\pi}\left[f_{L}\left(\omega-\mu\right)-f_{R}\left(\omega-\mu\right)\right]^{2}T_{LR}\left(\omega\right)\left[1-T_{LR}\left(\omega\right)\right],
\end{align}
where the $T_{LR}\left(\omega\right)$ are as in Eq.~\eqref{transmission}. The $T_{RL}\left(\omega\right)$ give the probability for electrons to pass from lead $R$ to lead $L$ and are sometimes written as the sum $T_{RL}\left(\omega\right)=\sum_i T_{i}\left(\omega\right)$ over a sum of discrete scattering channels, labelled by $i$.
Then in systems where two scattering channels dominate, experimentalists can fit their data on the first and second moments of the current to Eqs.~\eqref{eq:LBFormula}, \eqref{eq:thermal_noise} and~\eqref{eq:shot_noise} to determine the $T_{i}$ empirically \cite{nitzan2001electron,kim2011benzenedithiol, karimi2016shot}. The thermal noise term describes the transmission of electron--hole pairs across the nanojunction, and contains electron--hole distribution functions $\left[1-f\left(\omega-V_{i}-\mu\right)\right]f\left(\omega-V_{i}-\mu\right)$, for $i=L,R$. It corresponds to fluctuations in the particle number due to thermal excitations of the junction~\cite{blanter_shot_2000}, and vanishes when the temperature $1/\beta=0$. The shot noise term is also zero when the molecule is completely transparent ($T_{RL}=1$) or completely opaque ($T_{RL}=0$) to the propagating electrons, i.e. it is non-zero as a consequence of the partitioning of the propagating electronic states into transmitted and reflected wavepackets. When $V_{L}=V_{R}$, the shot noise term vanishes, as in this case there is no net drift of electrons across the junction.

\subsubsection{Phononic heat transport in the transient regime.}

The above formulation can be employed also in the non-interacting bosonic case, with the Green's function equations of motion in Eqs.~\eqref{KBED1} and~\eqref{KBED2}. In this situation, the many-body boson self-energy is zero, and for a phononic heat transport setup we look at the following partitioning for the one-particle Hamiltonian~\cite{Tuovinen2016PRB}:
\be\label{eq:phononpartition}
\mathbf{\Omega} = \begin{pmatrix}
            \mathbf{\Omega}_{LL} & \mathbf{\Omega}_{LC} & 0 \\ 
            \mathbf{\Omega}_{CL} & \mathbf{\Omega}_{CC} & \mathbf{\Omega}_{CR} \\ 
            0 & \mathbf{\Omega}_{RC} & \mathbf{\Omega}_{RR} 
            \end{pmatrix} ,
\ee
where the $L$, $C$, and $R$ denote the left reservoir, the central region, and the right reservoir, respectively. Similarly as in the electronic case, we project the equations of motion to the central-region block `$CC$' and obtain the embedded equation of motion
\begin{align}
(\im\mathbf{1}_{CC}\partial_z - \alphaBold_{CC}\mathbf{\Omega}_{CC})\DBold_{CC}(z,z') 
& = \alphaBold_{CC}\delta(z,z') + \alphaBold_{CC} \int_\gamma \ud \bar{z} \mathbf{\Pi}_{\text{em}}(z,\bar{z})\DBold_{CC}(\bar{z},z'), \label{eq:dprime-eom1} \\
\mathbf{\Pi}_{\text{em}}(z,z') & = \sum_{\lambda} \mathbf{\Omega}_{C \lambda}(z)\dBold_{\lambda \lambda}(z,z')\mathbf{\Omega}_{\lambda C}(z'), \label{eq:emb-def} \\
(\im\mathbf{1}_{\lambda\lambda}\partial_z - \alphaBold_{\lambda\lambda}\mathbf{\Omega}_{\lambda \lambda})\dBold_{\lambda \lambda}(z,z') & = \alphaBold_{\lambda\lambda} \delta(z,z'),\label{eq:small-d-eom} 
\end{align}
where $\dBold_{\lambda\lambda}$ is the isolated Green's function of the $\lambda$-th reservoir, and $\PiBold_{\text{em}}$ the phononic embedding self-energy. Note the appearance of the $\boldsymbol{\alpha}$ matrices [Eq.~\eqref{eq:bosAlpha}] for the equations of motion for bosonic Green's functions. From Eq.~\eqref{eq:dprime-eom1} and its adjoint equation at the equal-time limit, $z=t^f=t$, $z'=t^b=t^+$, we may derive the equation of motion for the one-particle phonon density matrix $\rhoBold_{\text{b}}(t) = \im\DBold^<(t,t)$ (we omit the subscript $CC$ from now on)
\be\label{eq:dlss-eom}
\im \frac{\ud}{\ud t}\DBold^<(t,t) - \left[\alphaBold \mathbf{\Omega} \DBold^<(t,t) - \DBold^<(t,t)\mathbf{\Omega} \alphaBold\right] = -\{[\DBold^r \cdot \PiBold_{\text{em}}^<] + [\DBold^< \cdot \PiBold_{\text{em}}^a] \}(t,t)\alphaBold + \text{h.c.},
\ee
where we use the short-hand notation in Eq.~\eqref{Convolution_dot}. Note that this is equivalent to the electron--boson GKBA case in Eq.~\eqref{eq:bosonMasterEquation} with the bosonic self-energy in the collision integral being solely due to the embedding. In this model, the partitioning procedure disregards the initial couplings in equilibrium, $\mathbf{\Omega}_{\lambda C} = 0$, so the integrations along the vertical track of the Keldysh contour on the right-hand side of Eq.~\eqref{eq:dlss-eom} are simply left out.

Similarly as in the electronic case, also Eq.~\eqref{eq:dlss-eom} can be solved analytically when a wide-band-like approximation is taken for the embedding self-energy $\PiBold_{\text{em}}$. The off-diagonal elements in Eq.~\eqref{eq:phononpartition} correspond to couplings only via the displacement contribution [recall the composite structure of the $\hat{\phi}$-fields in Eq.~\eqref{eq:nonInteractingBosons}], i.e., the embedding self-energy has a non-vanishing entry only in the displacement block:
\be\label{eq:pi-ret-mat}
\PiBold_{\text{em},\lambda}^r(\w) = \begin{pmatrix}\Pi_{\text{em},\lambda}^r(\w) & 0 \\ 0 & 0 \end{pmatrix} = \mathbf{\Lambda}_{\text{b},\lambda}(\w) - \frac{\im}{2}\mathbf{\Gamma}_{\text{b},\lambda}(\w),
\ee
where we introduced the bosonic level-shift and level-width matrices $\mathbf{\Lambda}_{\text{b}}=\sum_\lambda \mathbf{\Lambda}_{\text{b},\lambda}$ and $\mathbf{\Gamma}_{\text{b}} = \sum_\lambda \mathbf{\Gamma}_{\text{b},\lambda}$, respectively. In Ref.~\cite{Tuovinen2016PRB} a wide-band-like approximation was put forward
\begin{align}
\mathbf{\Lambda}_{\text{b},\lambda}(\w) & \approx \mathbf{\Lambda}_{\text{b},\lambda}(\w=0) \equiv \mathbf{\Lambda}_{0,\lambda}, \label{eq:linearized-re} \\ 
\mathbf{\Gamma}_{\text{b},\lambda}(\w) & \approx \w \left(\partial_\w \mathbf{\Gamma}_{\text{b},\lambda}\right)_{\w=0} \equiv \w \mathbf{\Gamma}_{0,\lambda}^\prime \label{eq:linearized-im} .
\end{align}
In contrast to the wide-band approximation in the electronic case [cf.~Eq.~\eqref{eq:width_WBLA}], now the retarded embedding self-energy in Eq.~\eqref{eq:pi-ret-mat} is not a purely imaginary constant. Instead, the real part is constant and the imaginary part is frequency dependent (linearized approximation). The imaginary part is not bounded when $|\w|\to\infty$, so the approximation for the retarded embedding self-energy is introduced with a cut-off frequency $\w_{c,\lambda}$:
\be\label{eq:pi-retarded-w}
\PiBold_{\text{em},\lambda}^r(\w) = \theta(\w_{c,\lambda} - |\w|)(\mathbf{\Lambda}_{0,\lambda} - \frac{\im\w}{2}\mathbf{\Gamma}_{0,\lambda}^\prime) .
\ee
The lesser self-energy is then given by the corresponding fluctuation--dissipation relation
\be\label{eq:pilss-step}
\PiBold_{\text{em},\lambda}^<(\w) = \theta(\w_{c,\lambda} - |\w|)[-\im f_\lambda(\w) \w \mathbf{\Gamma}_{0,\lambda}^\prime] 
\ee
with $f_\lambda(\w) = 1/(\ex^{\beta_\lambda \w} - 1)$ being the Bose function for the $\lambda$-th reservoir at inverse temperature $\beta_\lambda = (k_{\text{B}}T_\lambda)^{-1}$.
The cut-off frequency is a similar concept as the Debye temperature, and they are related by $\hbar\w_{c,\lambda} = k_{\text{B}}T_\lambda$\cite{Wang2007}.

The form of the embedding self-energy in Eqs.~\eqref{eq:pi-retarded-w} and~\eqref{eq:pilss-step} enables evaluating explicitly the time-convolutions in Eq.~\eqref{eq:dlss-eom} and then solving analytically the first-order differential equation for $\rhoBold_{\text{b}}(t)$ in a closed form
\be\label{eq:final-result}
\rhoBold_{\text{b}}(t) = \im \DBold_0^<(t,t) + \sum_{\lambda=L,R} \int_{-\w_{c,\lambda}}^{\w_{c,\lambda}} \frac{\ud \w}{2\pi} f_\lambda(\w)\left[\mathbf{1}-\ex^{\im(\w-\mathbf{\Omega}_{\text{eff}})t}\right]\mathbf{B}_\lambda(\w)\left[\mathbf{1}-\ex^{-\im(\w-\mathbf{\Omega}_{\text{eff}}^\dagger)t}\right] ,
\ee
where the effective Hamiltonian is defined as
\be\label{eq:eff-ham}
\mathbf{\Omega}_{\text{eff}} = \frac{1}{\alphaBold+\frac{\im}{2}\mathbf{\Gamma}_0^\prime}(\mathbf{\Omega}+\mathbf{\Lambda}_0),
\ee
which reduces to the uncoupled Hamiltonian, $\alphaBold\mathbf{\Omega}$, in the limit $\mathbf{\Lambda}_0,\mathbf{\Gamma}_0^\prime \to 0$. The initial condition in Eq.~\eqref{eq:final-result},
\be
\im \DBold_0^<(t,t) = \ex^{-\im \mathbf{\Omega}_{\text{eff}}t}\alphaBold f_C (\mathbf{\Omega}\alphaBold)\ex^{\im \mathbf{\Omega}_{\text{eff}}^\dagger t},
\ee
stems from the uncoupled lesser Green's function of the central region, where the distribution $f_C$ is defined via an equilibrium temperature for the central region before coupling. The spectral function $\mathbf{B}_\lambda(\w) \equiv \DBold^r(\w)\w\mathbf{\Gamma}_{0,\lambda}^\prime \DBold^a(\w)$ can be evaluated as
\begin{align}\label{eq:phononspectral}
\mathbf{B}_\lambda(\w) & = \frac{1}{\w(\alphaBold+\frac{\im}{2}\mathbf{\Gamma}_0^\prime)-\mathbf{\Omega}-\mathbf{\Lambda}_0}\w\mathbf{\Gamma}_{0,\lambda}^\prime \frac{1}{\w(\alphaBold-\frac{\im}{2}\mathbf{\Gamma}_0^\prime)-\mathbf{\Omega}-\mathbf{\Lambda}_0} \nonumber \\
& = \frac{1}{\w-\mathbf{\Omega}_{\text{eff}}}(\mathbf{\Gamma}_{0,\lambda}^\prime)_{\text{eff}}(\w)\frac{1}{\w-\mathbf{\Omega}_{\text{eff}}^\dagger}
\end{align}
with $(\mathbf{\Gamma}_{0,\lambda}^\prime)_{\text{eff}}(\w) = (\alphaBold+\im\mathbf{\Gamma}_0^\prime/2)^{-1}\w\mathbf{\Gamma}_{0,\lambda}^\prime(\alphaBold-\im\mathbf{\Gamma}_0^\prime/2)^{-1}$. 

The result in Eq.~\eqref{eq:final-result} is the time-dependent one-particle phonon density matrix. As was established for the electronic case, also this is a closed expression, i.e., no time propagation is needed for evaluating the time-dependent density matrix. The transient behavior is encoded in the exponentials: We find transient oscillation frequencies $\w_{jk} = |\text{Re} (\w_{j,\text{eff}}) - \text{Re}( \w_{k,\text{eff}})|$, where $\w_{\text{eff}}$ are the complex eigenvalues of the effective Hamiltonian $\mathbf{\Omega}_{\text{eff}}$, as transitions between the vibrational modes in the central region. Finally, $f_\lambda (\w)\mathbf{B}_\lambda(\w)$ is well-behaving at $\w=0$ (although $f_\lambda(\w)$ diverges at zero), and the cut-off frequency $\w_{c,\lambda}$ regulates the non-integrable behavior at $\w\to -\infty$.

We emphasize that even though this description is for non-interacting phonons, the central region is still coupled to the reservoirs, leading to dynamical effects accounted by the non-hermitian, effective Hamiltonian $\boldsymbol{\Omega}_{\text{eff}}$. The temperature gradient between the reservoirs transfers energy to the central region's molecular vibrations thus creating a heat wavefront propagating through the system. The exponential terms $\ex^{\im(\w-\boldsymbol{\Omega}_{\text{eff}})t}$ and $\ex^{-\im(\w-\boldsymbol{\Omega}_{\text{eff}}^\dagger)t}$ in Eq.~\eqref{eq:final-result} then resolve the associated `dephasing' and `decoherence' of the phononic signal.

Let us also look at how different limiting cases are recovered from the time-dependent phonon density matrix~\eqref{eq:final-result}. At $t=0$ the square brackets vanish and we are left with the uncoupled result, as should be the case due to the initial condition. This also happens if the systems remain uncoupled during the time evolution, i.e., $\mathbf{\Lambda}_0 = 0 = \mathbf{\Gamma}_0^\prime$. In this situation, we are left with the free evolution of the initial state as $\mathbf{\Omega}_{\text{eff}} \to \alphaBold\mathbf{\Omega}$ and $\mathbf{B}_\lambda(\w)\to 0$. The steady-state result comes from the limit $t\to\infty$, and the exponential terms vanish due to the non-hermitian structure of $\mathbf{\Omega}_{\text{eff}}$~\cite{Tuovinen2016PRB}:
\begin{align}\label{eq:ss-comp}
\rhoBold^{(\text{S})} = \sum_{\lambda=L,R}\int_{-\w_{c,\lambda}}^{\w_{c,\lambda}} \frac{\ud \w}{2\pi}  f_\lambda(\w) \frac{1}{\alphaBold \w-\mathbf{\Omega}-(\mathbf{\Lambda}_0-\frac{\im\w}{2}\mathbf{\Gamma}_0^\prime)}\w\mathbf{\Gamma}_{0,\lambda}^\prime \frac{1}{\alphaBold\w-\mathbf{\Omega}-(\mathbf{\Lambda}_0+\frac{\im\w}{2}\mathbf{\Gamma}_0^\prime)} .
\end{align}
Within the wide-band-like self-energy approximation, $\PiBold^{r/a}(\w) = \theta(\w_{c,\lambda}-|\w|)(\mathbf{\Lambda}_0 \mp \frac{\im\w}{2}\mathbf{\Gamma}_0^\prime)$, we may write Eq.~\eqref{eq:ss-comp} as
\be
\rhoBold^{(\text{S})} = \sum_{\lambda=L,R}\int_{-\infty}^{\infty} \frac{\ud \w}{2\pi} f_\lambda(\w) \DBold^r(\w)\mathbf{\Gamma}_{\lambda}(\w) \DBold^a(\w) ,
\ee
i.e., at $t\to\infty$ the time-dependent result reduces to the standard Landauer--B{\"u}ttiker type derived from various starting points\cite{Ozpineci2001, Dhar2006, Wang2006, Wang2008}.

\subsubsection{Driven systems.}\label{driven_systems}

\begin{figure}
  \begin{subfigure}{.5\linewidth}
  \centering
    \includegraphics[clip, width=\linewidth]{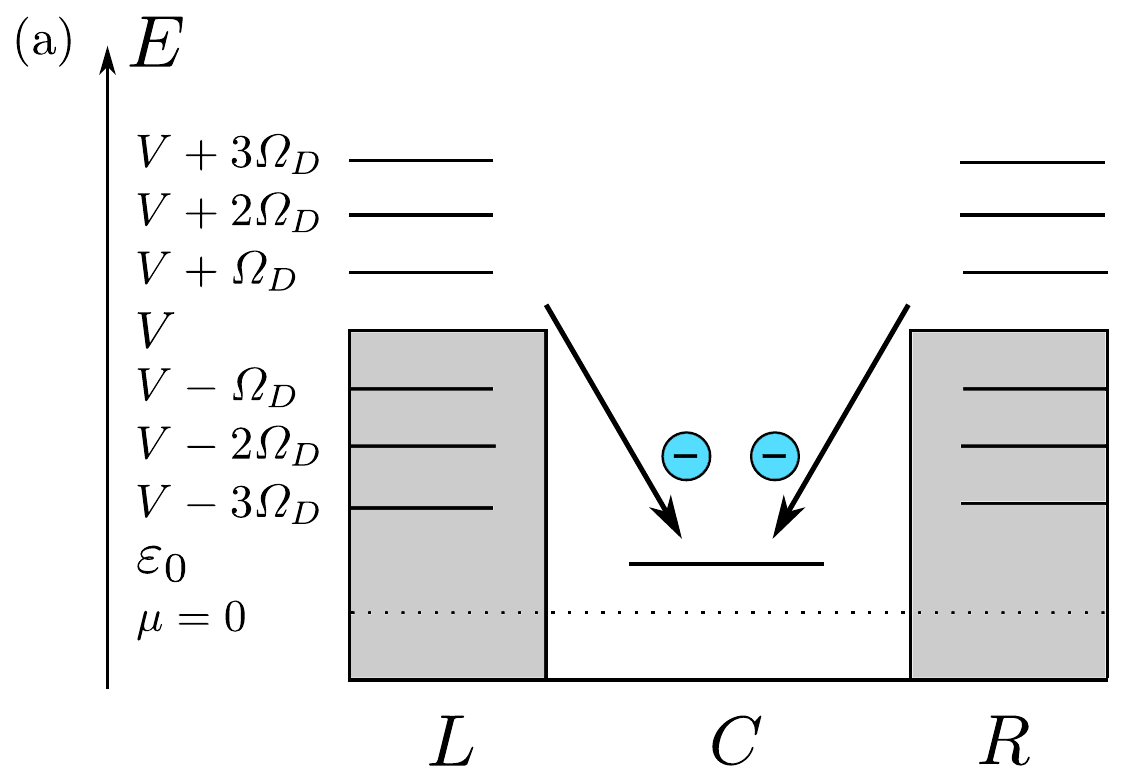}
  \end{subfigure}
  \begin{subfigure}{.5\linewidth}
  \centering
    \includegraphics[clip, width=\linewidth]{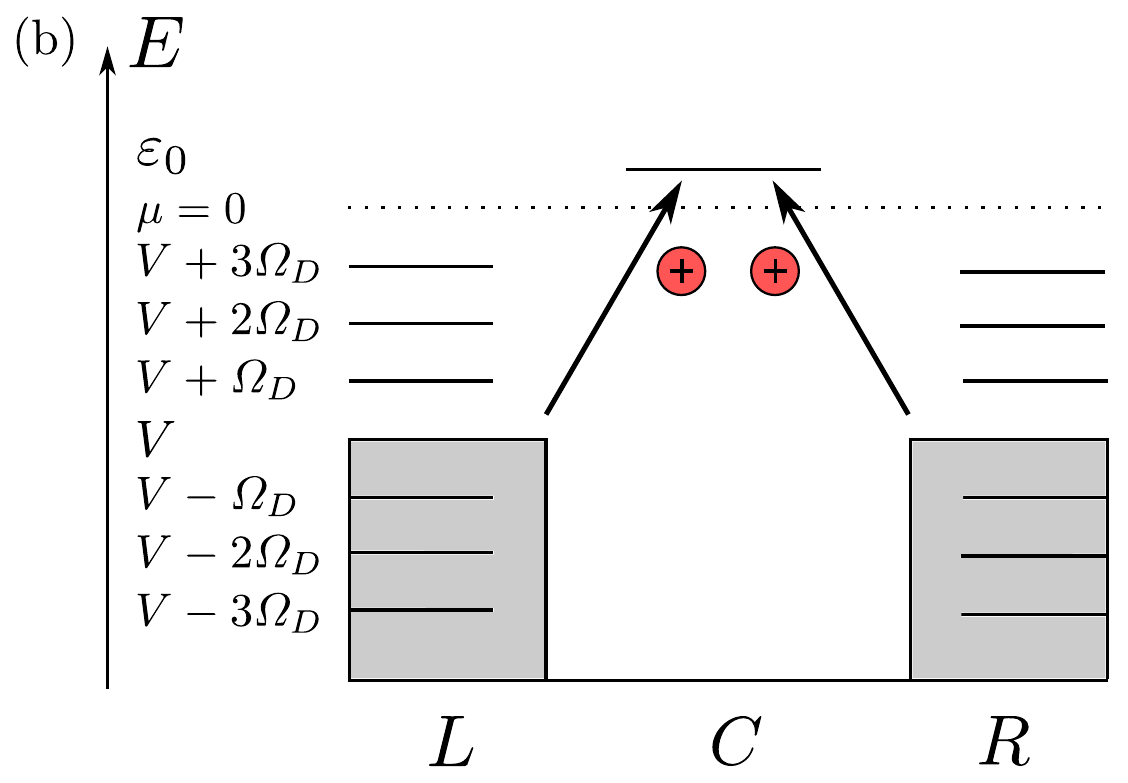}
  \end{subfigure}
\caption{Schematic for photon-assisted electron and hole transfer processes in the quantum dot case. (a) Electron transfer processes in the $V>\varepsilon_0$ case. (b) Hole transfer processes in the $V<\varepsilon_0$ case. From Ref.~\cite{Ridley2017GNR}.}
\label{fig:pumping_energy}
\end{figure}

We now turn to the important area of molecular junctions driven by a periodically-varying external field. In the simplest version of this, the energies of one of the leads $\alpha$ in a biased system are assumed to all be shifted in a spatially homogeneous manner by a periodic external field of driving frequency $\Omega_{\alpha}$, $V_{\alpha}\left(t\right)=V_{\alpha}\left(t+\frac{2\pi}{\Omega_{\alpha}}\right)$. Starting from the time-dependent Meir--Wingreen formula, Eq.~\eqref{eq:I_alpha2}, it can be shown that the period-averaged current reduces to an LB-type formula~\cite{platero2004photon}: 
\begin{align}\label{PAT_LB}
\lim_{t_{0}\to -\infty}\frac{\Omega_{D}}{2\pi}\int_0^{{\frac{2\pi}{\Omega_{D}}}}\ud\tau I_{\alpha}\left(\tau\right) = \frac{1}{\pi}\sum_\gamma & \int \ud\omega\,\left[f\left(\omega-V_{\alpha}-\mu\right)-f\left(\omega-V_{\gamma}-\mu\right)\right] \nonumber \\
& \times \sum_{n=-\infty}^{\infty}\left|c_{n}\right|^{2}T_{\gamma\alpha}\left(\omega+n\Omega_{D}\right).
\end{align}
Here, the usual transmission probability in Eq.~\eqref{transmission} is replaced by an effective photon-assisted tunneling (PAT) transmission function composed of a linear summation over regular transmission functions with peaks shifted by integer multiples of the driving frequency. This is equivalent to a splitting of the lead energies into sidebands of energy $\epsilon_{k\alpha}+V_{\alpha}+n\Omega_{\alpha}$, where the lead energies $\epsilon_{k\alpha}$ are defined in Eq.~\eqref{lead_energies} and $V_{\alpha}$ is a constant shift in the bias applied to lead $\alpha$. The situation is represented schematically in Fig.~\ref{fig:pumping_energy} for the case of a quantum dot molecule with dot energy equal to $\varepsilon_{0}$ lying (a) below and (b) above the lead Fermi level at $V$, respectively.

It is sometimes possible to induce a non-equilibrium process that breaks the spatial symmetry of current flow by introducing a term in the Hamiltonian that breaks time-reversal (TR) symmetry. Choosing $\Omega_{\alpha}=\Omega_{D}=\Omega_{\beta}$ to be the fundamental driving frequency of the periodic signal in the leads, we now define the pump current (also referred to as the DC component of the current in the literature~\cite{arrachea2005green}) at time $\tau$ after the switch-on time:
\begin{align}\label{pump_current}
    I_{\alpha\beta}^{\left(\mathrm{pump}\right)}\left(\tau\right)\equiv\frac{\Omega_{D}}{2\pi}\int_{\tau-t_{0}}^{\tau-t_{0}+\frac{2\pi}{\Omega_{D}}}\ud t\left(I_{\alpha}\left(t\right)-I_{\beta}\left(t\right)\right).
\end{align}
There may be conditions under which a finite pumping current exists in the transient regime following the switch-on, but not in the steady-state limit, when transient modes in the current have decayed to zero. In quantum pump setups, we would like to define general conditions under which this is not true, satisfying the following condition:
\begin{align}\label{eq:nonzero-pumpcurrent}
    \lim_{t_{0}\to-\infty}I_{\alpha\beta}^{\left(\mathrm{pump}\right)}\left(\tau\right)\neq 0.
\end{align}

Typically, a net directed current as in Eq.~\eqref{eq:nonzero-pumpcurrent} or a quantum pump may be achieved by the breaking of dynamical symmetries in the transport setup~\cite{Kohler2005, Denisov2014}, or varying the physical parameters of the nanojunction~\cite{Brouwer2001, Yuge2012}. These situations have been experimentally demonstrated for two-parameter charge pumping~\cite{DiCarlo2003, Blumenthal2007}. It has also been shown to be possible to introduce single-parameter charge pumping with a single periodic source generating a non-zero pump current~\cite{Moskalets2002, arrachea2005green, FoaTorres2005}. For the low-GHz frequency range, these predictions were experimentally confirmed in nanowires etched into semiconductor heterostructures~\cite{Kaestner2008, Fujikawa2008}.

In particular, one is interested in the effect on a pump current of breaking time reversal symmetry (invariance under the tranformation $t\rightarrow2t_{0}-t$) in parts of the Hamiltonian. Focusing on the case of the WBA TD-LB formalism, it was rigorously proven in Ref.~\cite{Ridley2017GNR} that the following theorem holds independently of the functional form of the driving bias:
\begin{tcolorbox}
\textbf{Quantum pump symmetry theorem}

\textit{If (i) $V_{\alpha}\left(t\right)$ and/or $V_{\beta}\left(t\right)$ is given by a sum of more than one harmonic with frequencies that are all integer multiples of $\Omega_{D}$, (ii) in at least one of the leads, TR symmetry is broken in at least one of the harmonics, and (iii) the TR symmetry-breaking is different in each lead, then there is a non-zero net pump current running between the $\alpha$ and $\beta$ leads.}
\end{tcolorbox}

We emphasize that the satisfaction of conditions (i), (ii) and (iii) together constitutes a sufficient, but not a necessary condition for the existence of a non-zero pump current, i.e. the stronger statement that the existence of a non-zero pump current requires (i), (ii) and (iii) to hold is not true. The quantum pump symmetry theorem however gives experimentalists a means of generating a net current per driving cycle with zero net bias per cycle and with no difference in the amplitude of driving signals across the terminals of a nanodevice. In the quantum ratchet effect, spatial asymmetry of the junction in addition to a periodic driving is often used to generate a pumped current \cite{gulyaev2020nanotransport}, but according to the quantum pump symmetry theorem proven here, the system may be completely spatially symmetric, so that $\mathbf{\Gamma}_{\alpha}=\mathbf{\Gamma}_{\beta}$, and still there will be a reliable rectified current if the purely dynamical conditions (i)-(iii) of this theorem are satisfied.

To satisfy all three conditions of this theorem, it is necessary to introduce a second harmonic in the bias with a time period that is some multiple of $2\pi/\Omega_{D}$:
\begin{align}\label{biharmonic_bias}
V_{\alpha}\left(t\right)=V_{\alpha}+A_{\alpha}^{\left(1\right)}\cos\left(p_{1}\Omega_{\alpha}\left(t-t_{0}\right)+\phi_{\alpha}\right)+A_{\alpha}^{\left(2\right)}\cos\left(p_{2}\Omega_{\alpha}\left(t-t_{0}\right)\right).
\end{align}
Here $p_{1}$, $p_{2}$ are any even integers.
Clearly the second higher frequency harmonic breaks the sinusoidal character of the bias when the amplitudes $A_{\alpha}^{\left(1\right)}$ and $A_{\alpha}^{\left(2\right)}$ are comparable. We exemplify these conditions on a non-zero net pump current in Sec.~\ref{sec:applications}; see Fig.~\ref{fig:nonintpanels}(d).

We then make use of a well-known expression for the generating function for the $n$-th order Bessel functions of the first kind $J_{r}\left(x\right)$:
\begin{align}\label{Bessel_ID}
    \ex^{\pm \im x\sin\left(z+\phi\right)}=\sum_{r=-\infty}^{\infty}J_{r}\left(x\right)\ex^{\pm \im r\left(z+\phi\right)}.
\end{align}
The biharmonic choice of bias then leads to the following representation of the exponential phase factor appearing in the Green's functions, currents and noise
\begin{align}\label{Bessel_biharmonic}
\ex^{\im\psi_{\alpha}\left(t_{1},t_{2}\right)}=\ex^{\im V_{\alpha}\left(t_{1}-t_{2}\right)}\sum_{r,r',s,s'} & J_{r}\left(\frac{A_{\alpha}^{\left(1\right)}}{p_{1}\Omega_{\alpha}}\right)J_{r'}\left(\frac{A_{\alpha}^{\left(1\right)}}{p_{1}\Omega_{\alpha}}\right)J_{s}\left(\frac{A_{\alpha}^{\left(2\right)}}{p_{2}\Omega_{\alpha}}\right)J_{s'}\left(\frac{A_{\alpha}^{\left(2\right)}}{p_{2}\Omega_{\alpha}}\right) \nonumber \\
& \times \ex^{\im \left(r-r'\right)\phi_{\alpha}}\ex^{\im\Omega_{\alpha}\left(p_{1}r+p_{2}s\right)\left(t_{1}-t_{0}\right)}\ex^{-\im\Omega_{\alpha}\left(p_{1}r'+p_{2}s'\right)\left(t_{2}-t_{0}\right)},
\end{align}
which enables the analytical removal of all time integrals appearing there, for instance, in the $\textbf{S}_{\alpha}$ matrix defined in Eq.~\eqref{S_alpha}. This leaves the problem of dealing with the frequency integrations appearing in the transport quantities of interest, which is the topic of the next subsection.

\subsection{Numerical approaches in the frequency domain}

In this section we give the interested reader the capability to carry out fast practical calculations of their own using the TD-LB framework. The expressions for the current in Eq.~\eqref{Current_alpha}, lesser/greater Green's functions in Eq.~\eqref{G_Lesser_Greater} and the quantum noise in Eq.~\eqref{Correlation_alpha_beta} are all given by frequency integrals which can be performed analytically. The full details of this can be found in
Refs.~\cite{Tuovinen2013,Tuovinen2014,Ridley2016JPC,Ridley2017GNR}. To proceed with a numerical implementation of these expressions one can introduce the right and left eigenproblems for the renormalized Hamiltonian matrix $\mathbf{h}_{CC}^{\mathrm{eff}}$~\cite{Tuovinen2014}:
\begin{align}\label{left_right_eenergies}
\mathbf{h}_{CC}^{\mathrm{eff}}\left|\varphi_{j}^{R}\right\rangle = \bar{\varepsilon}_{j}\left|\varphi_{j}^{R}\right\rangle \,\,\mbox{and}\,\,\left\langle \varphi_{j}^{L}\right|\mathbf{h}_{CC}^{\mathrm{eff}}=\bar{\varepsilon}_{j}\left\langle \varphi_{j}^{L}\right| .
\end{align}
The eigenenergies $\bar{\varepsilon}_{j}$ contain an imaginary part
that is strictly negative (as $\mathbf{\Gamma}$ is positive-definite),
and the same value of $\bar{\varepsilon}_{j}$ corresponds to each
of the left and right eigenvectors. Using the idempotency property
\begin{equation}
\underset{j}{\sum}\frac{\left|\varphi_{j}^{R}\right\rangle \left\langle \varphi_{j}^{L}\right|}{\left\langle \varphi_{j}^{L}\mid\varphi_{j}^{R}\right\rangle }=\mathbb{1}=\underset{j}{\sum}\frac{\left|\varphi_{j}^{L}\right\rangle \left\langle \varphi_{j}^{R}\right|}{\left\langle \varphi_{j}^{R}\mid\varphi_{j}^{L}\right\rangle }\;,
\end{equation}
we can map all matrices onto scalar integrals. Then, the Fermi function can be re-expressed with a series expansion whose terms possess a simple pole structure:
\begin{align}\label{PADE_FERMI}
f\left(x\right)=\frac{1}{\ex^{\beta x}+1}=\frac{1}{2}-\lim_{N_{p}\rightarrow\infty}\sum_{l=1}^{N_{p}}\eta_{l}\left(\frac{1}{\beta x+\im\zeta_{l}}+\frac{1}{\beta x-\im\zeta_{l}}\right).
\end{align}
When the parameter values are $\eta_{l}=1$ and $\zeta_{l}=\pi\left(2l-1\right)$, this is referred to as the Matsubara expansion:
\begin{align}\label{MATSUBARA_FERMI}
f\left(x\right)=\frac{1}{2}-\sum_{l=1}^{\infty}\left[\frac{1}{\beta x+\im\pi\left(2l-1\right)}+\frac{1}{\beta x-\im\pi\left(2l-1\right)}\right].
\end{align}
One can improve the convergence of this series for finite $N_{p}$ by expressing the Fermi function as a finite continued fraction, and then poles of the Fermi function can be found as the solution to the eigenproblem for a tridiagonal matrix, in the so-called Pad{\'e} approximation~\cite{ozaki2007continued, hu2010communication, hu2011pade, popescu2016efficient}. Using both expansions in Eqs.~\eqref{PADE_FERMI} and~\eqref{MATSUBARA_FERMI}, it is possible to replace all frequency integrals with special functions such as the digamma function, defined as the logarithmic derivative of the gamma function $\Psi\left(z\right)\equiv\frac{\ud\ln\Gamma\left(z\right)}{\ud z}$ and satisfying the property
\begin{align}\label{DIGAMMA}
    \Psi\left(z_{1}\right)-\Psi\left(z_{2}\right)	=	-\underset{n=0}{\overset{\infty}{\sum}}\left(\frac{1}{n+z_{1}}-\frac{1}{n+z_{2}}\right).
\end{align}
In addition, we will make use of the function
\begin{align}\label{HURWITZa}
\bar{\Phi}\left(\tau,\beta,z\right)\equiv\exp\left(-\frac{\pi\tau}{\beta}\right)\Phi\left(\ex^{-\frac{2\pi\tau}{\beta}},1,\frac{1}{2}+\frac{\beta z}{2\im \pi}\right)
\end{align}
where $\Phi$ is the so-called Hurwitz--Lerch Transcendent~\cite{lerch1887note}:
$\Phi\left(z,s,a\right)\equiv \sum_{n=0}^{\infty}\frac{z^{n}}{\left(n+a\right)^{s}}$.
We can, for instance, express the greater/lesser Green's functions in terms of a summation over these functions:  
\begin{align}\label{GF_left_right}
& \mathbf{G}^{\gtrless}\left(t_{1},t_{2}\right) \nonumber \\
& =\frac{1}{2\pi}\sum_{\gamma,k,j}\frac{\left|\varphi_{j}^{R}\right\rangle \left\langle \varphi_{j}^{L}\right|\mathbf{\Gamma}_{\gamma}\left|\varphi_{k}^{L}\right\rangle \left\langle \varphi_{k}^{R}\right|}{\left\langle \varphi_{j}^{L}\mid\varphi_{j}^{R}\right\rangle \left\langle \varphi_{k}^{R}\mid\varphi_{k}^{L}\right\rangle }\ex^{-\im\bar{\varepsilon}_{j}\left(t_{1}-t_{0}\right)}\ex^{\im\bar{\varepsilon}_{k}^{*}\left(t_{2}-t_{0}\right)} \nonumber \\
& \times\left\{ \frac{\im}{\bar{\varepsilon}_{k}^{*}-\bar{\varepsilon}_{j}}\left[\Psi\left(\frac{1}{2}+\frac{\beta}{2\im\pi}\left(\bar{\varepsilon}_{k}^{*}-\mu\right)\right)-\Psi\left(\frac{1}{2}-\frac{\beta}{2\im\pi}\left(\bar{\varepsilon}_{j}-\mu\right)\right)\right]\right.\nonumber \\
&\left.\pm\frac{\pi}{\bar{\varepsilon}_{k}^{*}-\bar{\varepsilon_{j}}}\left[\theta\left(t_{1}-t_{2}\right)\ex^{\im\left(\bar{\varepsilon}_{j}-\bar{\varepsilon}_{k}^{*}\right)\left(t_{2}-t_{0}\right)}+\theta\left(t_{2}-t_{1}\right)\ex^{\im\left(\bar{\varepsilon}_{j}-\bar{\varepsilon}_{k}^{*}\right)\left(t_{1}-t_{0}\right)}\right]\right.\nonumber\\
&\left.-\left(\int_{t_{0}}^{t_{1}}\ud\tau \ex^{\im\left(\bar{\varepsilon}_{j}-\mu\right)\left(\tau-t_{0}\right)}\ex^{-\im\psi_{\gamma}\left(\tau,t_{0}\right)}\bar{\Phi}\left(\tau-t_{0},\beta,\bar{\varepsilon}_{k}^{*}-\mu\right)-\mathrm{c.c.}_{\substack{j\leftrightarrow k \\ t_{1}\leftrightarrow t_{2}}}\right)\right. \nonumber \\
&\left.-\frac{2\pi}{\beta}\left[\theta\left(t_{1}-t_{2}\right)I_{\gamma}\left(t_{2},\beta,\mu,\bar{\varepsilon}_{j},\bar{\varepsilon}_{k}^{*}\right)+\theta\left(t_{2}-t_{1}\right)I_{\gamma}\left(t_{1},\beta,\mu,\bar{\varepsilon}_{j},\bar{\varepsilon}_{k}^{*}\right)\right]\right.\nonumber\\
&\left.-\frac{2\pi}{\beta}\sum_l\eta_{l} \! \left[\theta\left(t_{1}-t_{2}\right) \!\! \underset{t_2}{\overset{t_1}{\int}}\ud\tau \! \underset{t_0}{\overset{t_2}{\int}}\ud\bar{\tau}\ex^{\im\left(\bar{\varepsilon}_{j}-\mu+\im\frac{\zeta_{l}}{\beta}\right)\left(\tau-t_{0}\right)}\ex^{-\im\left(\bar{\varepsilon}_{k}^{*}-\mu+\im\frac{\zeta_{l}}{\beta}\right)\left(\bar{\tau}-t_{0}\right)}\ex^{-\im\psi_{\gamma}\left(\tau,\bar{\tau}\right)}-\mathrm{c.c.}_{\substack{j\leftrightarrow k\\t_{1}\leftrightarrow t_{2}}}\right] \! \right\}
\end{align}
where $\mathrm{c.c.}_{a\leftrightarrow b}$ denotes complex conjugation of the preceding term with the interchange of labels $a$ and $b$, and the following function was introduced for compactness:
\begin{align}\label{I_func}
& I_{\gamma}\left(t,\beta,\mu,\bar{\varepsilon}_{j},\bar{\varepsilon}_{k}^{*}\right) \nonumber \\
&=\frac{1}{2}\int_{t_{0}}^{t}\ud\tau\int_{t_{0}}^{t}\ud\bar{\tau}\ex^{\im\left(\bar{\varepsilon}_{j}-\mu\right)\left(\tau-t_{0}\right)}\ex^{-\im\left(\bar{\varepsilon}_{k}^{*}-\mu\right)\left(\bar{\tau}-t_{0}\right)}\ex^{-\im\psi_{\gamma}\left(\tau,\bar{\tau}\right)}\left.\mathrm{cosech}\left(\frac{\pi}{\beta}\left(\tau-\bar{\tau}\right)\right)\right|_{\tau\neq\bar{\tau}}.
\end{align}

Similar expressions can be derived for the current $I_\alpha\left(t\right)$ and the $\mathbf{\Lambda}_{\alpha}^{\pm}$ matrices [see Eqs. \eqref{eq:lambda_matrix_1} and \eqref{eq:lambda_matrix_2}] needed to evaluate the noise. When a specific functional form for the bias is assumed, all the time integrals can be performed such that the calculation of all dynamical properties in the molecular junction are `single shot' functions of time, making this an extremely efficient method for gaining physical insight into such systems. We include explicit formulas for the case of the biharmonic bias described for the pumping setup in Eq.~\eqref{biharmonic_bias} in~\ref{app:numerics}. These can be straightforwardly included in any software package and applied to a very wide range of transport setups, some of which are described in Sec.~\ref{sec:applications}.

\section{On the state-of-the-art numerical calculations}\label{sec:calc}

So far, we have established that the non-equilibrium Green's function approach is a state-of-the-art computational method for out-of-equilibrium many-body physics. While the non-linear integro-differential equations of Kadanoff and Baym were outlined already in the 1960s~\cite{Kadanoff1962}, the lack of computational capabilities for addressing the complicated double-time structure rendered them fairly impractical until their first numerical solutions were presented in 1984 by Danielewicz~\cite{Danielewicz1984b}. The double-time structure is very expensive for both computing time and for storing the objects in memory. The simplification brought by the generalized Kadanoff--Baym ansatz~\cite{Lipavsky1986} reduced the requirement of double-time propagation of the Green's function to that of a time-local density matrix, making it possible to address, e.g., carrier scattering dynamics and semi-conducting quantum well systems already during the 1990s~\cite{Haug1992, Bonitz1996, Jahnke1997}. Further numerical implementations of the full double-time equations appeared at the turn of the century~\cite{Kohler1999, Semkat1999, Kwong2000}, initiating a specialized community for NEGF based method development and progress~\cite{Bonitz2000, Bonitz2003, Bonitz2006, Bonitz2010, VanLeeuwen2013b, Verdozzi2016, Stefanucci2019}. Indeed, employing the NEGF approach during the past 20 years, a considerable amount of progress has been achieved from subatomic nuclear reactions~\cite{Rios2011} to atomic and molecular scales~\cite{Galperin2007a, Perfetto2015b}, further to condensed phase~\cite{Tuovinen2020, Schuler2020} and mesoscopic systems~\cite{Wang2014, Ludovico2016a}. We will look into these physical applications more in Sec.~\ref{sec:applications}. In this section, we provide a brief selection of practical state-of-the-art calculations with the NEGF approach, particularly focusing on the time-domain.

We have found in Sec.~\ref{sec:TDLB} that for non-interacting systems, the underlying Kadanoff--Baym equations for the electronic (and also phononic) Green's function can be solved analytically within the WBA for the embedding self-energy. For practical calculations, this means that the time-dependent problem is computationally not more demanding than the time-independent one. A stand-alone implementation is thus straightforward, and we have provided in~\ref{app:numerics} the explicit formulas. Frequency-dependent embedding self-energies, following a Lorentzian shape have been addressed, e.g., in Refs.~\cite{Zhu2005, Tang2014a}, without complicating the overall computational setting. More generally, beyond the WBA, the underlying equations for the non-interacting Green's functions and time-dependent currents~\cite{Wingreen1993,Jauho1994} form the basis of the state-of-the-art \texttt{TKWANT} software, simulating time-dependent quantum transport in nanoelectronic devices~\cite{Groth2014,Gaury2014,Kloss2021}.

For interacting systems, the calculation of the electron--electron and/or electron--boson collision integrals [see~Eqs.~\eqref{EOM_GCC1} and~\eqref{KBED1}] and the subsequent time-propagation of the Kadanoff--Baym equations require significant computational effort. Even in equilibrium, the self-consistent solution of the Dyson equation requires the calculation of the Matsubara Green’s function on the imaginary time axis~\cite{Dahlen2005}. This essentially boils down to recasting the integral equations in Tab.~\ref{tab:dyson-list} into a system of linear equations~\cite{Balzer2009, Schueler2018}. Very recently, efficient calculations of the imaginary time Green's functions, based on a low-rank decomposition of the spectral Lehmann representation, have been presented~\cite{Kaye2021b, Kaye2021c}.

Due to the compatibility of the underlying NEGF formalism, the accurate and self-consistent solution of the Dyson equation for the Matsubara Green's function is an ideal starting point to study the dynamics by the propagation of the real-time Dyson or Kadanoff--Baym equations~\cite{Dahlen2007}. Also in the non-equilibrium situation, discretizing the time variables transforms the real-time Dyson equations into a linear system of equations, which can be solved by means of standard operations such as matrix multiplications and inversions~\cite{Eckstein2010, Balzer2010, LoGullo2016, Talarico2019}. This direct inversion scheme may be expensive though. Recently, significant computational progress has been achieved by identifying compressed, low-rank representations of the underlying data structures~\cite{Kaye2021,Kaye2021a}. Within the NEGF formalism, a state-of-the-art \texttt{NESSi} software provides a versatile framework for the solution of various interacting many-body problems out of equilibrium~\cite{Schueler2020nessi}. For the description of time-resolved quantum transport, the embedded equations of motion (see Sec.~\ref{sec:Embedding}) are to be considered, but this does not change the approach for solving the underlying integro-differential equations~\cite{Myohanen2009}. At the GKBA level, including ionization and embedding effects, we mention the \texttt{CHEERS} code~\cite{Perfetto2018b}, which also enables interfacing with other \emph{ab initio} software.

Even though the acquired results from the above approaches can be trusted to be accurate within the limits of perturbation theory, still, the computational scaling of the time-propagation of the full Dyson or Kadanoff--Baym equations is not to be underestimated -- the number of time steps cubed. With recent developments, based on the GKBA for electrons~\cite{Schluenzen2020} and bosons~\cite{Karlsson2021}, it is possible to reformulate the integro-differential equations for the double-time Green's functions in terms of a system of coupled first-order ordinary differential equations for the single-time density matrices. On top of the original GKBA framework (quadratic scaling in the number of time steps), the ODE scheme brings about a speed-up of another order of magnitude, to linear scaling. Importantly, electron--electron and electron--boson interactions can be treated on equal footing without altering the time-linear scaling. One thing to note is that the time-linear form of the GKBA has to be established for each many-body approximation separately. In a very recent publication~\cite{Pavlyukh2021a}, a large class of these many-body approximations has been presented, and the associated diagrammatics have been explicitly outlined. Within this state-of-the-art methodology, all fundamental conservation laws are satisfied independently of the many-body approximation. In the same manner, development of the GKBA-ODE scheme for correlated quantum transport in the time-dependent setting is underway.

\begin{figure}[t]
\centering
\includegraphics[width=0.85\textwidth]{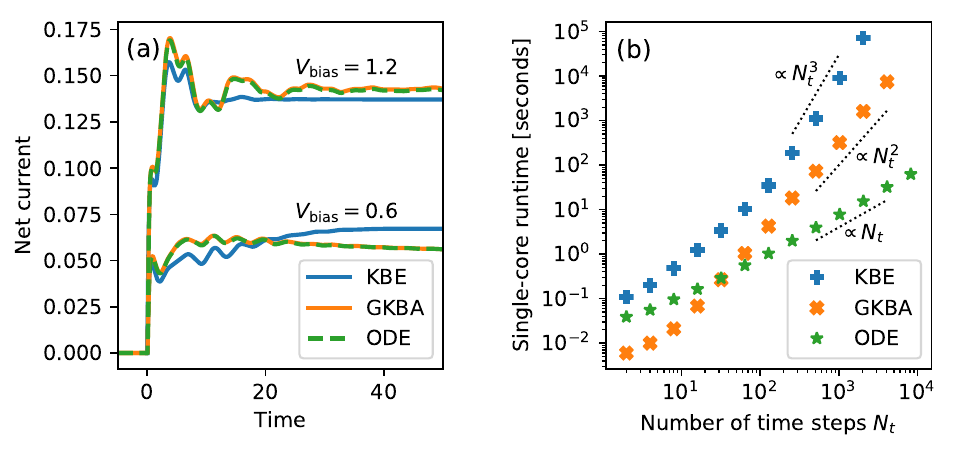}
\caption{An example calculation of time-dependent electronic transport in a four-site Hubbard cluster. (a) Time-dependent net current through the central region. (b) Calculation runtimes for varying number of time steps.}
\label{fig:4site}
\end{figure}

In Fig.~\ref{fig:4site}, we provide a simple yet pedagogic example calculation of time-dependent transport in a correlated quantum system. The system is a small Hubbard cluster, a four-site chain with on-site electron--electron interaction $U=1$ and hopping between the sites $t_C=-1$. The first and fourth sites are connected to left and right, non-interacting leads, respectively. The contact and lead hopping energies are $t_{\alpha C}=-0.5$ and $t_\alpha=-5$ (with $\alpha\in\{L,R\}$), respectively. A bias-voltage $V_{\text{bias}}=V_L=-V_R$ is applied with respect to the chemical potential, which is set in the middle of the highest occupied molecular orbital and the lowest unoccupied molecular orbital $\mu=U/2$ (particle--hole symmetry). The temperature is set by $\beta=100$. The interactions are described at the second-order Born level. For the KBE calculation, the initial equilibrium state (at $t=0$) is obtained by solving the Dyson equation on the imaginary-time axis for the Matsubara Green's function of the coupled and correlated system. For the GKBA (and its equivalent ODE representation), the initial equilibrium state is resolved by an adiabatic switching of correlations and contacts, with a time evolution from $t=-50$ to $t=0$ (extends over the figure frame) starting from the disconnected HF state. The time evolution with all methods is resolved with a uniform step size of $0.05$.

We plot the net current $I_R-I_L$ over the central region in Fig.~\ref{fig:4site}(a). Compared to the full KBE solution, we see that while the initial transient oscillations are well-captured by the GKBA (and its equivalent ODE representation), the steady-state current is underestimated by the GKBA in the smaller-bias case, and it is overestimated by the GKBA in the higher-bias case. This can be understood by the limited spectral features of the GKBA: Due to the exact condition $G^>-G^<=G^r-G^a$, the GKBA spectral function adheres to the form of the HF propagators in Eq.~\eqref{eq:propagator-gamma}. At the HF level, the spectral features are sharp~\cite{Myohanen2009,Tuovinen2021}, and at the lower-bias regime the GKBA solution does not contain as much spectral weight in the bias window as the more broadened KBE spectral function. On the other hand, at the higher-bias regime, the more broadened tails of the KBE spectral function are outside of the bias window, leading to a smaller current. Additionally, WBA is a very good approximation in this situation (utilized in the GKBA and ODE calculations), since the lead bandwidth, $4|t_\alpha|=20$, is considerably larger than the energy scales associated within the bias window. In Fig.~\ref{fig:4site}(b), we see the runtimes of the KBE, GKBA, and ODE calculations and their respective computational scaling with the number of time steps. Indeed, for longer simulation times, the difference in computing time becomes immense, particularly for the completely equivalent GKBA and ODE calculations.

Another aspect of a computational bottleneck we wish to point out here is the construction of various many-body self-energy approximations and, particularly, the associated scaling with respect to the basis size. Recently, this bottleneck has been significantly reduced by using stochastic methods~\cite{Neuhauser2014, Neuhauser2017, Vlcek2017, Vlcek2019}. As an example, we mention the work of Neuhauser, Baer, and Zgid who considered the 2B self-energy in an equilibrium setting and achieved a much more favorable quadratic scaling over the fifth power [cf.~Eq.~\eqref{eq:2b-expanded}] in the number of basis functions~\cite{Neuhauser2017}. While this stochastic sampling approach was performed in an equilibrium setting with a single $\tau$ axis (Matsubara), the non-equilibrium Green's functions depending on two times are more challenging to compute. Instead, such a stochastic-sampling approach may affect convergence or error propagation for the ever-expanding self-energies $\Sigma^\lessgtr(t,t')$ in the two-time plane. However, these issues have recently been addressed also in the real-time setting~\cite{Dou2019, Dou2022}.

\section{Applications}\label{sec:applications}

In this Section, we review applications of time-dependent quantum transport based on the NEGF approach. We set a milestone with the works of Jauho, Meir and Wingreen~\cite{Meir1992, Wingreen1993, Jauho1994}, and provide a thorough list onwards to the present date for various topics according to the logic of increasing complexity. This means starting from non-interacting electronic systems of quantum dots and single resonant levels, then including the spin degree-of-freedom discussing spintronics applications, and then concentrating on more generic tight-binding or DFT-based models for extended systems. Inter-particle interactions have been found to modify the transient features, and we then look at electron--electron interactions from the Anderson model level to extended correlated quantum systems. Electron--boson interactions, in turn, have been found to be important for, e.g., the description of heat transport using the Holstein model for electron--phonon interactions and for, e.g., cavity and radiation effects for the case of electron--photon interaction. We finally discuss purely phononic systems where lattice vibrations or the associated heat currents are being transported in harmonic systems, and how this picture is modified with phonon--phonon interactions.

\subsection{Electronic transport}

In general, a temporal response of a charge (or energy) current flowing through a quantum system when attached to an electrode circuitry rises rapidly after the junction has been switched on. The switch-on process could be, e.g., a fast contacting protocol~\cite{Caroli1971a, Caroli1971b, Pastawski1991, Wingreen1993, Jauho1994}, an application of a potential difference (voltage switch or pulse)~\cite{Cini1980, Stefanucci2004, Myohanen2008, Myohanen2009} or a thermal gradient~\cite{Eich2016, Covito2018} to a contacted system, or a combination of them~\cite{Stefanucci2004, Yang2015, Perfetto2015, Odashima2017, Ridley2017}. After the fast increase of the current signal, the transient behaviour often consists of oscillatory character depending on the internal structure of the quantum system and the coupling between the system and the environment~\cite{Wingreen1993, Jauho1994, Jauho1994b, Jauho1995, Sun1997b, Sun1998, You2000, Hou2006, Maciejko2006, Pan2009, Kalvova2013, Fukadai2018}. If dissipation and dephasing are energetically strong compared to other properties of the system, the decay envelope might completely mask the transient oscillations, and the current signal saturates fast to a stationary value. The other extreme case is that the system consists of bound states, which could lead to no unique stationary state when dissipation is weak~\cite{Dhar2006, Stefanucci2007, Khosravi2008, Khosravi2009}.

\subsubsection{Non-interacting systems.}

Transient current signals in non-interacting quantum-dot systems or single-level junctions have also been addressed systematically using a wavelet decomposition~\cite{Sasaoka2013}, where the transient signal is convolved with a set of basis functions called wavelets. This is useful for  distinguishing between rapid and slow current fluctuations which can, in turn, be addressed simultaneously at different frequencies and at different times. In these systems, not only the first moment of the single-particle observables but also current--current correlations and their associated frequency-dependent noise spectra have been considered~\cite{Feng2008, Joho2012, Yang2014, SeoaneSouto2017b, Carey2017} as they contain more useful information about the transient dynamics. Generally, it has been established that these fluctuations do not follow simple monotonic relaxation processes which could be estimated from their equilibrium or steady-state properties. Instead, the detailed relaxation dynamics are more complicated and depend on various factors such as the bandwidth of the leads and the bias voltage profile. Further information, such as electron waiting time distributions, queueing behaviour and time-dependent Fano factors, can also be gathered from the full-counting statistics (FCS)~\cite{Tang2014a, Tang2014b, Yu2016, Honeychurch2020}.

\begin{figure}[t]
\centering
\includegraphics[width=\textwidth]{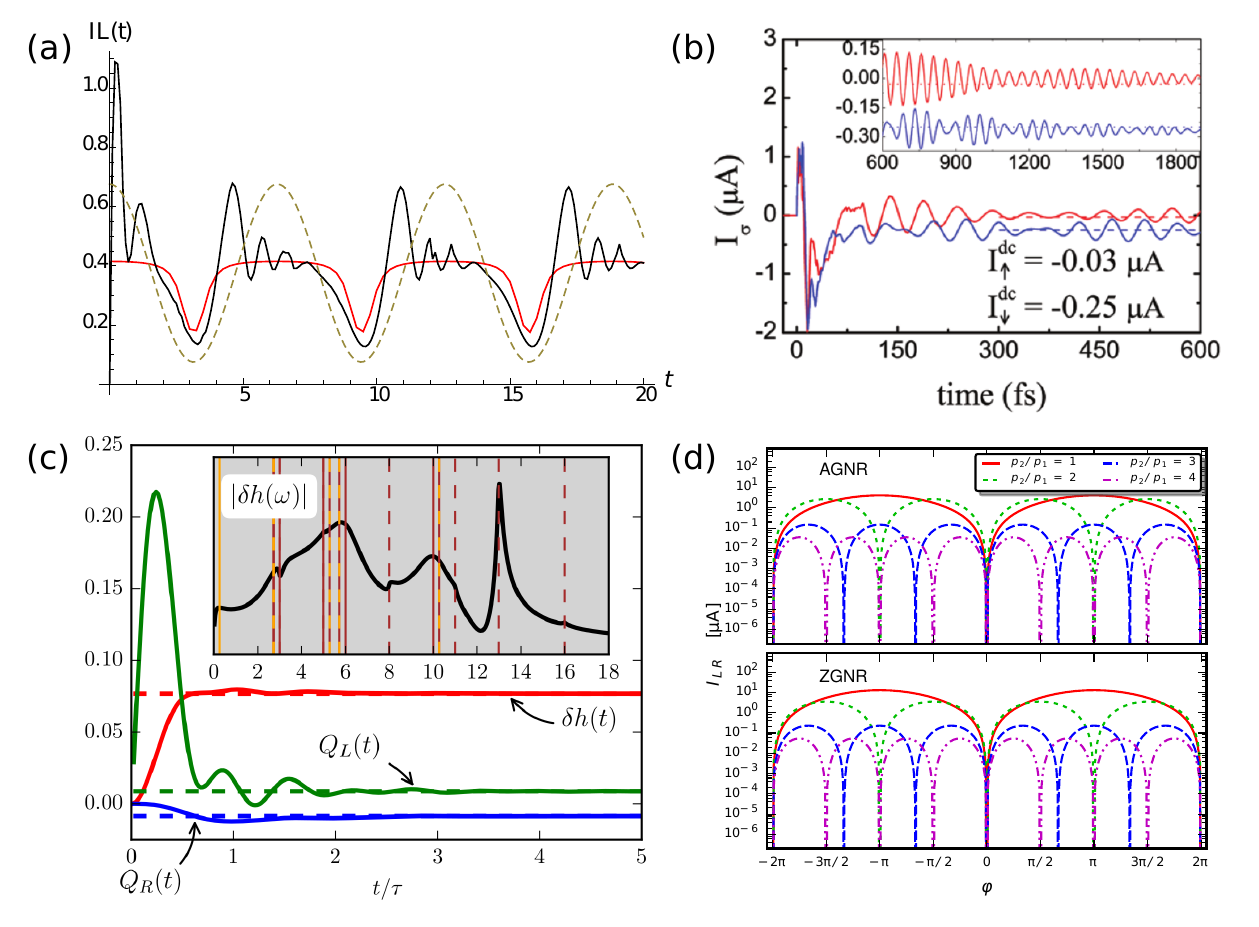}
\caption{(a) Left interface current of a single-level quantum dot coupled to two leads (solid black line), the normalized bias voltage profile (dotted line), and the instantaneous steady-state current corresponding to the bias (solid red line). Within each period, a `ringing' oscillation around a steady-state-like solution is observed. Adapted from Ref.~\cite{Ridley2015}. (b) Spin-polarized transient currents of a magnetic cobalt--graphene system, $I_\uparrow$ (solid red line), $I_\downarrow$ (solid blue line), and the corresponding steady-state currents $I_\sigma^{\text{dc}}$ (dashed lines). The inset shows slow, picosecond relaxation to the stationary state. Adapted from Ref.~\cite{Wang2015}. (c) Transient heat dynamics of a single-level impurity due to a thermal gradient: Local variation of the impurity energy, $\delta h(t)$, and the heat currents between left and right leads and the impurity, $Q_L(t), Q_R(t)$, respectively. The corresponding steady-state values are indicated by the dashed lines. The inset shows the Fourier transform of the energy variation; this power spectrum reflects the distribution of energy levels in the leads. Adapted from Ref.~\cite{Eich2016}. (d) Pump current versus the phase difference in armchair (top) and zigzag (bottom) graphene nanoribbons with different values of the frequency ratio $p_2/p_1$; see Eqs.~\eqref{pump_current} and~\eqref{biharmonic_bias}. The second harmonic of the bias drives the system with a frequency that is a multiple of the frequency of the first harmonic, causing additional nodes to form at $\pm n 2 \pi /p_2$. Adapted from Ref.~\cite{Ridley2017GNR}.}
\label{fig:nonintpanels}
\end{figure}

Already at the level of simple quantum-dot systems, AC driving presents another degree of freedom which makes the transient response more multifaceted~\cite{Prociuk2010b, Popescu2013, Ridley2015, Cuansing2017a, Chowdhury2021}~[see Fig.~\ref{fig:nonintpanels}(a)] with the possibility of, e.g., electron pumps that generate an ``uphill'' current opposing the potential drop~\cite{Ridley2017GNR}. Even though the description of these systems is purely electronic, these properties can generally be associated with photon-assisted tunneling in which irradiated tunnel junctions acquire additional peaks in their conductance spectra [see Fig.~\ref{fig:nonintpanels}(d)], as discussed in Sec.~\ref{driven_systems}. For these photoexcited junctions, extensions beyond the simple quantum-dot systems~\cite{Chen2014, Rahman2018, Beltako2019, Michelini2019, Bajpai2019b} can introduce additional transient effects due to a more complicated level structure, but they do not alter the description of the transient behaviour completely. Instead, these approaches constitute intuitive and technically accessible methods for modeling time-dependent transport phenomena in molecular junctions that are driven, e.g., by modulated electric fields or fluctuating environments.

For extended descriptions of the device region, the first step includes the spin degree of freedom. This makes it possible to address spin-transport phenomena and spintronics applications, such as magnetic tunnel junctions~\cite{Zhu2003, Perfetto2008, Petrovic2018, Bajpai2019, Suresh2021}. Thermal gradients have also been found to enhance spin currents in the transient regime~\cite{Chen2018}. The energy scales of the associated mechanisms are typically fairly small compared to, for example, electronic transitions. Thus, the transient spin-current signatures have characteristic oscillations in the pico- or even nano-second regime~\cite{Wang2015, Hammar2016, Hammar2017, Tao2017, Hammar2019}~[see Fig.~\ref{fig:nonintpanels}(b)]. Also, in the case of spin transport, FCS oscillations have been found to encode more detailed information about, e.g., local Rabi oscillations~\cite{Tang2014a}.

\begin{figure}
\centering
\includegraphics[width=\textwidth]{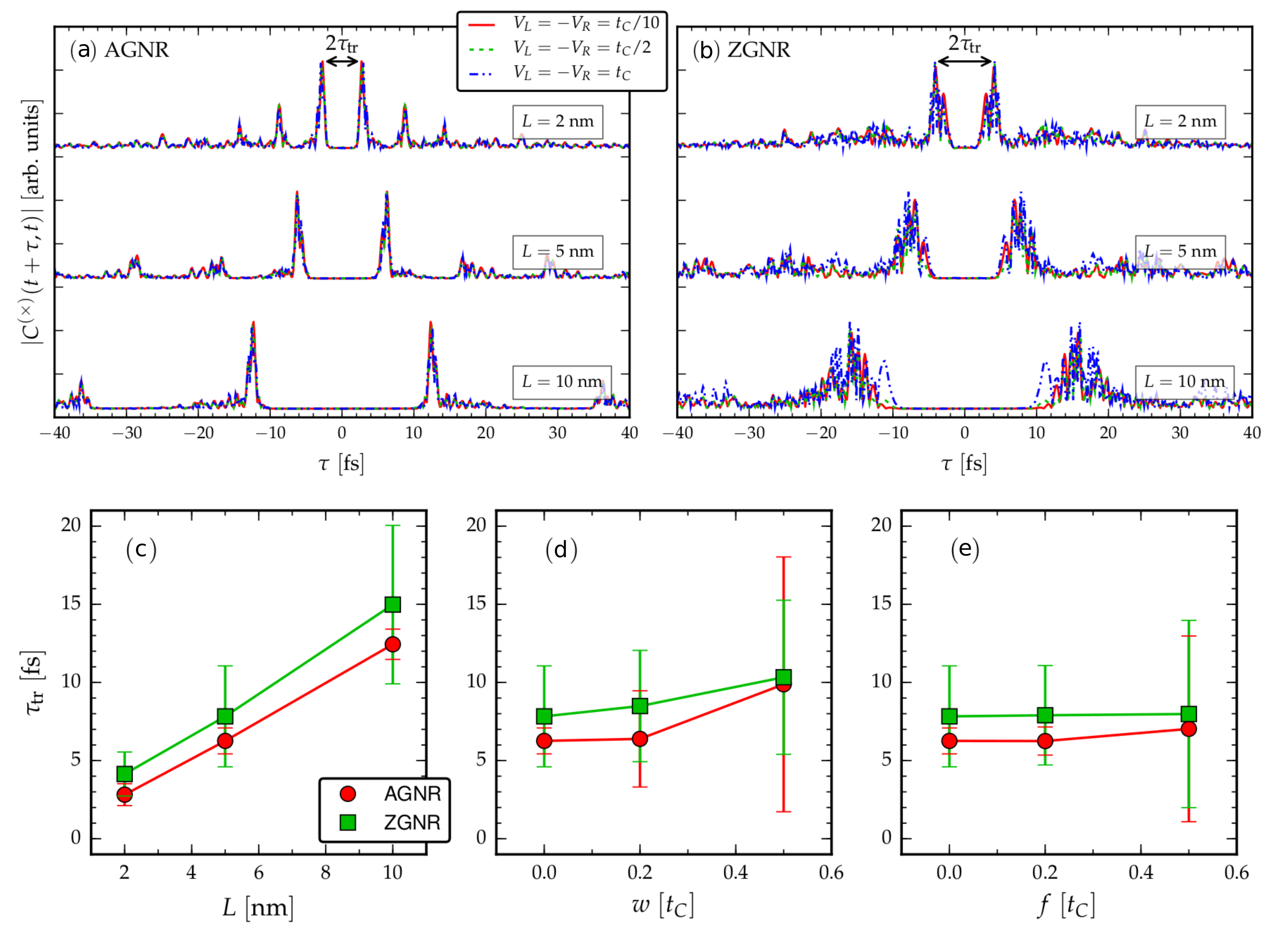}
\caption{Absolute values of the current cross-correlation [see Eq.~\eqref{eq:traversaltime}] at long observation times $t\to\infty$ for graphene nanoribbons and the associated traversal times $\tau_{\text{tr}}$. Panels (a-b) show undisordered armchair and zigzag geometries of different lengths and applies voltages. The current cross-correlations are mostly independent of the strength of the applied voltage, and there is a roughly linear increase of the time-difference between the first maxima with increasing lenghth, due to the time taken for the propagating electron wavefront to cross the structure. Panels (c-e) show the electron traversal times with respect to length, hopping disorder, and onsite disorder, respectively. These are estimated from the distance between the first maxima in the current cross-correlation and the error bars come from the cross-correlation peaks as the full width at half maximum. Adapted from Ref.~\cite{Ridley2019Entropy}.}
\label{fig:Extended_systems}
\end{figure}

The above logic for the description of the device region has also allowed for further extensions, particularly for the development of time-dependent Landauer--B{\"u}ttiker approach for arbitrary junction geometries. As was outlined in Section~\ref{sec:chap3}, some of the present authors have contributed to these developments for electronic transport in arbitrary junction shapes and sizes~\cite{Stefanucci2013Book, Tuovinen2013, Tuovinen2014}, time-dependent driving mechanisms~\cite{Ridley2015, Ridley2016, Ridley2016JPC, Ridley2017GNR}, and current-correlations~\cite{Ridley2017, Ridley2019Entropy}. What these approaches have in common is the use of the wide-band approximation (WBA) which makes it possible to analytically close the equations of motion for the lesser and greater Green's functions due to the time-local nature of the memory kernel, also discussed in Section \ref{sec:TDLB}. When the embedding self-energy has only a weak dependence on frequency around the biased Fermi level of the leads, the WBA is a very good approximation. In particular, this holds in the weak-coupling and small-bias regimes~\cite{Zhu2005, Verzijl2013, Covito2018}. Within the WBA, extended device geometries have been addressed in one-dimensional wires~\cite{Prociuk2008, Prociuk2010}, mesoscopic metallic rings threaded by a magnetic flux~\cite{Tatara2002, Arrachea2002}, disordered systems~\cite{Zhou2016,Ridley2019Entropy}, molecular junctions~\cite{Liu2011,bruch2016quantum,oz2019evaluation}, organic semiconductors~\cite{Leitherer2017}, double-quantum-dot Aharonov--Bohm interferometers~\cite{Dong2004, Tu2012, Tu2016, Yang2018}, graphene nanoribbons~\cite{Zheng2007, Rocha2015, Tuovinen2019nano}, and carbon nanotubes~\cite{Wen2011, Zhang2013}. These extended device structures generally modify the transient response of the junction when compared to the simple quantum-dot or single-level systems. Not only the bandwidth of the leads or the bias voltage profiles are responsible for the important transient mechanisms, but also intra-molecular transitions within the complex level structure of the device region contribute. Similarly, the localized character of external perturbations and the geometrical symmetries of the system, or the lack of them (disorder), give rise to specific transient signatures. Some of these results are reproduced in Fig.~\ref{fig:Extended_systems}.

While the WBA makes the computation of the time-resolved transport properties very efficient, it is not a critical approximation in the NEGF approach, and descriptions beyond the WBA are also possible to resolve numerically. For extended device descriptions, these developments include also one-dimensional wires~\cite{Zhu2005, He2007, Moldoveanu2007, Stefanucci2008, Xie2012, Popescu2013, Popescu2016, Weston2016b}, organic semiconductors~\cite{Stefanucci2008b}, systems on bulk surfaces or STM junctions~\cite{RWang2019, Wang2021}, graphene nanoribbons~\cite{Perfetto2010, Xie2013, Xie2014, Cheung2017}, and carbon nanotubes~\cite{BWang2010}. In general, descriptions beyond the WBA are essential when the interaction between the leads and the device region is not simple, and they introduce further intricacies for a quantitative analysis of transient dynamics of molecular devices.

Not only the charge transport at the nanoscale but also the energy and heat transport have received a lot of attention recently~\cite{Dubi2011}. The studies of efficient thermoelectric devices are naturally motivated by the idea of converting waste heat into usable energy. In this respect, transient thermal transport has been investigated in quantum-dot devices~\cite{Crepieux2011, WLiu2012, Dare2016, Ludovico2016a, Ludovico2016b, Eich2016, Covito2018, Yu2020}~[see Fig.~\ref{fig:nonintpanels}(c)], benzene-molecule junctions~\cite{Yu2014, Lehmann2018}, superconducting junctions~\cite{Arrachea2018}, single-molecule magnets~\cite{Hammar2019}, and nanoribbon quantum-point contacts~\cite{KaraSlimane2020}. Particularly, a partition-free approach with temperature gradients in the transient regime was outlined in Ref.~\cite{Eich2016}~[see Fig.~\ref{fig:nonintpanels}(c)] and applied in conducting nanowires. Regarding the transient behaviour, it is found that the propagation speed of the energy wavefronts is similar to the particle flow, and they are both insensitive to changes in the average electronic density.

The non-interacting picture can also be extended to describe superconducting systems. While the electron--phonon coupling is important for the formation of Cooper pairs, in the Bardeen--Cooper--Schrieffer~(BCS) representation there are strictly speaking no pairs but only an effective pairing field. This translates to a tractable single-particle picture using a Nambu-spinor representation.
This amounts to supplementing the electronic single-particle Hamiltonian in Eq.~\eqref{eq:intsysh0} with a pairing field $\varDelta$ for the anomalous components
\begin{equation}\label{eq:scham}
\hat{H}_0 = \sum_{ij,\sigma}h_{ij}\hat{d}_{i\sigma}^\dagger\hat{d}_{j\sigma} + \sum_{ij,\sigma\sigma'}\left(\varDelta_{ij,\sigma\sigma'}\hat{d}_{i\sigma}^\dagger\hat{d}_{j\sigma'}^\dagger + \varDelta^*_{ij,\sigma\sigma'}\hat{d}_{j\sigma'}\hat{d}_{i\sigma}\right).
\end{equation}
Introducing a spin$\otimes$particle-hole representation, $\hat{\varPhi}_i = (\hat{d}_{i\uparrow},\hat{d}_{i\downarrow}^\dagger,\hat{d}_{i\downarrow},\hat{d}_{i\uparrow}^\dagger)^T$, transforms the Hamiltonian in Eq.~\eqref{eq:scham} into an effective single-particle form $\hat{H}_0 = \sum_{ij}\hat{\varPhi}_i^\dagger\xi_{ij}\hat{\varPhi}_j$, where the matrix $\xi$ contains the same information as the terms in Eq.~\eqref{eq:scham} in a Bogoliubov--de Gennes form. However, it is worth noting that the size of the single-particle basis for, e.g., representing the Green's function, is quadrupled in this procedure. Similar procedure can be done with the lead and coupling Hamiltonians in Eqs.~\eqref{eq:intsyshlead} and~\eqref{eq:intsyshcoupl}.
This has been applied to Andreev-bound state~(ABS) dynamics~\cite{Perfetto2009, Stefanucci2010a, Stefanucci2010b, SeoaneSouto2016, SeoaneSouto2017}~[see Fig.~\ref{fig:interactionpanels}(a)], to the relaxation of Josephson junctions~\cite{Weston2016a}, hybrid Majorana-junction dynamics~\cite{Bondyopadhaya2019}, Majorana zero mode dynamics~\cite{Lai2018, Tuovinen2019NJP}, and to the charge-transfer Majorana operations~\cite{SeoaneSouto2020}. It has been found that characteristic transient oscillations can be attributed to the emergence of the zero-bias conductance peak associated with the Majorana zero modes. However, it is cautioned that also trivial zero-energy states due to, e.g., smooth confining potential around the superconducting island can give rise to similar transient oscillations. It is also interesting to note that the TD-LB formalism outlined in Section~\ref{sec:chap3} is readily applicable in normal metal--superconductor--normal metal junctions~\cite{Tuovinen2016PNGF, Tuovinen2019NJP}.

Even if a non-interacting description is used for the solution of the relevant Green's functions for the evaluation of transient currents, it is possible to combine the methodology with a predetermined \emph{interacting} description of the equilibrium state, for example, using DFT-combined modeling. This approach has been taken to study, e.g., gold--fullerene molecular devices~\cite{Kaun2005}, hydrogen chains~\cite{Ke2010}, single-electron transistors in the Kondo regime~\cite{Goker2010}, aluminum--carbon molecular devices~\cite{BWang2010, Yam2011, Zhang2012, LZhang2013, Yu2014}, double-stranded DNA molecules~\cite{Popescu2012, Kubar2017}, magnetic cobalt--graphene and cobalt--copper devices~\cite{Wang2015, Yu2017}~[see Fig.~\ref{fig:nonintpanels}(b)], black phosphorus transistors~\cite{Wang2019}, and excitation transport in quasi-one-dimensional quantum devices~\cite{Ho2019}. It is generally established that the transient signatures follow closely what one would expect from the underlying DFT-resolved electronic band structure even at the single-particle Green's function level. However, it is worth cautioning that this description might not remain completely accurate in strong out-of-equilibrium situations if, e.g., strong transient currents distort the density, the occupation, or the chemical potential, and a self-consistent potential profile should be included \cite{nitzan2001electron}.

\subsubsection{Electron--electron interactions.}

In general, electronic interactions can modify the transient features of non-interacting systems drastically. At the level of simple quantum dots or the Anderson impurity model, modifications to the transient current oscillations, or ringing due to AC driving, have been reported, even making it possible to transiently reverse the current direction in tunneling devices~\cite{Sun1997a, Fransson2002, Fransson2003, Vovchenko2013, Dong2015}. Due to scattering events, the transient currents can be suppressed~\cite{Wang2010}, and the quantum-dot populations or the consequent current--voltage characteristics may display highly nonlinear behaviour~\cite{Schmidt2008}. It has been cautioned that the fine-tuning of system parameters or the level of approximation for the interaction may enable multiple steady states and the transient switching between them~\cite{Uimonen2010, Uimonen2011}. Nonlinear response and the resonance oscillations in Kondo systems have also been reported~\cite{Goldin2000, Izmaylov2006, Goker2007}. Long-range interactions have been shown to cause a jamming effect drastically reducing the screening time and suppressing the zero-bias conductance~\cite{Perfetto2012}. In the regime of strong-correlations and low temperatures, FCS can be used as a probe for the transient buildup of Kondo features due to spin-flip scattering processes between the leads and the molecule~\cite{Ridley2018PRB}.

\begin{figure}[t]
\centering
\includegraphics[width=\textwidth]{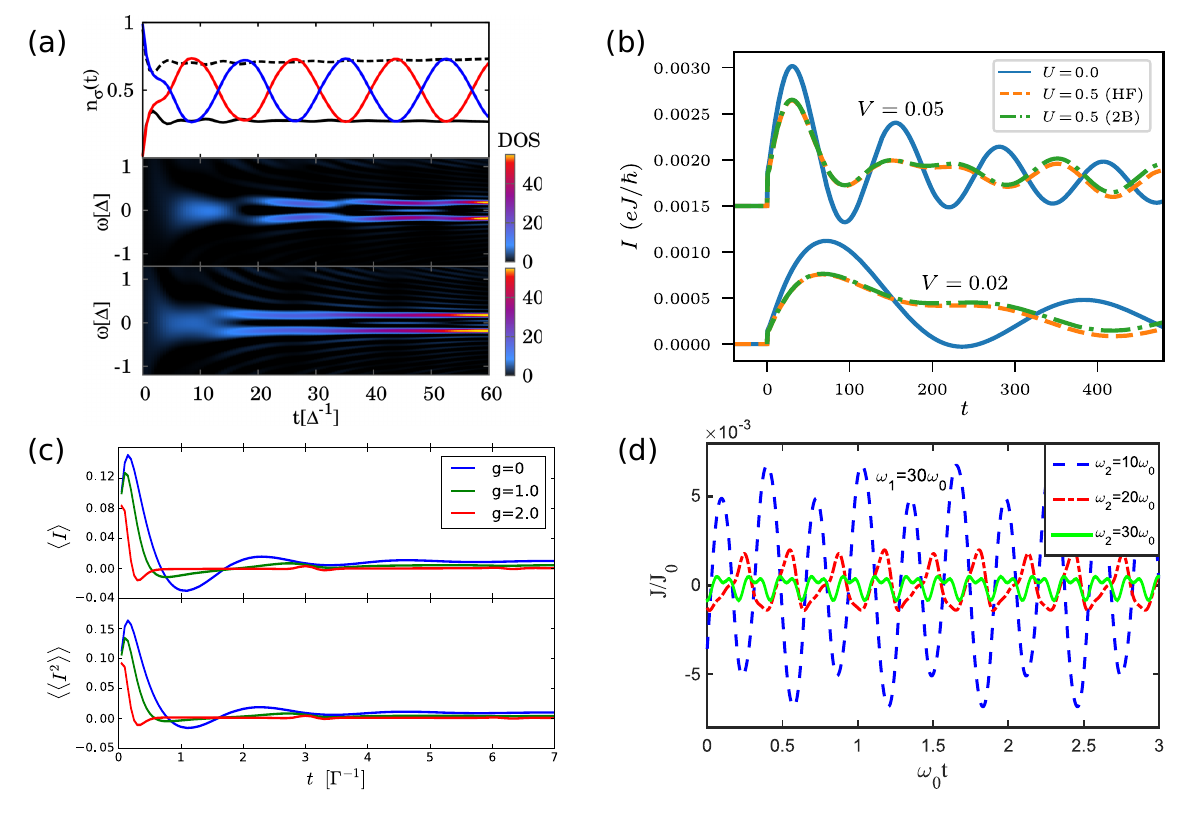}
\caption{(a) Single-level quantum dot population contacted to two superconducting electrodes with the gap parameter $\Delta_L=\Delta_R=\Delta$ (top). Initial values are $(n_\uparrow,n_\downarrow) = (0,0)$ (solid red), $(1,1)$ (solid blue) with $n_\uparrow(t) = n_\downarrow(t)$, and (0,1) [solid black for $n_\uparrow(t)$, dashed black for $n_\downarrow(t)$]. Transient occupied density of states (lower panels) and the associated ABS formation (in-gap peaks) for the $(0,0)$ and $(0,1)$ initial populations. Adapted from Ref.~\cite{SeoaneSouto2017}. (b) Current through a superconducting wire hosting Majorana zero modes (MZM). The solid blue line is the non-interacting limit~\cite{Tuovinen2019NJP}, the dashed green and orange include the on-site electron--electron interaction $U=0.5$ at the HF and 2B approximation, respectively. Electron correlation does not qualitatively destroy the main oscillation, resulting from transitions between the biased Fermi level of the leads and the MZM. Adapted from Ref.~\cite{Tuovinen2021}. (c) Current and noise for a quantum-dot molecular junction including electron--phonon interaction $g$. At $t=0$, a projective measurement is performed in the left lead for a non-equilibrium steady state. The initial `dip' positions of the current and noise shift towards smaller times with increasing $g$ due to polaron dynamics (see also Ref.~\cite{SeoaneSouto2015}). Adapted from Ref.~\cite{Tang2017b}. (d) Photon heat current through a series-coupled double mesoscopic Josephson junction (MJJ) biased by dc voltages. The MJJ device is coupled to photonic reservoirs. The amplitudes and frequencies of the associated photon heat current can be tuned by the MJJ parameters $\w_1,\w_2$, suppressing the sinusoidal form significantly. Adapted from Ref.~\cite{Lu2018}.}
\label{fig:interactionpanels}
\end{figure}

For quantum-dot devices, the study of thermal transport may bring additional insight of the electronic correlation effects~\cite{Goker2012, Goker2013, Tagani2014, Rossello2015, Talarico2020}. As the electronic scattering events redistribute not only the populations but also the energy in the interacting region, the direct exchange of energy of the lead environment is affected by the coupling of the leads through virtual processes involving the interacting region.

Extensions to the single-level quantum devices have included the spin degree of freedom for the study of spin currents, magnetism and torque~\cite{Kalvova2014, Kalvova2015, Kalvova2017, Sasaoka2014, Wang2015, Souza2007, Zhuravel2020}, establishing that, e.g., tunneling magnetoresistance may transiently develop negative values, and the electronic interactions may enhance this effect. In this regard, the GKBA reconstruction has been shown to capture the essential features of magnetic tunnel currents~\cite{Spicka2017, Kalvova2018, Kalvova2019, Spicka2021}.

Increasing the complexity of the tunneling device has resulted in the time-resolved simulation of electron-correlation effects in extended systems, such as correlated double quantum dots with initial correlations~\cite{Myohanen2008}, one-dimensional wires~\cite{Myohanen2009, Myohanen2010, Kwapinski2014}, magnetic skyrmions~\cite{Bostrom2019}, image-charge effects in molecular junctions~\cite{Myohanen2012}, graphene flakes~\cite{Tuovinen2021}, and carbon nanotubes~\cite{cosco_spectral_2020}. The Nambu-spinor representation [Eq.~\eqref{eq:scham}] has also been employed for the study of electron-correlation effects in superconducting nanowires in an out of equilibrium~\cite{Tuovinen2021b} (Fig.~\ref{fig:interactionpanels}(b)). Additional time-resolved features in extended systems have been investigated via DFT-combined modeling of organic molecules~\cite{Perfetto2018, Perfetto2019b, Perfetto2020, Mansson2021}, where ultrafast charge-transfer dynamics resolve faithfully the intricate, atomic details of the molecules, electronic correlations and polarization effects, and the multitude of possible ionization events due to external driving. A general finding in these more complex structures is that the transient features become more rich compared to the simple quantum-dot devices: The non-trivial energy-level or band structure gives rise to both intra- and intermolecular transient oscillations, which can be qualitatively affected by the interplay of electronic interactions, bias voltage or other external driving, or the bandwidth of the leads.

\begin{figure}[t]
\centering
\includegraphics[width=\textwidth]{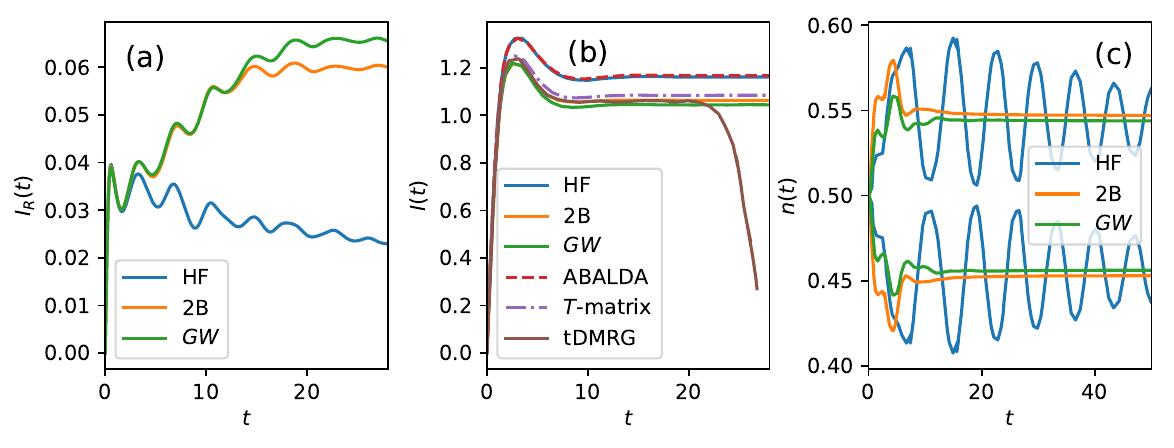}
\caption{Comparison of different levels of self-energy approximations. (a) Transient current flowing into the right lead in a four-site molecular junction with long-range interactions $v_{ij}=v_{ii}/(2|i-j|), v_{ii}=1.5$, and with the applied bias $V=0.8$. Data reproduced from Ref.~\cite{Myohanen2009}. (b) Transient current flowing through a one-site Anderson model with $U=0.5$ at the particle--hole symmetric point ($\epsilon_0=-U/2$) and applied bias $V=1.0$. The data includes an `adiabatic Bethe Ansatz local-density approximation' (ABALDA) calculation within TD-DFT~\cite{Kurth2010} and a numerically exact `time-dependent density matrix renormalization group' (tDMRG) calculation~\cite{HeidrichMeisner2009}. Notably, the tDMRG calculation, performed on a finite system, shows reflections at the system boundaries after sufficiently long propagation times. Data reproduced from Ref.~\cite{Uimonen2011}. (c) Time-dependent densities at the terminal sites of the left (bottom curve) and right (top curve) leads for a two-level molecular junction including an interaction $U=1.0$ between the molecular levels and the terminal sites of the leads. Data reproduced from Ref.~\cite{Myohanen2012}.}
\label{fig:mbpt}
\end{figure}

From the point of view of many-body perturbation theory, it has been established that an accurate description of transient effects in quantum transport requires beyond-HF treatment~\cite{Myohanen2008}, i.e., the inclusion of exchange and correlation effects in a conserving approximation scheme, such as 2B, $GW$, or the $T$-matrix. For molecular junctions, the 2B and $GW$ approximations are generally in agreement with each other at all times for moderate interaction strengths (magnitude similar to that of the hopping integrals), see Fig.~\ref{fig:mbpt}(a). This indicates that the second order bubble diagram (common to both the 2B and $GW$) capture the essential correlation effects~\cite{Myohanen2009}. Benchmark calculations for the Anderson model with broad parameter ranges have shown that densities and currents at the 2B, $GW$, and $T$-matrix level, both for the transient and steady-state regimes, are in good agreement with the time-dependent density matrix renormalization group~\cite{Uimonen2011}, see Fig.~\ref{fig:mbpt}(b). In addition, the mean-field HF approximation fails to capture polarization effects both in and out of equilibrium, leading to the electronic levels of the molecular region not being renormalized by the electronic correlations. Since both 2B and $GW$ contain polarization diagrams, they properly account for the screening of the charged molecular region (and even better with the explicitly calculated screened interaction in the $GW$ approximation), and describe accurately the formation of image charges~\cite{Myohanen2012}, see Fig.~\ref{fig:mbpt}(c).

\subsubsection{Electron--phonon interactions.}

A simple quantum-dot device can also incorporate the coupling of electrons and nuclear motion or lattice vibrations via the (Hubbard--)Holstein model [cf.~Eqs.~\eqref{eq:intsysh0} and~\eqref{eq:intsysv}]. For transport in molecular junctions, this entails the development of many-body perturbation theories for the inclusion of phonon effects~\cite{Mitra2004, Galperin2006a, Ness2011, Wilner2014, Souto2018, Honeychurch2019}. In general, the phonon-induced renormalization of the density of states on the quantum dot and the phonon-induced renormalization of the dot--lead coupling are important. Distinct time-dependent characteristics have been related to polaron effects~\cite{Yamada2009, Albrecht2013a}, tunneling switch and nonadiabaticity~\cite{Riwar2009, Ding2016}, nanomechanical oscillations~\cite{Tahir2010, Biggio2014}, optical excitations~\cite{Park2011, Park2012, Cuansing2017b}, Kondo resonance~\cite{Goker2011, Goker2012, Albrecht2013b}, image-charge effect~\cite{Perfetto2013} and also in higher-order cumulants, waiting-time distributions, FCS and noise~\cite{SeoaneSouto2015, Agarwalla2015, Tang2017b, Tang2017a} (Fig.~\ref{fig:interactionpanels}(c)). In this context, also the partitioning protocols between the dot and the leads have been investigated~\cite{Perfetto2015} disentangling dynamical charge rearrangements due to the dot--lead contacting. Phonons or vibrons may also induce or mediate dynamical effects, such as spin-flip processes~\cite{Maslova2019}, heating~\cite{Pei2012, Liu2013} or electron correlations~\cite{Avriller2019}

Modeling the electron--phonon coupling can also be extended to larger systems. Nuclear dynamics in a quantum transport setup of atomic-ring structures were investigated in Ref.~\cite{Verdozzi2006}, establishing the importance of nonadiabatic effects for ultrafast phenomena in nanodevices. In this context, phonons may also induce decoherence or quantum interference in extended systems such as tight-binding wires~\cite{pastawski2002electron,Zhang2013, Zhang2015}. Quantum interference can be dynamically suppressed by the electron--phonon interaction, and the phase coherence destroyed by phonon scattering. Vibronic dephasing in molecular wires has been characterized in Ref.~\cite{Rahman2019}, where the effect of the thermal environment could be understood as fluctuating site energies of the wire.

\subsubsection{Electron--photon interactions.}

Already at the simple quantum-dot level, Raman spectroscopy has been shown to be a promising diagnostic and control tool in molecular junctions~\cite{Park2011, Park2012}, establishing that the time-dependent optical response of molecular junctions is correlated with the associated electron transport. In this setting, optical fields have also been shown to dynamically interact with a thermal environment~\cite{Manirul2017}. Depending on the system--environment coupling strength, the transient correlations of photon statistics behave differently: A smooth transition from antibunching to bunching in the weak-coupling regime, and persisting two-time correlations up to the steady-state limit in the strong-coupling case.

For extended descriptions of the device region, also here, spin transport has been studied~\cite{Bi2019} with the additional influence of non-classical light, establishing the importance of a quantum-correlated description for phenomena occurring at the interface of spintronics and circuit quantum electrodynamics. Extended device geometries include Josephson devices~\cite{Lu2017a, Lu2017b, Lu2018}, where the time-dependent photon heat current has been investigated (Fig.~\ref{fig:interactionpanels}(d)), and benzene-molecule junctions~\cite{Zhang2020}, where the current-induced angular momentum radiation patterns have been analyzed. In general, the inherently time-dependent mechanisms between electrons and (quantum) photons has been found to be crucial for the accurate description of time-resolved transport properties.

\subsection{Phononic transport}

The Landauer--B{\"u}ttiker and Meir--Wingreen approaches are not limited to electronic transport only, but they can also be formulated in the context of thermal transport for phononic systems~\cite{Ozpineci2001, Galperin2004, Yamamoto2006, Galperin2007b, Wang2008, Wang2014}. Here, we concentrate on these developments from the NEGF perspective.

\subsubsection{Non-interacting systems.}

\begin{figure}[t]
\centering
\includegraphics[width=\textwidth]{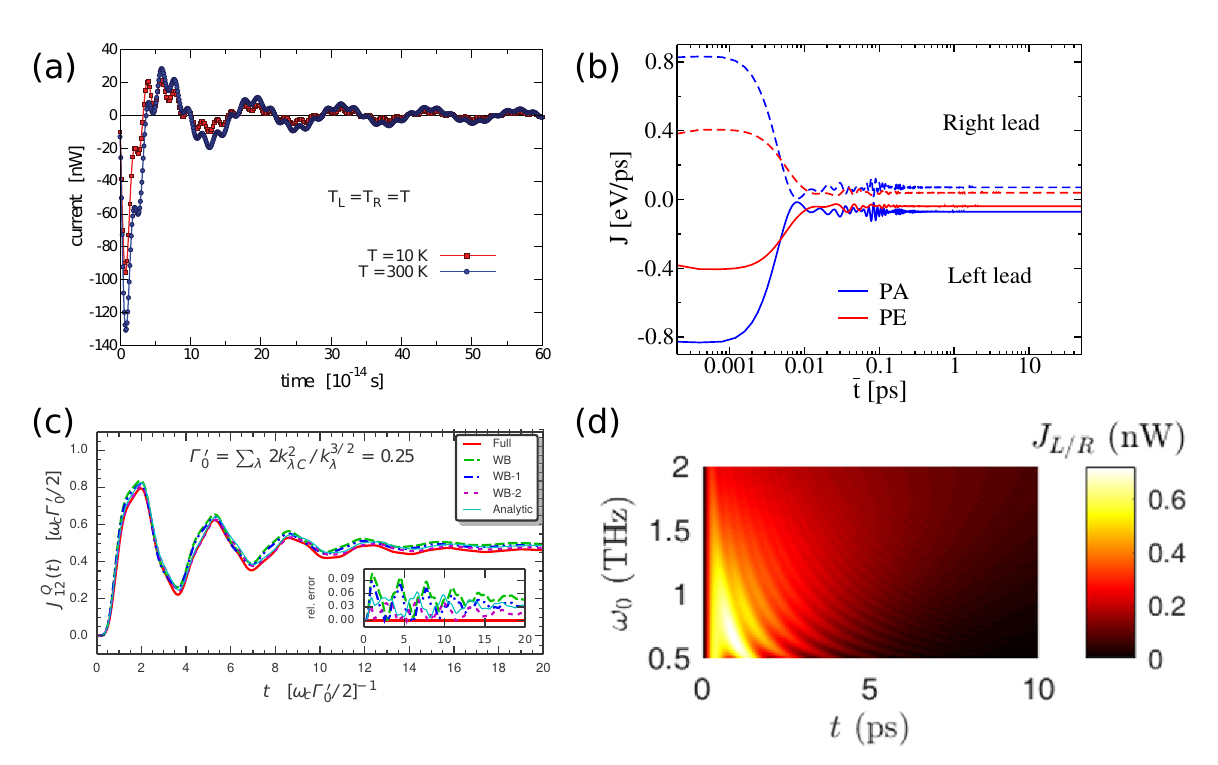}
\caption{(a) Time-dependent heat current through a thermal-switch junction, consisting of left and right leads that are initially uncoupled. The left and right leads have the same temperature: $T=10$~K (red) and $T=300$~K (blue). Adapted from Ref.~\cite{Cuansing2010a}. (b) Dynamics of the heat current for left and right leads, for molecular junctions made of polyacetylene (PA) and polyethylene (PE) dimers, after applying a temperature bias of $\Delta T = 60$~K at $T_0 = 300$~K base temperature. Adapted from Ref.~\cite{MedranoSandonas2018}. (c) Heat current transients through a dimer molecule connected to two reservoirs with weak coupling and narrow spectrum. Thick solid red line is the numerical solution to the full Kadanoff--Baym equation~\eqref{eq:dprime-eom1}, dashed and dotted lines refer to different levels of approximation, and the thin solid cyan line describes the analytic result at the wide-band-like approximation~\eqref{eq:final-result}. The inset shows the relative error between the approximate and full solutions. Adapted from Ref.~\cite{Tuovinen2016PRB}. (d) Transient thermal currents for one-dimensional chains connected to one-atom center with varying characteristic frequency $\w_0$. The oscillation frequencies and the decay times increase by increasing $\w_0$, showing picosecond-scale relaxation. Adapted from Ref.~\cite{Chen2021}.}
\label{fig:phononpanels}
\end{figure}

In the absence of interactions, phononic transport has been described as progating vibrational modes in harmonic lattice systems. This includes the study of heat currents due to thermal gradients~\cite{Cuansing2010a, Cuansing2010b, Cuansing2012, Tuovinen2016PRB, Chen2021} (Fig.~\ref{fig:phononpanels}(a)) and also higher-order cumulants such as FCS~\cite{Wang2011, Agarwalla2012}. Notably, the heat current may transiently flow against the thermal gradient, although at the stationary state the current settles into flowing from the hotter to the colder bath. In addition to the strength of the thermal gradients, the saturation times for this to occur depend on the atomic details of the harmonic lattice~\cite{Cuansing2010a, Tuovinen2016PRB}, on the coupling strength to the leads~\cite{Cuansing2010b, Tuovinen2016PRB, Chen2021} (Fig.~\ref{fig:phononpanels}(c,d)), and also on possible pinning-potentials~\cite{Cuansing2012} applied on the lattice. A combination of these effects may also enable the design of tunable heat pumps~\cite{Cuansing2010b}. 

A step towards more complex device structures has been taken in Refs.~\cite{MedranoSandonas2018, MedranoSandonas2019}, where DFT-combined modeling was carried out for polyacetylene and polyethylene dimers (Fig.~\ref{fig:phononpanels}(b)). Also, the FCS of transient heat currents in graphene junctions~\cite{Wang2011} show that the coupling mechanism between the device and lead intricately connects to the direction of energy flow, even at the stationary state.

\subsubsection{Interacting systems.}

As has already been outlined in the electronic and electron--boson cases, also phononic interactions may play a crucial role, particularly for the description of phonon transport at higher temperatures. As the phononic interaction contributes additional self-energies for the equation of motion for the device-region Green's function, the problem again needs to be considered self-consistently, considerably increasing the computational effort.

Already at the stationary state, nonlinear phononic interactions have been addressed in the calculation of thermal conductance in a one-dimensional chain and carbon nanotube junctions~\cite{Wang2006, Wang2007}. It has been established that higher-order phonon scattering mechanisms are indeed important, particularly for diffusive heat transport. In this context, we also mention a few studies concentrating on heat transport in molecular junctions where the electron--phonon interaction is considered~\cite{Galperin2007b, Lu2007}, as the self-consistent interplay of electrons and phonons effectively generates a mediated phononic interaction through virtual processes involving the electrons.

\section{Summary}\label{sec:summary}

We have presented a thorough review of non-equilibrium Green's function based methodologies for simulating time-resolved quantum transport. This entailed a many-body approach from the transient regime to the stationary state, i.e., we have discussed inter-particle interactions via diagrammatic techniques and system--bath couplings via embedding/inbedding techniques in the fully time-dependent picture. We have also outlined how the underlying NEGF formalism elegantly reduces to well-known quantum-transport results, e.g., the Landauer--B{\"uttiker} formula, in the steady-state limit as special cases.

Even though the NEGF formalism has profound mathematical foundations put forward some 60 years ago, practical simulations of time-dependent transport properties, particularly in experimentally relevant setups, have only recently spiked in activity. The main reason for this is related to the computational effort for solving the integro--differential equations of motion for the one-particle Green's function, i.e. the Kadanoff--Baym equations. Present-day implementations of the KBE are well-established both in accuracy (many-body self-energy approximations) and in computational tractability (time-diagonal reconstruction via the GKBA), and they allow for the simultaneous description of fermion and boson dynamics. This makes the NEGF formalism highly attractive compared to other many-body methods as the computational cost is brought down to the same level and from the viewpoint of the theoretical construction, there are no limitations for the system's dimensionality, the strength of external perturbations, zero or elevated temperatures, or between the transient regime and the stationary state.

State-of-the-art time-resolved transport measurements are pushing the temporal resolution down to the femtosecond regime~\cite{Cocker2013, McIver2020, delaTorre2021}, where the quantum effects of coupled electrons and bosons could be observed in real time. On the other hand, the theoretical progress with the NEGF and GKBA approaches is very topical and timely since it dramatically increases the number of possible out-of-equilibrium systems to be addressed, directly comparable with experiments in real materials. For reasonable comparisons, there is an increasing demand for scalable and accurate computational tools to describe the coupled out-of-equilibrium dynamics of electrons and bosons in quantum systems.

We have outlined the calculation of not only the first statistical moments of the time-dependent density matrix (charge and energy currents) but also current--current correlations and noise. We have pedagogically shown how the time-dependent current formulas including many-body interactions and initial correlations via the solution of the full KBE reduce to the Meir--Wingreen formula in the stationary limit. Furthermore, restricting the description to non-interacting systems, we have formulated the governing equations of the time-dependent Landauer--B{\"u}ttiker approach, also for dynamically driven systems, and shown how this reduces to the traditional Landauer--B{\"u}ttiker formula in the stationary limit. We have established similar correspondence for the static limit of the current correlations and the experimentally accessible power spectrum and thermal and shot noise. For completeness, we have provided explicit expressions for all necessary objects to be utilized in any software package employing the TD-LB formalism.

The NEGF formalism has been successfully applied in various electron--electron and electron--boson systems of quantum-technological relevance. We have presented a thorough literature overview for recent progress in, e.g., time-resolved spintronics, topological superconductors, DFT-combined modeling of organic molecules, and quantum interference with phonon-induced dynamical effects. As the methodology itself is very general and with recent developments also computationally feasible for larger, experimentally relevant setups, we strongly believe that particularly the NEGF+GKBA approach (with the time-linear ODE scheme) will have a significant impact for the field of time-resolved quantum transport. While recent reports do cast GKBA as a trustworthy method when compared to experimental data~\cite{Perfetto2020, Mansson2021}, more research is still required for further validation, since the approach is, to some extent, based on approximative methods (diagrammatic perturbation theory and the time-diagonal reconstruction). Additionally, we envisage further developments of the time-linear GKBA scheme for open quantum systems, photoionization processes, symmetry broken states, to name just a few possible future research directions.

\ack

M.R. wishes to acknowledge support from the Israel Science Foundation Grant No. 2064/19 and the National Science Foundation--US-Israel Binational Science Foundation Grant No. 735/18.
N.W.T. acknowledges financial support from the Academy of Finland via the Centre of Excellence program (Project no. 312058) and from the Finnish Cultural Foundation via PoDoCo program (Project no. 00210085).
D.K. likes to thank the Academy of Finland for funding under Project No. 308697.
R.T. would like to thank the Academy of Finland for financial support under the Project No. 321540, 345007.

\appendix
\addcontentsline{toc}{section}{\ref{app:langreth} Langreth rules}
\addtocontents{toc}{\protect\setcounter{tocdepth}{0}}
\section{Langreth rules}\label{app:langreth}
The integro--differential equations of motion for Green's functions defined on the Konstantinov--Perel' contour $C_{\gamma}=C_{f}\oplus C_{b}\oplus C_{M}$ contain convolution integrals of functions with the following structure:
\begin{equation}
\mathbf{C}\left(z_{1},z_{2}\right)=\int_{\gamma}\ud\bar{z}\mathbf{A}\left(z_{1},\bar{z}\right)\mathbf{B}\left(\bar{z},z_{2}\right).\label{eq: (28)-1}
\end{equation}
To compute specific time-dependent quantum statistical variables, it is necessary to map integrals of this form onto definite integrals on the real number line. This is done using the notational conventions for integrals on the horizontal and vertical contour from Eqs.~\eqref{Convolution_dot} and~\eqref{Convolution_star}. Next, we re-write Eq.~\eqref{eq: (28)-1} in a matrix of `time blocks', for example $\mathbf{A}^{fM}\left(z_{1},\bar{z}\right)$ corresponds to the `block' of $\mathbf{A}$ with $z_{1}$ on $C_{f}$ and $\bar{z}$ on $C_{M}$:
\begin{align}
\mathbf{C}\left(z_{1},z_{2}\right) & =\left(\begin{array}{ccc}
\mathbf{C}^{ff} & \mathbf{C}^{fb} & \mathbf{C}^{fM}\\
\mathbf{C}^{bf} & \mathbf{C}^{bb} & \mathbf{C}^{bM}\\
\mathbf{C}^{Mf} & \mathbf{C}^{Mb} & \mathbf{C}^{MM}
\end{array}\right)\left(z_{1},z_{2}\right)\nonumber\\
 & =\int_{\gamma}\ud\bar{z}\left(\begin{array}{ccc}
\mathbf{A}^{ff} & \mathbf{A}^{fb} & \mathbf{A}^{fM}\\
\mathbf{A}^{bf} & \mathbf{A}^{bb} & \mathbf{A}^{bM}\\
\mathbf{A}^{Mf} & \mathbf{A}^{Mb} & \mathbf{A}^{MM}
\end{array}\right)\left(z_{1},\bar{z}\right)\left(\begin{array}{ccc}
\mathbf{B}^{ff} & \mathbf{B}^{fb} & \mathbf{B}^{fM}\\
\mathbf{B}^{bf} & \mathbf{B}^{bb} & \mathbf{B}^{bM}\\
\mathbf{B}^{Mf} & \mathbf{B}^{Mb} & \mathbf{B}^{MM}
\end{array}\right)\left(\bar{z},z_{2}\right).\label{eq:timeblock}
\end{align}
Carrying out the matrix multiplication, where the integrated time $\bar{z}$ runs over values in the contour branch labelled by the second superscript of $\mathbf{A}$ and the first index of $\mathbf{B}$, we find for instance that the `$fM$' time block of $\mathbf{C}$ is given by
\begin{equation}
\mathbf{C}^{fM}\left(t_{1},\tau_{2}\right)=\left[\mathbf{A}^{ff}\cdot\mathbf{B}^{fM}-\mathbf{A}^{fb}\cdot\mathbf{B}^{bM}+\mathbf{A}^{fM}\star\mathbf{B}^{MM}\right]\left(t_{1},\tau_{2}\right),\label{eq:C-M}
\end{equation}
where the minus sign in the second term arises from the fact that time runs anti-chronologically along $C_{b}$. In similar fashion we find that the general rule for the `$\mu\nu$' time block may be written compactly
\begin{equation}
\mathbf{C}^{\mu\nu}\left(z_{1},z_{2}\right)=\left[\mathbf{A}^{\mu f}\cdot\mathbf{B}^{f\nu}-\mathbf{A}^{\mu b}\cdot\mathbf{B}^{b\nu}+\mathbf{A}^{\mu M}\star\mathbf{B}^{M\nu}\right]\left(z_{1},z_{2}\right).\label{eq:Cmunu}
\end{equation}
Generic two-point correlators are matrices whose elements are structures of the form
\begin{equation}
C\left(z_{1},z_{2}\right)=\textrm{Tr}\left[\hat{\rho}{\mathrm T}_{\gamma}\left[\hat{O}_{1}\left(z_{1}\right)\hat{O}_{2}\left(z_{2}\right)\right]\right]\label{eq:twopointcorr}
\end{equation}
In Ref. \cite{Stefanucci2013Book}, \textit{Keldysh space} is introduced as the space of all two-point correlators which can be represented as a sum of singular, `greater' and `lesser' components
\begin{equation}
C\left(z_{1},z_{2}\right)=C^{\delta}\left(z_{1}\right)\delta\left(z_{1},z_{2}\right)+\theta\left(z_{1},z_{2}\right)C^{>}\left(z_{1},z_{2}\right)+\theta\left(z_{2},z_{1}\right)C^{<}\left(z_{1},z_{2}\right).\label{eq:keldyshspace}
\end{equation}
Here, the two-point correlator is also assumed to be defined with respect to a Hamiltonian that is the same on each of the two horizontal branches, that is
\begin{align}
C\left(z_{1}\in C_{f},z_{2}\right) & =C\left(z_{1}\in C_{b},z_{2}\right),\label{eq:branchindependence1}\\
C\left(z_{1},z_{2}\in C_{f}\right) & =C\left(z_{1},z_{2}\in C_{b}\right).\label{eq:branchindependence2}
\end{align}
This may not always be the case, and is an assumption that is violated by generating functional approaches to the full counting statistics which use a branch-dependent auxiliary Hamiltonian~\cite{gogolin_towards_2006, Tang2014a, Tang2017a, Ridley2018PRB}. Following the convention of Ref.~\cite{Stefanucci2013Book} for the Dirac delta function, and using the generalized contour step function (see below Eq.~\eqref{time_ordering}), we list the complete decompositions of each component:
\begin{align}
C^{ff}\left(t_{1},t_{2}\right)&=C^{\delta}\left(t_{1}^{f}\right)\delta\left(t_{1}-t_{2}\right)+\theta\left(t_{1}-t_{2}\right)C^{>}\left(t_{1}^{f},t_{2}^{f}\right)+\theta\left(t_{2}-t_{1}\right)C^{<}\left(t_{1}^{f},t_{2}^{f}\right),\label{eq:minusminus}\\
C^{fb}\left(t_{1},t_{2}\right)&=C^{<}\left(t_{1}^{f},t_{2}^{b}\right),\label{eq:minusplus}\\
C^{bf}\left(t_{1},t_{2}\right)&=C^{>}\left(t_{1}^{b},t_{2}^{f}\right),\label{eq:plusminus}\\
C^{bb}\left(t_{1},t_{2}\right)&=-C^{\delta}\left(t_{1}^{b}\right)\delta\left(t_{1}-t_{2}\right)+\theta\left(t_{2}-t_{1}\right)C^{>}\left(t_{1}^{b},t_{2}^{b}\right)+\theta\left(t_{1}-t_{2}\right)C^{<}\left(t_{1}^{b},t_{2}^{b}\right),\label{eq:plusplus}\\
C^{fM}\left(t_{1},\tau_{2}\right)&=C^{<}\left(t_{1}^{f},t_{0}-\im\tau_{2}\right),\label{eq:minusM}\\
C^{bM}\left(t_{1},\tau_{2}\right)&=C^{<}\left(t_{1}^{b},t_{0}-\im\tau_{2}\right),\label{eq:plusM}\\
C^{Mf}\left(\tau_{1},t_{2}\right)&=C^{>}\left(t_{0}-\im\tau_{1},t_{2}^{f}\right),\label{eq:Mminus}\\
C^{Mb}\left(\tau_{1},t_{2}\right)&=C^{>}\left(t_{0}-\im\tau_{1},t_{2}^{b}\right),\label{eq:Mplus}\\
C^{MM}\left(\tau_{1},\tau_{2}\right)&=iC^{\delta}\left(t_{0}-\im\tau_{1}\right)\delta\left(\tau_{1}-\tau_{2}\right)+\theta\left(\tau_{1}-\tau_{2}\right)C^{>}\left(t_{0}-\im\tau_{1},t_{0}-\im\tau_{2}\right)\nonumber\\
&+\theta\left(\tau_{2}-\tau_{1}\right)C^{<}\left(t_{0}-\im\tau_{1},t_{0}-\im\tau_{2}\right).\label{eq:MM}
\end{align}
The `greater' and `lesser' parts of this function therefore have a tridiagonal structure in the `time block' representation
\begin{equation}
C^{>}\left(t_{1},t_{2}\right)=\left(\begin{array}{ccc}
C^{>}\left(t_{1}^{f},t_{2}^{f}\right) & 0 & 0\\
C^{>}\left(t_{1}^{b},t_{2}^{f}\right) & C^{>}\left(t_{1}^{b},t_{2}^{b}\right) & 0\\
C^{>}\left(t_{0}-\im\tau_{1},t_{2}^{f}\right) & C^{>}\left(t_{0}-\im\tau_{1},t_{2}^{b}\right) & C^{>}\left(t_{0}-\im\tau_{1},t_{0}-\im\tau_{2}\right)
\end{array}\right),\label{eq:cgreatblockdiag}
\end{equation}
\begin{equation}
C^{<}\left(t_{1},t_{2}\right)=\left(\begin{array}{ccc}
C^{<}\left(t_{1}^{f},t_{2}^{f}\right) & C^{<}\left(t_{1}^{f},t_{2}^{b}\right) & C^{<}\left(t_{1}^{f},t_{0}-\im\tau_{2}\right)\\
0 & C^{<}\left(t_{1}^{b},t_{2}^{b}\right) & C^{<}\left(t_{1}^{b},t_{0}-\im\tau_{2}\right)\\
0 & 0 & C^{<}\left(t_{0}-\im\tau_{1},t_{0}-\im\tau_{2}\right)
\end{array}\right).\label{eq:clessblockdiag}
\end{equation}
Generalized Langreth rules for Hamiltonians which vary between the branches $C^f$ and $C^b$ may be found in Ref.~\cite{kantorovich_generalized_2020}. Here, we assume a branch-independent Hamiltonian, such that the elements in these matrices with both times on $C_{f}\oplus C_{b}$ are given by an identical function, and we can make the identifications
\begin{align}
C^{>}\left(t_{1},t_{2}\right) & = C^{bf}\left(t_{1},t_{2}\right),\label{eq:cgreatplusminus}\\
C^{<}\left(t_{1},t_{2}\right) & = C^{fb}\left(t_{1},t_{2}\right).\label{eq:clessminusplus}
\end{align}
In this case, one can therefore talk of a single `greater' or `lesser' function without specifying the contour branches of the operators $\hat{O}_{1}\left(z_{1}\right)$, $\hat{O}_{2}\left(z_{2}\right)$ in Eq. (\ref{eq:twopointcorr}). Similarly, we can talk of a single `right' function
\begin{equation}
C^{\urcorner}\left(t_{1},\tau_{2}\right)=C^{<}\left(t_{1}^{f/b},t_{0}-\im\tau_{2}\right),\label{eq:rightdef}
\end{equation}
and a single `left' function
\begin{equation}
C^{\ulcorner}\left(\tau_{1},t_{2}\right)=C^{>}\left(t_{0}-\im\tau_{1},t_{2}^{f/b}\right).\label{eq:leftdef}
\end{equation}
These functions are very important in the context of NEGF. We can understand their physical meaning in terms of processes connecting times before and after the quench time. The `Matsubara' component is then given by $C^{M}\left(\tau_{1},\tau_{2}\right)\equiv C^{MM}\left(\tau_{1},\tau_{2}\right)$, where we can now drop one of the `$M$'s for notational convenience. Since the greater and lesser functions now satisfy the properties~\eqref{eq:branchindependence1} and~\eqref{eq:branchindependence2}, we can introduce the `retarded' and `advanced' functions of two real times
\begin{align}
C^{r}\left(t_{1},t_{2}\right) & =  C^{\delta}\left(t_{1}\right)\delta\left(t_{1}-t_{2}\right)+\theta\left(t_{1}-t_{2}\right)\left[C^{>}\left(t_{1},t_{2}\right)-C^{<}\left(t_{1},t_{2}\right)\right],\label{eq:retarded}\\
C^{a}\left(t_{1},t_{2}\right) & =  C^{\delta}\left(t_{1}\right)\delta\left(t_{1}-t_{2}\right)-\theta\left(t_{2}-t_{1}\right)\left[C^{>}\left(t_{1},t_{2}\right)-C^{<}\left(t_{1},t_{2}\right)\right].\label{eq:advanced}
\end{align}
Evidently, these functions satisfy the relation
\begin{equation}
C^{r}\left(t_{1},t_{2}\right)-C^{a}\left(t_{1},t_{2}\right)=C^{>}\left(t_{1},t_{2}\right)-C^{<}\left(t_{1},t_{2}\right).\label{eq:fluctuation-dissipation}
\end{equation}
The greater, lesser, left, right, Matsubara, retarded, and advanced components of a two-time correlator may be collectively referred to as the \textit{Keldysh components}. The problem of deriving analogous expressions to Eq. (\ref{eq:Cmunu}) for the Keldysh components of $\mathbf{C}$ in terms of Keldysh components of $\mathbf{A}$ and $\mathbf{B}$ was solved by Langreth in Ref. \cite{Langreth1976}. One can perform a quick derivation of these so-called \textit{Langreth Rules} by noting the following
\begin{eqnarray}
\mathbf{C}^{r}\left(t_{1},t_{2}\right) & = & \mathbf{C}^{bf}\left(t_{1},t_{2}\right)-\mathbf{C}^{bb}\left(t_{1},t_{2}\right)=\mathbf{C}^{ff}\left(t_{1},t_{2}\right)-\mathbf{C}^{fb}\left(t_{1},t_{2}\right),\label{eq:retcont}\\
\mathbf{C}^{a}\left(t_{1},t_{2}\right) & = & \mathbf{C}^{fb}\left(t_{1},t_{2}\right)-\mathbf{C}^{bb}\left(t_{1},t_{2}\right)=\mathbf{C}^{ff}\left(t_{1},t_{2}\right)-\mathbf{C}^{bf}\left(t_{1},t_{2}\right).\label{eq:advcont}
\end{eqnarray}
We then use the identities~\eqref{eq:cgreatplusminus}, \eqref{eq:clessminusplus}, \eqref{eq:retcont}, and~\eqref{eq:advcont} to get the Langreth Rule for the greater component directly as the `$bf$' time block in Eq.~\eqref{eq:Cmunu}
\begin{align}
\mathbf{C}^{>}\left(t_{1},t_{2}\right) & = \left[\mathbf{A}^{>}\cdot\left(\mathbf{B}^{<}-\mathbf{B}^{bb}+\mathbf{B}^{>}\right)-\left(\mathbf{A}^{<}-\mathbf{A}^{ff}+\mathbf{A}^{>}\right)\cdot\mathbf{B}^{>}+\mathbf{A}^{\urcorner}\star\mathbf{B}^{\ulcorner}\right]\left(t_{1},t_{2}\right)\nonumber \\
 & = \left[\mathbf{A}^{>}\cdot\mathbf{B}^{a}+\mathbf{A}^{r}\cdot\mathbf{B}^{>}+\mathbf{A}^{\urcorner}\star\mathbf{B}^{\ulcorner}\right]\left(t_{1},t_{2}\right).\label{eq:Langrethgreater}
\end{align}
We use Eqs.~\eqref{eq:retcont} and~\eqref{eq:Cmunu} to derive the Langreth Rule for the retarded component
\begin{align}
\mathbf{C}^{r}\left(t_{1},t_{2}\right) & =  \left[\mathbf{A}^{>}\cdot\left(\mathbf{B}^{ff}-\mathbf{B}^{<}\right)-\mathbf{A}^{bb}\cdot\left(\mathbf{B}^{>}-\mathbf{B}^{bb}\right)+\mathbf{A}^{\urcorner}\star\mathbf{B}^{\ulcorner}-\mathbf{A}^{\urcorner}\star\mathbf{B}^{\ulcorner}\right]\left(t_{1},t_{2}\right)\nonumber \\
& = \left[\mathbf{A}^{r}\cdot\mathbf{B}^{r}\right]\left(t_{1},t_{2}\right).\label{eq:Langrethretarded}
\end{align}
The right component is obtained from Eqs.~\eqref{eq:C-M}, \eqref{eq:rightdef}, and~\eqref{eq:retcont}
\begin{align}
\mathbf{C}^{\urcorner}\left(t_{1},\tau_{2}\right) & = \left[\left(\mathbf{A}^{<}-\mathbf{A}^{bb}+\mathbf{A}^{>}\right)\cdot\mathbf{B}^{\urcorner}-\mathbf{A}^{<}\cdot\mathbf{B}^{\urcorner}+\mathbf{A}^{\urcorner}\star\mathbf{B}^{\ulcorner}\right]\left(t_{1},\tau_{2}\right)\nonumber \\
 & = \left[\mathbf{A}^{r}\cdot\mathbf{B}^{\urcorner}+\mathbf{A}^{\urcorner}\star\mathbf{B}^{M}\right]\left(t_{1},\tau_{2}\right).\label{eq:Langrethright}
\end{align}
All the other Langreth Rules are similarly obtained, and we collect them in Tab.~\ref{tab:Langreth-list} for convenience.

\begin{table}
\centering
\begin{tcolorbox}
\begin{tabular}{ c|c }
Component & Langreth rule \\
\hline
\hline \\
Ret/Adv &  $\mathbf{C}^{r/a}\left(t_{1},t_{2}\right)=\left[\mathbf{A}^{r/a}\cdot\mathbf{B}^{r/a}\right]\left(t_{1},t_{2}\right)$\\ \\
Lss/Gtr & $\mathbf{C}^{\lessgtr}\left(t_{1},t_{2}\right)  =\left[\mathbf{A}^{\lessgtr}\cdot\mathbf{B}^{a}+\mathbf{A}^{r}\cdot\mathbf{B}^{\lessgtr}+\mathbf{A}^{\urcorner}\star\mathbf{B}^{\ulcorner}\right]\left(t_{1},t_{2}\right)$ \\ \\
Right & $\mathbf{C}^{\urcorner}\left(\tau_{1},t_{2}\right)=\left[\mathbf{A}^{\urcorner}\cdot\mathbf{B}^{a}+\mathbf{A}^{M}\star\mathbf{B}^{\urcorner}\right]\left(\tau_{1},t_{2}\right)$ \\ \\
Left & $\mathbf{C}^{\ulcorner}\left(\tau_{1},t_{2}\right)=\left[\mathbf{A}^{\ulcorner}\cdot\mathbf{B}^{a}+\mathbf{A}^{M}\star\mathbf{B}^{\ulcorner}\right]\left(\tau_{1},t_{2}\right)$ \\ \\
Matsubara & $\mathbf{C}^{M}\left(\tau_{1},\tau_{2}\right)  =  \left[\mathbf{A}^{M}\star\mathbf{B}^{M}\right]\left(\tau_{1},\tau_{2}\right)$ \\ 
\end{tabular}
\caption{Langreth rules for the convolution integrals appearing in the Kadanoff--Baym equations.}
\label{tab:Langreth-list}
\end{tcolorbox}
\end{table}

These rules provide us with a concise means of writing down equations of motion for specific physical quantities that are described by parts of the full one-particle Green's Function. Usually they are derived in a much more tedious fashion, by fixing the time arguments of $\mathbf{C}$ to isolate one of the Keldysh components, then expanding the integral~\eqref{eq: (28)-1} into integrals along each of the three time branches of the Konstantinov--Perel' contour. When these rules are applied to the equation of motion for the Green's function (cf.~Secs.~\ref{sec:spgf} and~\ref{sec:kbet2}), one obtains the full set of coupled differential equations describing the many-body dynamics, the \emph{Kadanoff--Baym equations}~\cite{Kadanoff1962}.

\addtocontents{toc}{\protect\setcounter{tocdepth}{1}}
\addcontentsline{toc}{section}{\ref{app:numerics} Explicit expressions for TD-LB with biharmonic driving}
\addtocontents{toc}{\protect\setcounter{tocdepth}{0}}
\section{Explicit expressions for TD-LB with biharmonic driving}\label{app:numerics}

Here, we provide explicit expressions for the calculation of lesser and greater (double time) Green's functions, charge current, and current correlations in the TD-LB formalism with the biharmonic driving. All time and frequency integrals can be represented in terms of known special functions, so the presented expressions are extremely fast to evaluate numerically. Notably, these same expressions can be used for the monoharmonic drive, i.e., when the amplitude of the second harmonic $A^{(2)}$ in Eq.~\eqref{biharmonic_bias} is zero. In this situation, the corresponding nested summations may be optimized.

When we substitute Eq.~\eqref{biharmonic_bias} into Eq.~\eqref{GF_left_right}, we obtain the following result for the greater and lesser Green's functions, also published in Refs.~\cite{Ridley2017, Ridley2017GNR}:
\begin{align}\label{GF_biharmonic}
& \mathbf{G}_{CC}^{\gtrless}\left(t_{1},t_{2}\right) \nonumber \\ & =\frac{1}{2\pi}\sum_{\gamma,k,j}\frac{\left|\varphi_{j}^{R}\right\rangle \left\langle \varphi_{j}^{L}\right|\mathbf{\Gamma}_{\gamma}\left|\varphi_{k}^{L}\right\rangle \left\langle \varphi_{k}^{R}\right|}{\left\langle \varphi_{j}^{L}\mid\varphi_{j}^{R}\right\rangle \left\langle \varphi_{k}^{R}\mid\varphi_{k}^{L}\right\rangle }\left[\tcircle{1}_G + \im\sum_{r,s}J_{r}\left(\frac{A_{\gamma}^{\left(1\right)}}{p_{1}\Omega_{\gamma}}\right)J_{s}\left(\frac{A_{\gamma}^{\left(2\right)}}{p_{2}\Omega_{\gamma}}\right)\left(\tcircle{2}_G + \tcircle{3}_G \right)\right. \nonumber \\
& \left.+\im\sum_{r,r',s,s'}J_{r}\left(\frac{A_{\gamma}^{\left(1\right)}}{p_{1}\Omega_{\gamma}}\right)J_{r'}\left(\frac{A_{\gamma}^{\left(1\right)}}{p_{1}\Omega_{\gamma}}\right)J_{s}\left(\frac{A_{\gamma}^{\left(2\right)}}{p_{2}\Omega_{\gamma}}\right)J_{s'}\left(\frac{A_{\gamma}^{\left(2\right)}}{p_{2}\Omega_{\gamma}}\right)\right.\nonumber \\
& \left.\times \frac{\ex^{-\im\left(r-r'\right)\phi_{\gamma}}}{\bar{\varepsilon}_{j}-\bar{\varepsilon}_{k}^{*}-\Omega_{\gamma}\left(p_{1}\left(r-r'\right)+p_{2}(s-s')\right)}\left(\tcircle{4}_G + \tcircle{5}_G + \tcircle{6}_G + \tcircle{7}_G\right)\right]
\end{align}
with the introduced abbreviations
\begin{align}
\tcircle{1}_G & = \pm\frac{\pi\left[\theta\left(t_{1}-t_{2}\right)\ex^{-\im\bar{\varepsilon}_{j}\left(t_{1}-t_{2}\right)}+\theta\left(t_{2}-t_{1}\right)\ex^{\im\bar{\varepsilon}_{k}^{*}\left(t_{2}-t_{1}\right)}\right]}{\bar{\varepsilon}_{k}^{*}-\bar{\varepsilon_{j}}}\nonumber\\
& + \frac{\im\ex^{-\im\bar{\varepsilon}_{j}\left(t_{1}-t_{0}\right)}\ex^{\im\bar{\varepsilon}_{k}^{*}\left(t_{2}-t_{0}\right)}}{\bar{\varepsilon}_{k}^{*}-\bar{\varepsilon}_{j}}\left[\Psi\left(\frac{1}{2}+\frac{\beta}{2\pi\im}\left(\bar{\varepsilon}_{k}^{*}-\mu\right)\right)-\Psi\left(\frac{1}{2}-\frac{\beta}{2\pi\im}\left(\bar{\varepsilon}_{j}-\mu\right)\right)\right],
\end{align}
\begin{align}
\tcircle{2}_G & = \frac{\ex^{-\im r\phi_{\gamma}}\ex^{\im\frac{A_{\gamma}^{\left(1\right)}}{p_{1}\Omega_{\gamma}}\sin\phi_{\gamma}}}{\bar{\varepsilon}_{j}-\bar{\varepsilon}_{k}^{*}-V_{\gamma}-\Omega_{\gamma}\left(p_{1}r+p_{2}s\right)}\left[\ex^{-\im\bar{\varepsilon}_{j}\left(t_{1}-t_{0}\right)}\ex^{\im\bar{\varepsilon}_{k}^{*}\left(t_{2}-t_{0}\right)}\right.\nonumber\\
& \left.\times\left[\Psi\left(\frac{1}{2}+\frac{\beta}{2\pi \im}\left(\bar{\varepsilon}_{k}^{*}-\mu\right)\right)-\Psi\left(\frac{1}{2}+\frac{\beta}{2\pi \im}\left(\bar{\varepsilon}_{j}-\mu-V_{\gamma}-\Omega_{\gamma}\left(p_{1}r+p_{2}s\right)\right)\right)\right]\right. \nonumber \\
& \left.+\ex^{\im\bar{\varepsilon}_{k}^{*}\left(t_{2}-t_{0}\right)}\ex^{-\im\left(\mu+V_{\gamma}+\Omega_{\gamma}\left(p_{1}r+p_{2}s\right)\right)\left(t_{1}-t_{0}\right)}\left[\bar{\Phi}\left(t_{1}-t_{0},\beta,\bar{\varepsilon}_{k}^{*}-\mu\right)\right.\right.\nonumber \\
& \left.\left.-\bar{\Phi}\left(t_{1}-t_{0},\beta,\bar{\varepsilon}_{j}-\mu-V_{\gamma}-\Omega_{\gamma}\left(p_{1}r+p_{2}s\right)\right)\right]\right],
\end{align}
\begin{align}
\tcircle{3}_G & = \frac{\ex^{\im r\phi_{\gamma}}\ex^{-\im\frac{A_{\gamma}^{\left(1\right)}}{p_{1}\Omega_{\gamma}}\sin\phi_{\gamma}}}{\bar{\varepsilon}_{k}^{*}-\bar{\varepsilon}_{j}-V_{\gamma}-\Omega_{\gamma}\left(p_{1}r+p_{2}s\right)}\left[\ex^{-\im\bar{\varepsilon}_{j}\left(t_{1}-t_{0}\right)}\ex^{\im\bar{\varepsilon}_{k}^{*}\left(t_{2}-t_{0}\right)}\right. \nonumber \\
& \left.\times\left[\Psi\left(\frac{1}{2}-\frac{\beta}{2\pi \im}\left(\bar{\varepsilon}_{j}-\mu\right)\right)-\Psi\left(\frac{1}{2}-\frac{\beta}{2\pi \im}\left(\bar{\varepsilon}_{k}^{*}-\mu-V_{\gamma}-\Omega_{\gamma}\left(p_{1}r+p_{2}s\right)\right)\right)\right]\right. \nonumber \\
& \left.+\ex^{-\im\bar{\varepsilon}_{j}\left(t_{1}-t_{0}\right)}\ex^{\im\left(\mu+V_{\gamma}+\Omega_{\gamma}\left(p_{1}r+p_{2}s\right)\right)\left(t_{2}-t_{0}\right)}\left[\bar{\Phi}\left(t_{2}-t_{0},\beta,-\left(\bar{\varepsilon}_{j}-\mu\right)\right)\right.\right.\nonumber \\
& \left.\left.-\bar{\Phi}\left(t_{2}-t_{0},\beta,-\left(\bar{\varepsilon}_{k}^{*}-\mu-V_{\gamma}-\Omega_{\gamma}\left(p_{1}r+p_{2}s\right)\right)\right)\right]\right],
\end{align}
\begin{align}
\tcircle{4}_G & = \ex^{-\im\bar{\varepsilon}_{j}\left(t_{1}-t_{0}\right)}\ex^{\im\bar{\varepsilon}_{k}^{*}\left(t_{2}-t_{0}\right)}\left[\Psi\left(\frac{1}{2}+\frac{\beta}{2\pi \im}\left(\bar{\varepsilon}_{j}-\mu-V_{\gamma}-\Omega_{\gamma}\left(p_{1}r+p_{2}s\right)\right)\right)\right.\nonumber \\
& \left.-\Psi\left(\frac{1}{2}-\frac{\beta}{2\pi \im}\left(\bar{\varepsilon}_{k}^{*}-\mu-V_{\gamma}-\Omega_{\gamma}\left(p_{1}r'+p_{2}s'\right)\right)\right)\right],
\end{align}
\begin{align}
\tcircle{5}_G & = \ex^{\im\bar{\varepsilon}_{k}^{*}\left(t_{2}-t_{0}\right)}\ex^{-\im\left(\mu+V_{\gamma}+\Omega_{\gamma}\left(p_{1}r+p_{2}s\right)\right)\left(t_{1}-t_{0}\right)}\left[\bar{\Phi}\left(t_{1}-t_{0},\beta,\bar{\varepsilon}_{j}-\mu-V_{\gamma}-\Omega_{\gamma}\left(p_{1}r+p_{2}s\right)\right)\right. \nonumber \\
& \left.-\bar{\Phi}\left(t_{1}-t_{0},\beta,\bar{\varepsilon}_{k}^{*}-\mu-V_{\gamma}-\Omega_{\gamma}\left(p_{1}r'+p_{2}s'\right)\right)\right] \nonumber \\
& +\ex^{-\im\bar{\varepsilon}_{j}\left(t_{1}-t_{0}\right)}\ex^{\im\left(\mu+V_{\gamma}+\Omega_{\gamma}\left(p_{1}r'+p_{2}s'\right)\right)\left(t_{2}-t_{0}\right)}\left[\bar{\Phi}\left(t_{2}-t_{0},\beta,-\left(\bar{\varepsilon}_{j}-\mu-V_{\gamma}-\Omega_{\gamma}\left(p_{1}r+p_{2}s\right)\right)\right)\right. \nonumber \\
& \left.-\bar{\Phi}\left(t_{2}-t_{0},\beta,-\left(\bar{\varepsilon}_{k}^{*}-\mu-V_{\gamma}-\Omega_{\gamma}\left(p_{1}r'+p_{2}s'\right)\right)\right)\right],
\end{align}
\begin{align}
\tcircle{6}_G & = \theta\left(t_{1}-t_{2}\right)\left[\ex^{-\im\Omega_{\gamma}\left(p_{1}r+p_{2}s\right)\left(t_{1}-t_{0}\right)}\ex^{\im\Omega_{\gamma}\left(p_{1}r'+p_{2}s'\right)\left(t_{2}-t_{0}\right)}\ex^{-\im\left(\mu+V_{\gamma}\right)\left(t_{1}-t_{2}\right)}\right. \nonumber \\
& \left.\times\left[\bar{\Phi}\left(t_{1}-t_{2},\beta,\bar{\varepsilon}_{k}^{*}-\mu-V_{\gamma}-\Omega_{\gamma}\left(p_{1}r'+p_{2}s'\right)\right)\right.\right.\nonumber \\
& \left.\left.-\bar{\Phi}\left(t_{1}-t_{2},\beta,\bar{\varepsilon}_{j}-\mu-V_{\gamma}-\Omega_{\gamma}\left(p_{1}r+p_{2}s\right)\right)\right]\right. \nonumber \\
& \left.+\ex^{-\im\bar{\varepsilon}_{j}\left(t_{1}-t_{2}\right)}\ex^{-\im\Omega_{\gamma}\left(p_{1}\left(r-r'\right)+p_{2}\left(s-s'\right)\right)\left(t_{2}-t_{0}\right)}\left[\Psi\left(\frac{1}{2}-\frac{\beta}{2\pi \im}\left(\bar{\varepsilon}_{j}-\mu-V_{\gamma}-\Omega_{\gamma}\left(p_{1}r+p_{2}s\right)\right)\right)\right.\right.\nonumber \\
& \left.\left.-\Psi\left(\frac{1}{2}+\frac{\beta}{2\pi \im}\left(\bar{\varepsilon}_{j}-\mu-V_{\gamma}-\Omega_{\gamma}\left(p_{1}r+p_{2}s\right)\right)\right)\right]\right],
\end{align}
\begin{align}
\tcircle{7}_G & = \theta\left(t_{2}-t_{1}\right)\left[\ex^{-\im\Omega_{\gamma}\left(p_{1}r+p_{2}s\right)\left(t_{1}-t_{0}\right)}\ex^{\im\Omega_{\gamma}\left(p_{1}r'+p_{2}s'\right)\left(t_{2}-t_{0}\right)}\ex^{-\im\left(\mu+V_{\gamma}\right)\left(t_{1}-t_{2}\right)}\right. \nonumber \\
& \left.\times\left[\bar{\Phi}\left(t_{2}-t_{1},\beta,-\left(\bar{\varepsilon}_{k}^{*}-\mu-V_{\gamma}-\Omega_{\gamma}\left(p_{1}r'+p_{2}s'\right)\right)\right)\right.\right.\nonumber \\
& \left.\left.-\bar{\Phi}\left(t_{2}-t_{1},\beta,-\left(\bar{\varepsilon}_{j}-\mu-V_{\gamma}-\Omega_{\gamma}\left(p_{1}r+p_{2}s\right)\right)\right)\right]\right.\nonumber \\
& \left.+\ex^{\im\bar{\varepsilon}_{k}^{*}\left(t_{2}-t_{1}\right)}\ex^{-\im\Omega_{\gamma}\left(p_{1}\left(r-r'\right)+p_{2}\left(s-s'\right)\right)\left(t_{1}-t_{0}\right)}\left[\Psi\left(\frac{1}{2}-\frac{\beta}{2\pi \im}\left(\bar{\varepsilon}_{k}^{*}-\mu-V_{\gamma}-\Omega_{\gamma}\left(p_{1}r'+p_{2}s'\right)\right)\right)\right.\right.\nonumber \\
& \left.\left.-\Psi\left(\frac{1}{2}+\frac{\beta}{2\pi \im}\left(\bar{\varepsilon}_{k}^{*}-\mu-V_{\gamma}-\Omega_{\gamma}\left(p_{1}r'+p_{2}s'\right)\right)\right)\right]\right].
\end{align}
We note that the digamma function $\Psi$ for complex arguments is readily available in the GSL numerical library~\cite{gsl}. The $\bar{\Phi}$ function can, instead, be related to the hypergeometric function ${}_2 F_1$ by~\cite{Tuovinen2019NJP}
\[\bar{\Phi}(\tau,\beta,z) = 2\pi\ex^{-\pi\tau/\beta}\frac{1}{\pi -\im\beta z}{}_2F_1\left(1, \frac{1}{2} - \frac{\im\beta z}{2\pi}, \frac{3}{2} - \frac{\im\beta z}{2\pi}, \ex^{-2\pi\tau/\beta}\right),\]
for which there are efficient numerical implementations available~\cite{hypf-impl}. Alternatively, it is possible to directly expand the digamma and the Hurwitz--Lerch functions in terms of the Pad{\'e} poles, see Eq.~\eqref{PADE_FERMI}.

The formula for the current in lead $\alpha$ can be similarly derived by inserting Eq.~\eqref{biharmonic_bias} into Eq.~\eqref{eq:I_alpha}, yielding the following expression~\cite{Ridley2017GNR}:
\begin{align}\label{Current_Bessel}
& I_{\alpha}\left(t\right) \nonumber \\
& = \frac{1}{\pi}\underset{j}{\sum}\left\{ 2\textrm{Re}\left[\frac{\left\langle \varphi_{j}^{\text{L}}\right|\mathbf{\Gamma}_{\alpha}\left|\varphi_{j}^{\text{R}}\right\rangle }{\left\langle \varphi_{j}^{\text{L}}\mid\varphi_{j}^{\text{R}}\right\rangle}\im\ex^{\im\frac{A_{\alpha}^{\left(1\right)}}{p_{1}\Omega_{\alpha}}\left(\sin\left(p_{1}\Omega_{\alpha}\left(t-t_{0}\right)+\phi_{\alpha}\right)-\sin\phi_{\alpha}\right)}\ex^{\im\frac{A_{\alpha}^{\left(2\right)}}{p_{2}\Omega_{\alpha}}\sin\left(p_{2}\Omega_{\alpha}\left(t-t_{0}\right)\right)}\right.\right.\nonumber \\
& \left.\left.\times\underset{r,s}{\sum}J_{r}\left(\frac{A_{\alpha}^{\left(1\right)}}{p_{1}\Omega_{\alpha}}\right)J_{s}\left(\frac{A_{\alpha}^{\left(2\right)}}{p_{2}\Omega_{\alpha}}\right)\ex^{-\im r\phi_{\alpha}}\left(\tcircle{1}_I + \tcircle{2}_I\right)\right]\right. \nonumber \\
& \left.-\underset{\gamma,k}{\sum}\frac{\left\langle \varphi_{k}^{\text{R}}\right|\mathbf{\Gamma}_{\alpha}\left|\varphi_{j}^{\text{R}}\right\rangle \left\langle \varphi_{j}^{\text{L}}\right|\mathbf{\Gamma}_{\gamma}\left|\varphi_{k}^{\text{L}}\right\rangle}{\left\langle \varphi_{j}^{\text{L}}\mid\varphi_{j}^{\text{R}}\right\rangle \left\langle \varphi_{k}^{\text{R}}\mid\varphi_{k}^{\text{L}}\right\rangle}\left[\tcircle{3}_I+\underset{r,s}{\sum}J_{r}\left(\frac{A_{\gamma}^{\left(1\right)}}{p_{1}\Omega_{\gamma}}\right)J_{s}\left(\frac{A_{\gamma}^{\left(2\right)}}{p_{2}\Omega_{\gamma}}\right)\left(\tcircle{4}_I + \tcircle{5}_I\right)\right.\right. \nonumber \\
& \left.\left.+\underset{r,r',s,s'}{\sum}J_{r}\left(\frac{A_{\gamma}^{\left(1\right)}}{p_{1}\Omega_{\gamma}}\right)J_{r'}\left(\frac{A_{\gamma}^{\left(1\right)}}{p_{1}\Omega_{\gamma}}\right)J_{s}\left(\frac{A_{\gamma}^{\left(2\right)}}{p_{2}\Omega_{\gamma}}\right)J_{s'}\left(\frac{A_{\gamma}^{\left(2\right)}}{p_{2}\Omega_{\gamma}}\right)\right.\right.\nonumber \\
& \left.\left.\times\frac{\ex^{-\im\left(r-r'\right)\phi_{\gamma}}}{\bar{\varepsilon}_{j}-\bar{\varepsilon}_{k}^{*}-\Omega_{\gamma}\left(p_{1}\left(r-r'\right)+p_{2}(s-s')\right)}\left(\tcircle{6}_I + \tcircle{7}_I\right)\right]\right\} 
\end{align}
with the introduced abbreviations
\begin{align}
\tcircle{1}_I & = \ex^{-\im\left(\bar{\varepsilon}_{j}-\mu-V_{\alpha}\right)\left(t-t_{0}\right)}\left[\bar{\Phi}\left(t-t_{0},\beta,-\left(\bar{\varepsilon}_{j}-\mu-V_{\alpha}-\Omega_{\alpha}\left(p_{1}r+p_{2}s\right)\right)\right)\right.\nonumber \\
& \left.-\bar{\Phi}\left(t-t_{0},\beta,-\left(\bar{\varepsilon}_{j}-\mu\right)\right)\right],
\end{align}
\begin{align}
\tcircle{2}_I & = \ex^{-\im\Omega_{\alpha}\left(p_{1}r+p_{2}s\right)\left(t-t_{0}\right)}\Psi\left(\frac{1}{2}-\frac{\beta}{2\pi \im}\left(\bar{\varepsilon}_{j}-\mu-V_{\alpha}-\Omega_{\alpha}\left(p_{1}r+p_{2}s\right)\right)\right),
\end{align}
\begin{align}
\tcircle{3}_I & = \frac{\ex^{-\im\left(\bar{\varepsilon}_{j}-\bar{\varepsilon}_{k}^{*}\right)\left(t-t_{0}\right)}}{\bar{\varepsilon}_{k}^{*}-\bar{\varepsilon}_{j}}\left[\Psi\left(\frac{1}{2}+\frac{\beta}{2\pi \im}\left(\bar{\varepsilon}_{k}^{*}-\mu\right)\right)-\Psi\left(\frac{1}{2}-\frac{\beta}{2\pi \im}\left(\bar{\varepsilon}_{j}-\mu\right)\right)\right],
\end{align}
\begin{align}
\tcircle{4}_I & = \frac{\ex^{-\im r\phi_{\gamma}}\ex^{\im\frac{A_{\gamma}^{\left(1\right)}}{p_{1}\Omega_{\gamma}}\sin\phi_{\gamma}}}{\bar{\varepsilon}_{j}-\bar{\varepsilon}_{k}^{*}-V_{\gamma}-\Omega_{\gamma}\left(p_{1}r+p_{2}s\right)} \left\{\ex^{-\im\left(\bar{\varepsilon}_{j}-\bar{\varepsilon}_{k}^{*}\right)\left(t-t_{0}\right)}\right.\nonumber\\
& \left.\times\left[\Psi\left(\frac{1}{2}+\frac{\beta}{2\pi \im}\left(\bar{\varepsilon}_{k}^{*}-\mu\right)\right)-\Psi\left(\frac{1}{2}+\frac{\beta}{2\pi \im}\left(\bar{\varepsilon}_{j}-\mu-V_{\gamma}-\Omega_{\gamma}\left(p_{1}r+p_{2}s\right)\right)\right)\right]\right.\nonumber \\
& \left.+\ex^{\im\left(\bar{\varepsilon}_{k}^{*}-\mu-V_{\gamma}-\Omega_{\gamma}\left(p_{1}r+p_{2}s\right)\right)\left(t-t_{0}\right)} \right.\nonumber \\
& \left.\times\left[\bar{\Phi}\left(t-t_{0},\beta,\bar{\varepsilon}_{k}^{*}-\mu\right)-\bar{\Phi}\left(t-t_{0},\beta,\bar{\varepsilon}_{j}-\mu-V_{\gamma}-\Omega_{\gamma}\left(p_{1}r+p_{2}s\right)\right)\right]\right\},
\end{align}
\begin{align}
\tcircle{5}_I & = \frac{\ex^{\im r\phi_{\gamma}}\ex^{-\im\frac{A_{\gamma}^{\left(1\right)}}{p_{1}\Omega_{\gamma}}\sin\phi_{\gamma}}}{\bar{\varepsilon}_{k}^{*}-\bar{\varepsilon}_{j}-V_{\gamma}-\Omega_{\gamma}\left(p_{1}r+p_{2}s\right)}\left\{\ex^{-\im\left(\bar{\varepsilon}_{j}-\bar{\varepsilon}_{k}^{*}\right)\left(t-t_{0}\right)}\right.\nonumber \\
&\left.\times\left[\Psi\left(\frac{1}{2}-\frac{\beta}{2\pi \im}\left(\bar{\varepsilon}_{j}-\mu\right)\right)-\Psi\left(\frac{1}{2}-\frac{\beta}{2\pi \im}\left(\bar{\varepsilon}_{k}^{*}-\mu-V_{\gamma}-\Omega_{\gamma}\left(p_{1}r+p_{2}s\right)\right)\right)\right]\right.\nonumber \\
& \left.+\ex^{-\im\left(\bar{\varepsilon}_{j}-\mu-V_{\gamma}-\Omega_{\gamma}\left(p_{1}r+p_{2}s\right)\right)\left(t-t_{0}\right)}\right.\nonumber \\
&\left.\times\left[\bar{\Phi}\left(t-t_{0},\beta,-\left(\bar{\varepsilon}_{j}-\mu\right)\right)-\bar{\Phi}\left(t-t_{0},\beta,-\left(\bar{\varepsilon}_{k}^{*}-\mu-V_{\gamma}-\Omega_{\gamma}\left(p_{1}r+p_{2}s\right)\right)\right)\right]\right\},
\end{align}
\begin{align}
\tcircle{6}_I & = \ex^{-\im\Omega_{\gamma}\left(p_{1}\left(r-r'\right)+p_{2}\left(s-s'\right)\right)\left(t-t_{0}\right)}\left[\Psi\left(\frac{1}{2}-\frac{\beta}{2\pi \im}\left(\bar{\varepsilon}_{j}-\mu-V_{\gamma}-\Omega_{\gamma}\left(p_{1}r+p_{2}s\right)\right)\right)\right.\nonumber \\
& \left.-\Psi\left(\frac{1}{2}+\frac{\beta}{2\pi \im}\left(\bar{\varepsilon}_{k}^{*}-\mu-V_{\gamma}-\Omega_{\gamma}\left(p_{1}r'+p_{2}s'\right)\right)\right)\right]\nonumber \\
& +\ex^{-\im\left(\bar{\varepsilon}_{j}-\bar{\varepsilon}_{k}^{*}\right)\left(t-t_{0}\right)}\left[\Psi\left(\frac{1}{2}+\frac{\beta}{2\pi \im}\left(\bar{\varepsilon}_{j}-\mu-V_{\gamma}-\Omega_{\gamma}\left(p_{1}r+p_{2}s\right)\right)\right)\right.\nonumber \\
&\left.-\Psi\left(\frac{1}{2}-\frac{\beta}{2\pi \im}\left(\bar{\varepsilon}_{k}^{*}-\mu-V_{\gamma}-\Omega_{\gamma}\left(p_{1}r'+p_{2}s'\right)\right)\right)\right],
\end{align}
\begin{align}
\tcircle{7}_I & = \ex^{\im \left(\bar{\varepsilon}_{k}^{*}-\mu-V_{\gamma}-\Omega_{\gamma}\left(p_{1}r+p_{2}s\right)\right)\left(t-t_{0}\right)}\left[\bar{\Phi}\left(t-t_{0},\beta,\bar{\varepsilon}_{j}-\mu-V_{\gamma}-\Omega_{\gamma}\left(p_{1}r+p_{2}s\right)\right)\right.\nonumber \\
& \left.-\bar{\Phi}\left(t-t_{0},\beta,\bar{\varepsilon}_{k}^{*}-\mu-V_{\gamma}-\Omega_{\gamma}\left(p_{1}r'+p_{2}s'\right)\right)\right] \nonumber \\
& +\ex^{-\im\left(\bar{\varepsilon}_{j}-\mu-V_{\gamma}-\Omega_{\gamma}\left(p_{1}r'+p_{2}s'\right)\right)\left(t-t_{0}\right)}\left[\bar{\Phi}\left(t-t_{0},\beta,-\left(\bar{\varepsilon}_{j}-\mu-V_{\gamma}-\Omega_{\gamma}\left(p_{1}r+p_{2}s\right)\right)\right)\right.\nonumber \\
& \left.-\bar{\Phi}\left(t-t_{0},\beta,-\left(\bar{\varepsilon}_{k}^{*}-\mu-V_{\gamma}-\Omega_{\gamma}\left(p_{1}r'+p_{2}s'\right)\right)\right)\right].
\end{align}

To evaluate the two-time current correlation function in Eq.~\eqref{Correlation_alpha_beta} for the biharmonic driving model, it is necessary to evaluate the frequency integrals in the $\boldsymbol{\Lambda}$ matrix expression of Eq.~\eqref{eq:lambda_matrix_1}~\cite{Ridley2017}:
\begin{align}\label{Lamba_plus}
& \mathbf{\Lambda}_{\beta}^{+}\left(t_{2},t_{1}\right) \nonumber \\
& =\underset{j}{\sum}\frac{\mathbf{\Gamma}_{\beta}\left|\varphi_{j}^{L}\right\rangle \left\langle \varphi_{j}^{R}\right|}{\left\langle \varphi_{j}^{R}\right|\left.\varphi_{j}^{L}\right\rangle }\left\{\phantom{-\frac{\im}{2\pi}\ex^{\im\bar{\varepsilon}_{j}^{*}\left(t_{1}-t_{0}\right)}}\right.\nonumber \\
&\left.-\frac{\im}{2\pi}\ex^{\im\bar{\varepsilon}_{j}^{*}\left(t_{1}-t_{0}\right)}\ex^{-\im\left(\mu+V_{\beta}\right)\left(t_{2}-t_{0}\right)}\ex^{-\im\frac{A_{\beta}^{\left(1\right)}}{p_{1}\Omega_{\beta}}\sin\left(p_{1}\Omega_{\beta}\left(t_{2}-t_{0}\right)+\phi_{\beta}\right)}\ex^{-\im\frac{A_{\beta}^{\left(2\right)}}{p_{2}\Omega_{\beta}}\sin\left(p_{2}\Omega_{\beta}\left(t_{2}-t_{0}\right)\right)}\right. \nonumber \\
& \left.\times\left[\tcircle{1}_{\Lambda^+}+\underset{r,s}{\sum}J_{r}\left(\frac{A_{\beta}^{\left(1\right)}}{p_{1}\Omega_{\beta}}\right)J_{s}\left(\frac{A_{\beta}^{\left(2\right)}}{p_{2}\Omega_{\beta}}\right)\ex^{\im r\phi_{\beta}}\left( \tcircle{2}_{\Lambda^+} + \tcircle{3}_{\Lambda^+} \right) \right] -\theta\left(t_{1}-t_{2}\right)\frac{\ex^{\im\bar{\varepsilon}_{j}^{*}\left(t_{1}-t_{2}\right)}}{2}\right\},
\end{align}
with the introduced abbreviations
\begin{align}
\tcircle{1}_{\Lambda^+} & = \ex^{\im\frac{A_{\beta}^{\left(1\right)}}{p_{1}\Omega_{\beta}}\sin\left(\phi_{\beta}\right)}\bar{\Phi}\left(\beta,t_{2}-t_{0},\bar{\varepsilon}_{j}^{*}-\mu\right),
\end{align}
\begin{align}
\tcircle{2}_{\Lambda^+} & = \theta\left(t_{1}-t_{2}\right)\ex^{-\im\left(\bar{\varepsilon}_{j}^{*}-\mu-V_{\beta}-\Omega_{\beta}\left(p_{1}r+p_{2}s\right)\right)\left(t_{2}-t_{0}\right)} \nonumber \\
& \times\left[\Psi\left(\frac{1}{2}-\frac{\beta}{2\pi \im}\left(\bar{\varepsilon}_{j}^{*}-\mu-V_{\beta}-\Omega_{\beta}\left(p_{1}r+p_{2}s\right)\right)\right)\right.\nonumber \\
& \left.-\Psi\left(\frac{1}{2}+\frac{\beta}{2\pi \im}\left(\bar{\varepsilon}_{j}^{*}-\mu-V_{\beta}-\Omega_{\beta}\left(p_{1}r+p_{2}s\right)\right)\right)\right]\nonumber \\
& -\bar{\Phi}\left(t_{2}-t_{0},\beta,\bar{\varepsilon}_{j}^{*}-\mu-V_{\beta}-\Omega_{\beta}\left(p_{1}r+p_{2}s\right)\right),
\end{align}
\begin{align}
\tcircle{3}_{\Lambda^+} & = \ex^{-\im\left(\bar{\varepsilon}_{j}^{*}-\mu-V_{\beta}-\Omega_{\beta}\left(p_{1}r+p_{2}s\right)\right)\left(t_{1}-t_{0}\right)} \nonumber \\
& \times\left[\theta\left(t_{1}-t_{2}\right)\bar{\Phi}\left(t_{1}-t_{2},\beta,-\left(\bar{\varepsilon}_{j}^{*}-\mu-V_{\beta}-\Omega_{\beta}\left(p_{1}r+p_{2}s\right)\right)\right)\right. \nonumber \\
& \left.+\theta\left(t_{2}-t_{1}\right)\bar{\Phi}\left(t_{2}-t_{1},\beta,\bar{\varepsilon}_{j}^{*}-\mu-V_{\beta}-\Omega_{\beta}\left(p_{1}r+p_{2}s\right)\right)\right].
\end{align}
Similarly for Eq.~\eqref{eq:lambda_matrix_2}:
\begin{align}\label{Lamba_minus}
& \mathbf{\Lambda}_{\alpha}^{-}\left(t_{1},t_{2}\right) \nonumber \\
& = \underset{j}{\sum}\frac{\boldsymbol{\Gamma}_{\alpha}\left|\varphi_{j}^{L}\right\rangle \left\langle \varphi_{j}^{R}\right|}{\left\langle \varphi_{j}^{R}\right|\left.\varphi_{j}^{L}\right\rangle }\left\{ \phantom{\frac{i}{2\pi}\ex^{\im\bar{\varepsilon}_{j}^{*}\left(t_{2}-t_{0}\right)}} \right.\nonumber \\
& \left.-\frac{\im}{2\pi}\ex^{\im\bar{\varepsilon}_{j}^{*}\left(t_{2}-t_{0}\right)}\ex^{-\im\left(\mu+V_{\alpha}\right)\left(t_{1}-t_{0}\right)}\ex^{-\im\frac{A_{\alpha}^{\left(1\right)}}{p_{1}\Omega_{\alpha}}\sin\left(p_{1}\Omega_{\alpha}\left(t_{1}-t_{0}\right)+\phi_{\alpha}\right)}\ex^{-\im\frac{A_{\alpha}^{\left(2\right)}}{p_{2}\Omega_{\alpha}}\sin\left(p_{2}\Omega_{\alpha}\left(t_{1}-t_{0}\right)\right)}\right. \nonumber \\
& \left.\times\left[\tcircle{1}_{\Lambda^-} +\underset{r,s}{\sum}J_{r}\left(\frac{A_{\alpha}^{\left(1\right)}}{p_{1}\Omega_{\alpha}}\right)J_{s}\left(\frac{A_{\alpha}^{\left(2\right)}}{p_{2}\Omega_{\alpha}}\right)\ex^{\im r\phi_{\alpha}}\left( \tcircle{2}_{\Lambda^-}+\tcircle{3}_{\Lambda^-}\right) \right] +\theta\left(t_{2}-t_{1}\right)\frac{\ex^{\im\bar{\varepsilon}_{j}^{*}\left(t_{2}-t_{1}\right)}}{2}\right\}
\end{align}
with the introduced abbreviations
\begin{align}
\tcircle{1}_{\Lambda^-} & = \ex^{\im\frac{A_{\alpha}^{\left(1\right)}}{p_{1}\Omega_{\alpha}}\sin\left(\phi_{\alpha}\right)}\bar{\Phi}\left(\beta,t_{1}-t_{0},\bar{\varepsilon}_{j}^{*}-\mu\right),
\end{align}
\begin{align}
\tcircle{2}_{\Lambda^-} & = \theta\left(t_{2}-t_{1}\right)\ex^{-\im\left(\bar{\varepsilon}_{j}^{*}-\mu-V_{\alpha}-\Omega_{\alpha}\left(p_{1}r+p_{2}s\right)\right)\left(t_{1}-t_{0}\right)} \nonumber \\
& \times\left[\Psi\left(\frac{1}{2}-\frac{\beta}{2\pi \im}\left(\bar{\varepsilon}_{j}^{*}-\mu-V_{\alpha}-\Omega_{\alpha}\left(p_{1}r+p_{2}s\right)\right)\right)\right.\nonumber \\
& \left.-\Psi\left(\frac{1}{2}+\frac{\beta}{2\pi \im}\left(\bar{\varepsilon}_{j}^{*}-\mu-V_{\alpha}-\Omega_{\alpha}\left(p_{1}r+p_{2}s\right)\right)\right)\right] \nonumber \\
& -\bar{\Phi}\left(t_{1}-t_{0},\beta,\bar{\varepsilon}_{j}^{*}-\mu-V_{\alpha}-\Omega_{\alpha}\left(p_{1}r+p_{2}s\right)\right),
\end{align}
\begin{align}
\tcircle{3}_{\Lambda^-} & = \ex^{-\im\left(\bar{\varepsilon}_{j}^{*}-\mu-V_{\alpha}-\Omega_{\alpha}\left(p_{1}r+p_{2}s\right)\right)\left(t_{2}-t_{0}\right)}\nonumber \\
&\times\left[\theta\left(t_{1}-t_{2}\right)\bar{\Phi}\left(t_{1}-t_{2},\beta,\bar{\varepsilon}_{j}^{*}-\mu-V_{\alpha}-\Omega_{\alpha}\left(p_{1}r+p_{2}s\right)\right)\right.\nonumber \\
& \left.+\theta\left(t_{2}-t_{1}\right)\bar{\Phi}\left(t_{2}-t_{1},\beta,-\left(\bar{\varepsilon}_{j}^{*}-\mu-V_{\alpha}-\Omega_{\alpha}\left(p_{1}r+p_{2}s\right)\right)\right)\right].
\end{align}

\section*{References}
\bibliographystyle{iopart-num}

\providecommand{\newblock}{}

\end{document}